\documentclass[a4paper,11pt]{article}
\pdfoutput=1 

\usepackage{jheppub} 
\usepackage[dvipsnames]{xcolor}
\usepackage[T1]{fontenc} 
\usepackage{cancel}
\usepackage{amsmath,amssymb,epsfig,ulem,color,slashed}
\usepackage{arydshln} 
\usepackage{tikz}
\usepackage{pgfplots}
\usepackage{textcomp}
\usepackage{yfonts}
\allowdisplaybreaks  
\allowbreak
\preprint{CERN-TH-2018-170, MITP/18-066}
\definecolor{nicered}{rgb}{0.7,0.1,0.1}
\definecolor{nicegreen}{rgb}{0.1,0.5,0.1}

\newcommand{\beq}{\begin{equation} }
\newcommand{\eeq}{\end{equation}} 
\DeclareMathOperator{\diag}{diag}

\title{\boldmath A clockwork solution to the flavor puzzle}
\dedicated{
\begin{center}\includegraphics[width=.45\textwidth]{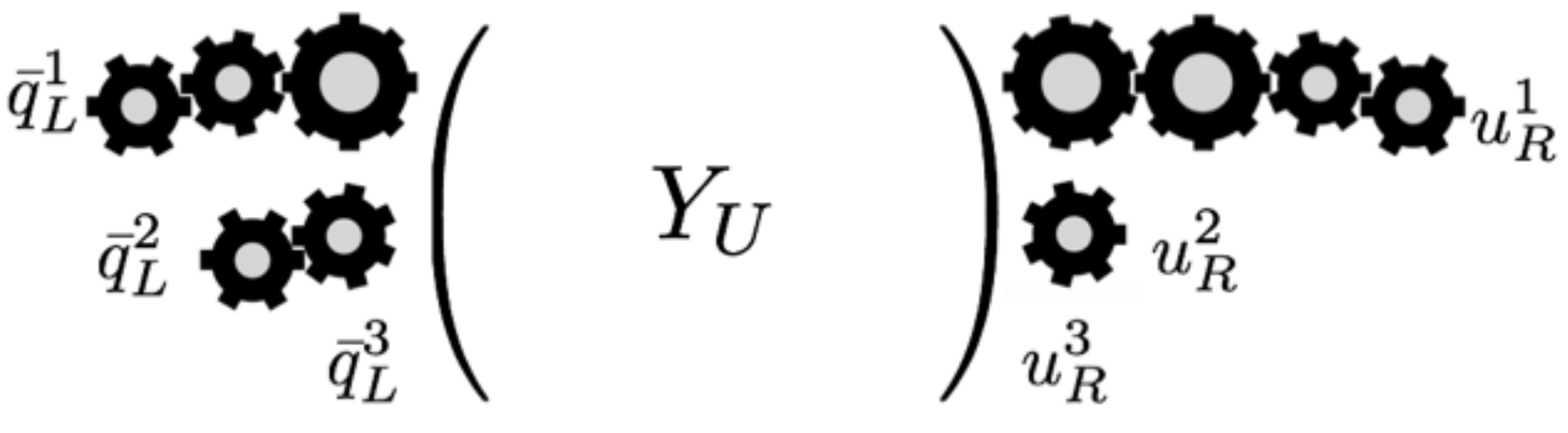}\hfill
\includegraphics[width=.45\textwidth]{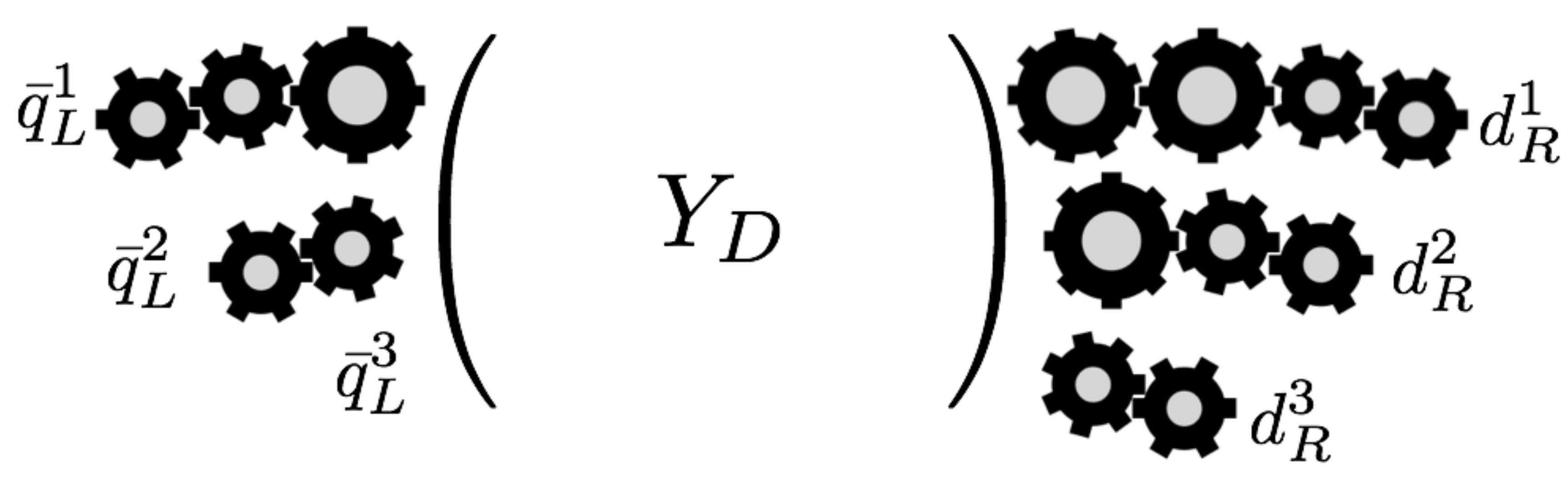}\end{center}
}

\def\Cincy{Department of Physics, University of Cincinnati, Cincinnati, Ohio 45221,USA} 
\def\CERN{Theoretical Physics Department, CERN, 1211 Geneva 23, Switzerland}
\def\IJS{Jo\v zef Stefan Institute, Jamova 39, 1000 Ljubljana, Slovenia}
\def\FMF{Faculty of Mathematics and Physics, University of Ljubljana, Jadranska 19, 1000 Ljubljana, Slovenia}
\def\Plymouth{Centre for Mathematical Sciences, Plymouth University, PL4-8AA Plymouth, UK}
\def\Mainz{PRISMA Cluster of Excellence \& Mainz Institute for Theoretical Physics, Johannes Gutenberg University, 55099 Mainz, Germany}
\def\RockyHeaven{Instituto de Astrof\'isica de Canarias, C/ V\'ia L\'actea, s/n E38205 - La Laguna (Tenerife), Espa\~na} 

\author[a]{\textbf{Rodrigo Alonso,}} 
\author[b]{\textbf{Adrian Carmona,}} 
\author[c]{\textbf{Barry M. Dillon,}} 
\author[d,e]{\textbf{Jernej F. Kamenik,}} 
\author[a,f]{\textbf{Jorge Martin Camalich,}} 
\author[g]{\textbf{and Jure Zupan}} 

\affiliation[a]{\CERN}
\affiliation[b]{\Mainz}
\affiliation[c]{\Plymouth}
\affiliation[d]{\IJS}
\affiliation[e]{\FMF}
\affiliation[f]{\RockyHeaven}
\affiliation[g]{\Cincy}

\emailAdd{rodrigo.alonso@cern.ch}
\emailAdd{adrian.carmona@uni-mainz.de}
\emailAdd{barry.dillon@plymouth.ac.uk}
\emailAdd{jernej.kamenik@ijs.si}
\emailAdd{jorge.martin.camalich@cern.ch}
\emailAdd{zupanje@ucmail.uc.edu}

\abstract{
We introduce a set of clockwork models of flavor that can naturally explain the large hierarchies of the Standard Model quark masses and mixing angles. Since the clockwork only contains chains of new vector-like fermions without any other dynamical fields, the flavor constraints allow for relatively light new physics scale. For two benchmarks with gear masses just above 1 TeV, allowed by flavor constraints, we discuss the collider searches and the possible ways of reconstructing gear spectra at the LHC. We also examine the similarities and differences with the other common solutions to the SM flavor puzzle, i.e., with the Froggatt-Nielsen models, where we identify a new {\it clockworked } version, and with the Randall-Sundrum models. 
}

\begin{document} 

\maketitle

\flushbottom

\section{Introduction}
\label{sec:introduction}
An outstanding puzzle in the Standard Model (SM) of particle physics is the origin of the observed hierarchies in the fermion masses and mixing angles, the so-called {\it SM flavor puzzle}.
There have been many attempts to address the SM flavor puzzle, either alone or in conjunction with solving the hierarchy problem, i.e., how to stabilize the Higgs mass against its sensitivity to a New Physics (NP) scale. Among the latter, more ambitious, models, the relevant examples include the Randall-Sundrum (RS) models~\cite{Randall:1999ee, Grossman:1999ra}, and the 4D dual approximately conformal models of composite Higgs  with partial compositeness~\cite{Dimopoulos:1981xc, Kaplan:1983fs, Kaplan:1991dc, Contino:2003ve, Agashe:2004rs}. 
These models typically exploit the fact that the mass hierarchy between the lightest and the heaviest SM fermion is exponentially large, as is  the hierarchy  between the Planck and the electroweak (EW) scales. The  canonical representative of the models that explain only the SM flavor structure, and do not solve the hierarchy problem, are the Froggatt-Nielsen (FN) models~\cite{Froggatt:1978nt, Leurer:1992wg, Leurer:1993gy}, based on horizontal abelian flavor symmetries.

The main purpose of this work is to explore how the SM flavor puzzle can be solved within the framework of a {\it clockwork theory}.
Originally presented in the context of axion physics~\cite{Choi:2015fiu, Kaplan:2015fuy}, and later generalized to a broader context in Ref.~\cite{Giudice:2016yja}, the clockwork provides a natural mechanism for obtaining large hierarchies in couplings or scales, and has already been successfully applied to the hierarchy problem~\cite{Giudice:2016yja, Craig:2017cda, Giudice:2017suc, Giudice:2017fmj}. Even though the clockwork mechanism itself is four-dimensional, 
it can, in some cases, be viewed as a deconstructed version of a higher dimensional theory. For example, the model addressing the hierarchy problem can be related to the five-dimensional (5D) linear dilaton model~\cite{Antoniadis:2001sw, Antoniadis:2011qw, Baryakhtar:2012wj, Cox:2012ee}, motivated by the six-dimensional strongly coupled duals~\cite{Aharony:1998ub, Giveon:1999px} of Little String Theory ~\cite{Berkooz:1997cq, Seiberg:1997zk}. The clockwork mechanism has also been used in contexts extending beyond the hierarchy problem, see e.g. Refs.~\cite{Kehagias:2016kzt, Ahmed:2016viu, Coy:2017yex, Hong:2017tel, Park:2017yrn, Lee:2017fin, Ibanez:2017vfl, Kehagias:2017grx, Ibarra:2017tju, Patel:2017pct, Choi:2017ncj, Teresi:2018eai, Kim:2018xsp, Niedermann:2018lhx, Agrawal:2018mkd,Goudelis:2018xqi}.

In what follows, we show that the clockwork mechanism can also successfully address the SM flavor puzzle. It can reproduce the hierarchy of quark masses and mixing angles with anarchic Yukawa couplings thanks to the hierarchical `overlaps' of the chiral fermion modes with the Higgs field.  We identify two particular limits in which this solution shares some similarities with the existing FN and RS solutions to the flavor puzzle. We show that a certain limit of clockwork may correspond to a novel realization of the FN mechanism in which the chiral fermions do not carry horizontal charges while the hierarchy of  the flavon vevs and the Dirac mass parameters is reversed.
On the other hand, the flavor clockwork model has no 5D continuous limit so that the connection with the RS model is only very approximate, at best at the level of first fermionic KK states.

The relevance of clockwork for flavor physics has been explored before, in Refs.~\cite{Ibarra:2017tju,Patel:2017pct}, though with little overlap with the present work. Ref.~\cite{Ibarra:2017tju} only dealt with neutrino masses, while we focus on the quark sector.  Ref.~\cite{Patel:2017pct}, while focusing on the charged fermion sector, considered a direction orthogonal to the one explored in this work, closer to the investigation of the relevance of random matrix theory for flavor~\cite{vonGersdorff:2017iym}. Furthermore, it did not consider phenomenological consequences -- neither at colliders nor in low energy experiments, which constitute a major part of our work.

The remainder of the paper is organized as follows. In Sec.~\ref{sec:flavor:hierarchy} we show how flavor hierarchies can arise from the discrete clockwork mechanism, starting with a single fermion and then generalizing to three generations, including the discussion of how the clockwork mechanism could arise dynamically. This section also addresses the matter of gauge Landau poles, as well as the perturbativity and stability of the Higgs potential in presence of a large number of additional fermions coupled to the SM. In Sec.~\ref{sec:flavor} we derive the flavor constraints on the clockwork models of flavor, while Sec.~\ref{sec:collider} contains the collider physics considerations, both the present constraints on the clockwork gears, as well as a tentative proposal for how the gear spectra could be reconstructed in case of a discovery. Our conclusions are given in Sec.~\ref{sec:conclusions}, while appendices contain a detailed discussion of the phenomenological challenges with the continuum limit of the clockwork mechanism (App.~\ref{sec:continuum}), the details on the matching of the dynamical fermionic clockwork onto the SM effective field theory (App.~\ref{sec:matching}),

\section{Flavor hierarchies from a discrete clockwork}
\label{sec:flavor:hierarchy}
We begin with a discrete version of clockwork and show how this can lead to hierarchical SM fermion mass parameters (challenges facing a continuum version are discussed in App.~\ref{sec:continuum}). We then discuss differences and similarities with two other mechanisms of generating quark flavor hierarchies -- the FN and the RS models of flavor.

\subsection{Clockworking a single fermion}\label{sec:1fcw}
 \begin{figure}[t]
\begin{center}
\includegraphics[width=7.5cm]{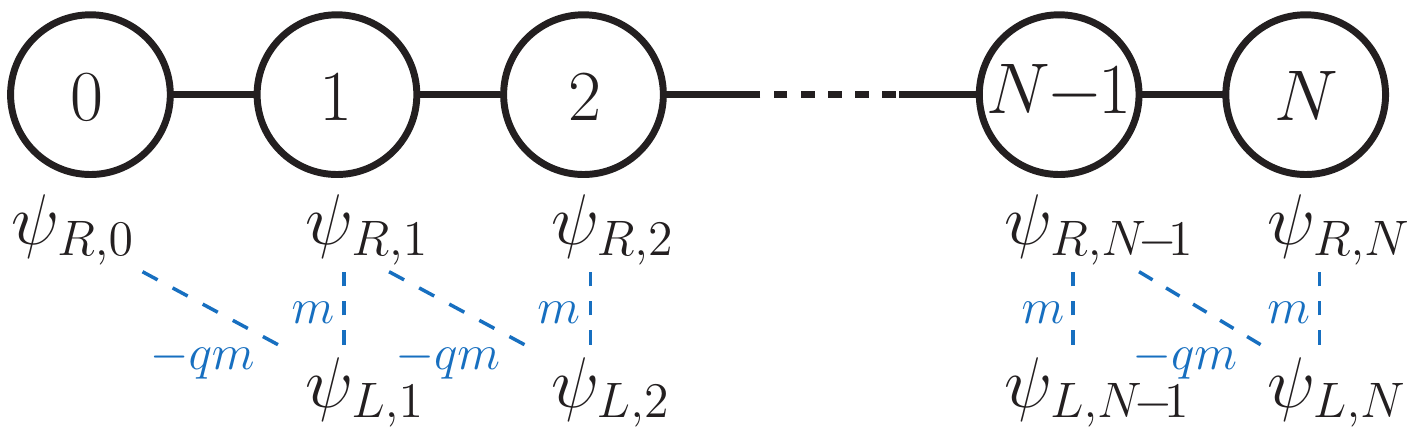}
\caption{\label{fig:single:chain} A single clockwork chain with a chiral fermion, $\psi_{R,0}$, on the $0$-th node, and vector-like fermion pairs, $\psi_{R,i}, \psi_{L,i}$, on the other $N$  nodes. The pattern of mass couplings is denoted in blue. }
\end{center}
\end{figure}

We start by reviewing the clockwork mechanism for a single right-handed chiral fermion, $\psi_R$ (see also Fig.~\ref{fig:single:chain}). The fermion $\psi_R$ interacts with an $N$-node  chain of vector-like fermions with mass terms, $m$, on each of the nodes, and a series of nearest neighbour mass terms, $q m$, between the nodes, 
\beq
\label{eq:L_psiR}
{\cal L}_{\psi_R}= i\sum_{j=0}^{N} \bar \psi_{R,j} \slashed D \psi_{R,j} + i\sum_{j=1}^{N} \bar \psi_{L,j} \slashed D \psi_{L,j} - m \sum_{j=1}^{N}\big(\bar \psi_{L,j} \psi_{R,j}- q\bar \psi_{L,j} \psi_{R,j-1}\big) +{\rm h.c.}, 
\eeq 
where for notational simplicity we have identified $\psi_{R,0}\equiv \psi_R$. The chains of fermions $\psi_{R,j}$ and $\psi_{L,j}$  carry the same gauge quantum numbers as $\psi_{R,0}$. 
The covariant derivatives are thus the same for fermions on all the nodes. For successful clockworking one requires $q>1$. 

The $N\times(N+1)$ mass matrix, 
\begin{align}
	\label{eq:matrix}
{\cal M}_{\psi}=m\,
 \begin{pmatrix}
 -q & 1   & 0 &  \ldots & 0 \\ 
 0 & -q & 1 &   \ldots & 0 \\  
 \vdots & \ddots&  \ddots & \ddots & 0 \\
 0 & \cdots &0 & -q &  1\\ 
\end{pmatrix},
\end{align}
is diagonalized by the unitary rotations, $\diag(0,M_1,\ldots,M_N)=(V^{L})^T {\cal M}_\psi V^R$. This gives one zero mode -- a right-handed chiral fermion $\psi_{R,0}'$ with mass $M_0=0$,
\beq
\label{eq:singleVR0}
\psi_{R,0}'=\sum_{j=0}^{N}V_{j0}^{R} \psi_{R,j},
\eeq
 and $N$ Dirac fermion mass-eigenstates -- the gears,  
 \beq
 \label{eq:field:rotation}
 \psi_{R,k}'=\sum_{j=0}^{N} V_{jk}^{R} \psi_{R,j}, \qquad \psi_{L,k}'=\sum_{j=1}^{N} V_{jk}^{L} \psi_{L,j}, \qquad k=1,\ldots,N,
 \eeq
  with nonzero masses 
\begin{align}
	M_{k}^2=m^2\Big(1+q^2-2q\cos\Big(\frac{k\pi}{N+1}\Big)\Big).
\end{align}
For $q \gtrsim \mathcal O(1)$ there is an ${\mathcal O}(m)$ mass gap between the  gears and the zero mode, with the mass splittings between two adjacent gears scaling in the large $N$ limit as $\mathcal{O}(m/N)$. More precisely, in the large $N$ limit the mass of the first gear is $M_1\simeq m(q-1)$, while the mass splitting between the heaviest and the lightest gear is, $M_N-M_1\simeq 2m$. This means that for  $q\to 1$ the clockwork chain contains modes much lighter than $M$, 
with the mass of the first gear $M_1\to 0$. On the other hand, when $q\gg 1$, all the gears have masses of roughly ${\mathcal O}(qm)$. In this case the spectrum of the gears is compressed in a $2m$ band around $qm$, with $(M_N-M_1)\ll M_1$.

The $N\times N$ left-handed rotation matrix in Eq.~\eqref{eq:field:rotation}, $V^L$,  is given by
\beq
V^L_{jk}=-\sqrt{\frac{2}{N+1}}\sin\frac{(N-j+1)k\pi}{N+1},~~j,k=1,\ldots,N,
\label{eq:VLjk}
\eeq
while the $(N+1)\times (N+1)$ right-handed rotation matrix in Eq.~\eqref{eq:field:rotation}, $V^R$, has the following entries, for $ j=0,\ldots,N$,
\begin{align}
V^R_{j0}&=\frac{\mathcal{N}_0}{q^{N-j}},
\label{eq:VRj0}
\\
V^R_{jk}&=\mathcal{N}_k\left( q\sin\frac{(N-j)k\pi}{N+1}-\sin\frac{(N-j+1)k\pi}{N+1}  \right),~~~ \,k=1,\ldots,N,
\label{eq:VRjk}
\end{align}
where the pre-factors are given by
\begin{align}
	\mathcal{N}_0=\sqrt{\frac{q^2-1}{q^2-q^{-2N}}},~~~~~\mathcal{N}_k=\sqrt{\frac{2}{(N+1)}}\,\frac{M}{M_k}.
\end{align}

 \begin{figure}[t]
\begin{center}
\includegraphics[width=8.cm]{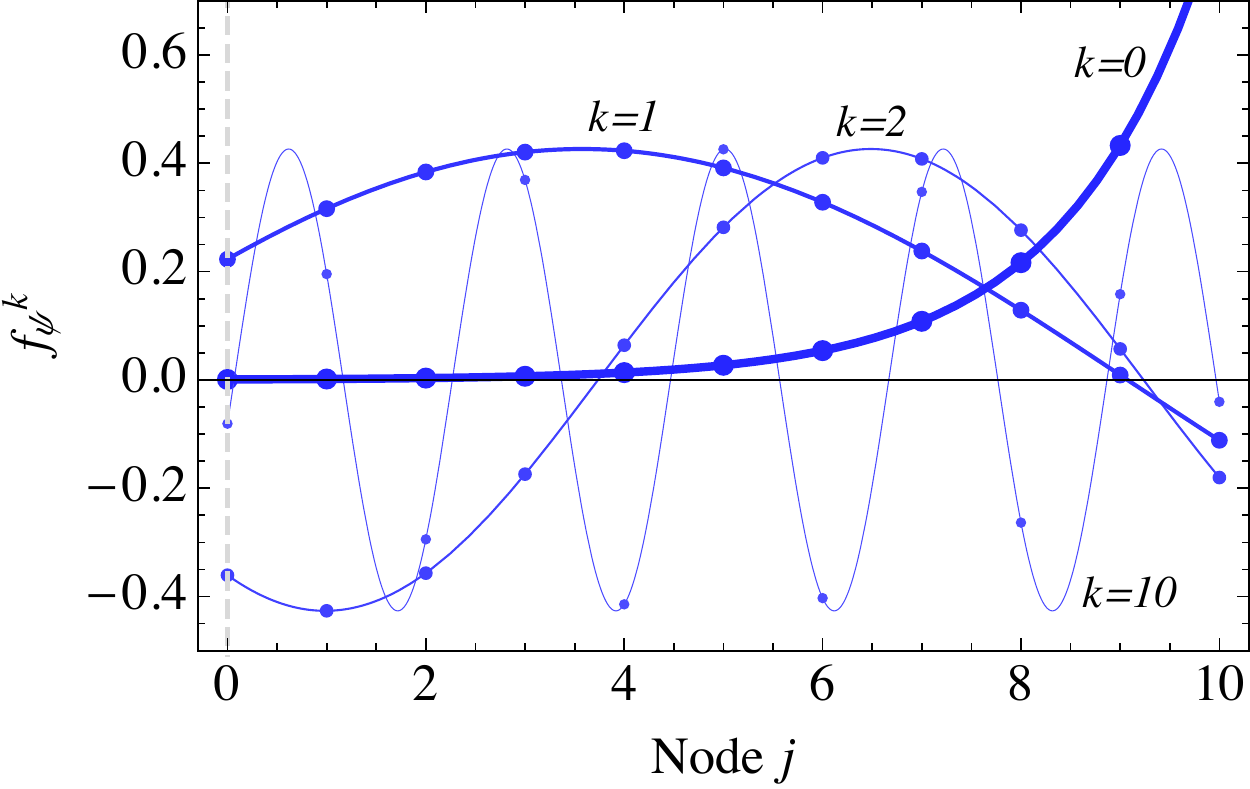}
\caption{\label{fig:profiles} The profiles of the zero mode, $\psi_{R,0}'$ ($k=0$), and the clockwork gears, $\psi'_{R,k}$ ($k=1,2,10$ from thicker to thiner lines),  in the case of clockworking a single fermion, $\psi_R$, for $N=10$ nodes with $q=2$. The values of the profiles on each of the $N+1$ nodes are denoted with a blue dot. 
}
\end{center}
\end{figure}

The entries in the $0$-th column of the $V^R$ rotation matrix, $V^R_{j0}$, can be interpreted as the profile of the zero mode $\psi_{R,0}'$ on the $j$-th node. For $q>1$ the profile of the zero mode is monotonically increasing from $j=0$ to $j=N$, see Fig.~\ref{fig:profiles}.  For future reference we denote the value of the zero mode on the $0$-th node as $f_\psi$. For $q,N\gg1 $ it is exponentially suppressed,
\beq
\label{eq:fpsi^0}
f_\psi\equiv V_{00}^R=
\left\{
\begin{matrix}
&\sim {1}/{q^N}, &\qquad q\gg 1;
\\
&\frac{1}{\sqrt{1+N}},  &\qquad q\to1.
\end{matrix}
\right.
\eeq
 This suppression will  be the origin of the SM quark mass hierarchy once we introduce the SM Higgs which is confined to couple only to the $0$-th node. Similarly, the $j$-th entry in the $k$-th column of the $V^R$ rotation matrix, $V^R_{jk}$, gives the profile of the $k$-th clockwork gear on $j$-th node. In particular, the profile of the $k$-th clockwork takes the following value on the $0$-th node
 \beq
 f_\psi^k\equiv V_{0k}^R
 =\sqrt{\frac{2}{N+1}}q \sin\Big(\frac{\pi k}{1+1/N}\Big) \frac{1}{|q- e^{i \pi k/(N+1)}|} \overset{\overset{N\gg k}{ q\gg 1}}{=}(-1)^{k+1}\sqrt{\frac{2}{N}}\frac{\pi k}{N}.
 \eeq
Unlike the zero mode, the profiles of the gears are not exponentially suppressed on the $0$-th node, even when $q,N\gg 1$. A useful relation that the profiles of these gears fulfill is the unitarity relation
\beq
\label{eq:unitarity}
\sum_{k=1}^{N} (f_\psi^k)^2=1-(f_\psi)^2=1-\mathcal{O}(1/q^{2N}).
\eeq

Clockworking  a single left-handed fermion, $\psi_L$, proceeds along exactly the same lines, but exchanging  $L\leftrightarrow R$ everywhere. For instance, one has now $N+1$ left-handed $\psi_{L,j}$ fields, where $j=0,\ldots,N$, identifying $\psi_{L,0}\equiv \psi_L$. There are $N$ right-handed fields, $\psi_{R,j}$, where $j=1,\ldots, N$, so that on the $N$ nodes one has vector-like fermions. 
After diagonalization the left-handed zero mode profile is given by $V_{j0}^L$ with the entries given in Eq.~\eqref{eq:VRj0}. The profile of the $k$-th left-(right-)handed gear is given by $V_{jk}^{L(R)}$ with entries given in Eq.~\eqref{eq:VRjk} (in Eq.~\eqref{eq:VLjk}).

\subsection{Three generations and the solution to the SM flavor puzzle}\label{CW3gen}
 \begin{figure}[t]
\begin{center}
\includegraphics[width=11.cm]{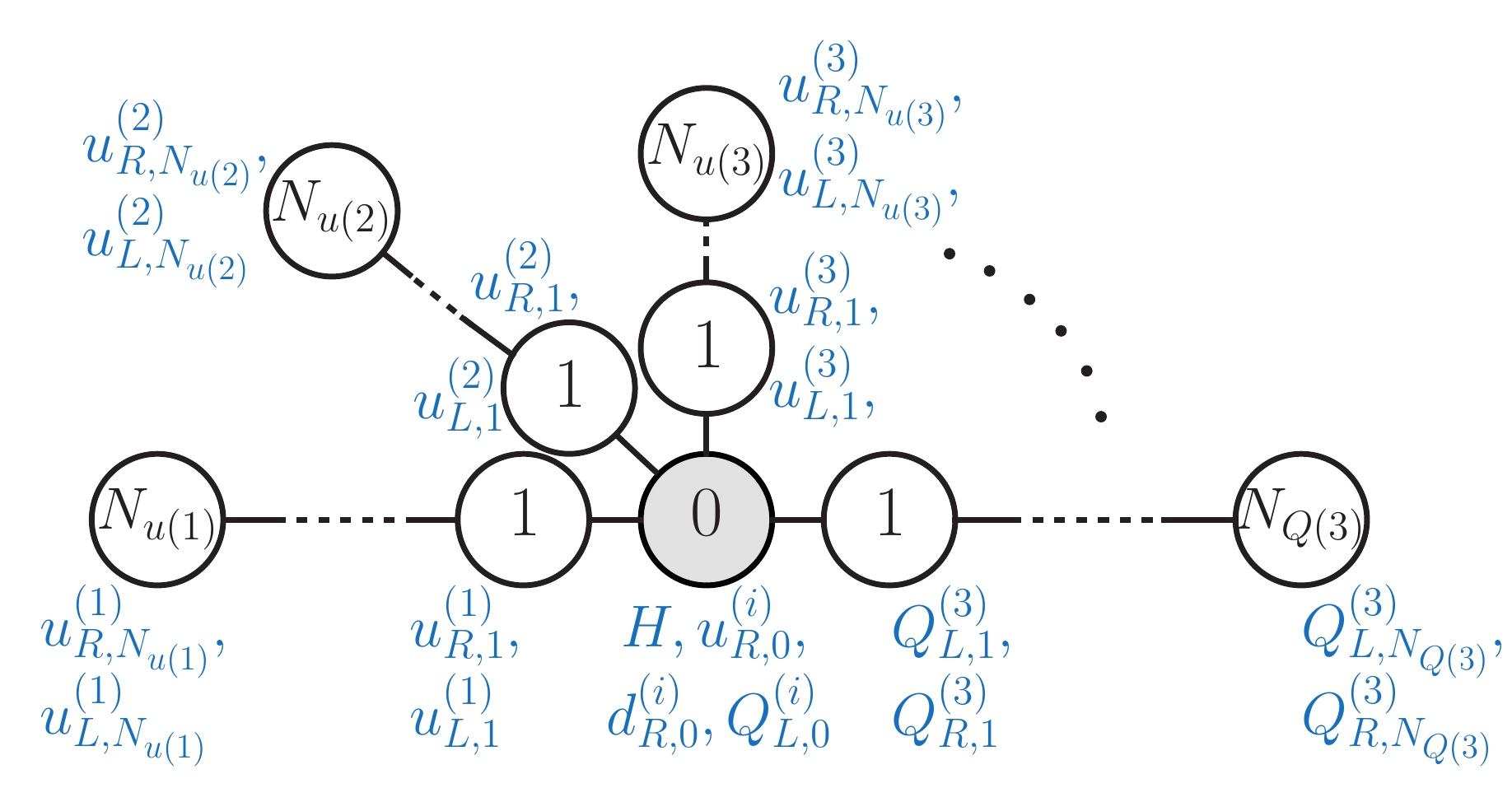}
\caption{\label{fig:clockwork:diagram} The clockwork chains of vector-like fermions for each of the flavors meet at the central node, the only one that contains the Higgs.  The field content of each node is denoted in blue.}
\end{center}
\end{figure}

We are now ready to introduce the set-up that explains the hierarchy of SM quark masses through the clockworking mechanism. Each of the SM fermions, $\psi_i$, where $i=1,2,3$, is the generation index, is supplemented by an $N_{\psi_i}$-node chain of 
vector-like fermions with the same quantum numbers. That is, for each SM $\psi_i$ one has a clockwork Lagrangian as in Eq. \eqref{eq:L_psiR}. In addition, the SM Higgs resides on the $0$-th node, coupling the fermions on the $0$-th node through 
Yukawa interactions, see Fig. \ref{fig:clockwork:diagram}.
For instance, the three families of right-handed up quarks, $u_{R}^{(i)}\equiv u_{R,0}^{(i)} $, $i=1,2,3$, residing on the $0$-th node,  are supplemented by the corresponding vector-like partners $u_{R,k}^{(i)}, u_{L,k}^{(i)}$, on the nodes $k=1,\ldots,N_{u(i)}$. Similarly, the right-handed down quarks, $d_R^{(i)}$ and the left-handed doublets, $Q_L^{(i)}$ are supplemented by their own vector-like chains. In general, the chains are of different lengths, $N_{\psi^{(i)}}$.

The Lagrangian for three generations is thus given by
\beq
\begin{split}
\label{eq:LagSMCK0}
{\cal L}=&\sum_{i=1}^3 \Big({\cal L}_{u_R^{(i)}}+{\cal L}_{d_R^{(i)}}+{\cal L}_{Q_L^{(i)}}\Big)\\
&-\sum_{i,j=1}^3\Big[\big(Y_D\big)_{ij}\,\bar Q_{L,0}^{( i)} H \,d_{R,0}^{(j)}+\big(Y_U\big)_{i j} \,\bar Q_{L,0}^{(i)} \tilde H u_{R,0}^{(j)}+\text{h.c.}\Big],
\end{split}
\eeq
where ${\cal L}_{u_R^{(i)}},{\cal L}_{d_R^{(i)}},{\cal L}_{Q_L^{(i)}}$ are given in Eq.~\eqref{eq:L_psiR} with obvious replacements in the notation. Each of the clockworking Lagrangians ${\cal L}_{u_R^{(i)}},{\cal L}_{d_R^{(i)}},{\cal L}_{Q_L^{(i)}}$ comes with a separate mass gap parameter, $m_{u(i)}, m_{d(i)}, m_{Q(i)}$ and the clockworking factor, $q_{u(i)}, q_{d(i)}, q_{Q(i)}$.\footnote{This is not the most general possibility as the masses and clockworking factors can be non-universal within a chain, and also have off-diagonal entries, a possibility that we briefly discuss in the conclusions, Sec. \ref{sec:conclusions}.} 
In the following we keep the clockworking factors $q_{\psi^{(i)}}$ and lengths of the chains, $N_{\psi(i)}$,
flavor-dependent and study  the different possibilities to induce flavor hierarchies in the quark sector.

After electroweak symmetry breaking the Yukawa interactions lead to a mass term for the zero modes. We use the unitary gauge, $H=\big(0, (v+h)/\sqrt2\big)$, with $v=246$ GeV. The zero modes are identified with the SM fermions. To leading order in $v^2/M^2$ expansion the SM Higgs Yukawa matrices are given by the products of zero mode overlaps with the $0$-th node, $f_\psi$,
\begin{align}
\label{eq:YuSM}
\left(Y_u^{\rm SM}\right)_{ij}&=f_{Q(i)}\left(Y_U\right)_{ij}f_{u(j)} \sim q^{-N_{Q(i)}}_{Q(i)}\left(Y_U\right)_{ij}q^{-N_{u(j)}}_{u(j)}, 
\\
\label{eq:YdSM}
\left(Y_d^{\rm SM}\right)_{ij}&= f_{Q(i)} \left(Y_D\right)_{ij}f_{d(j)} \sim q^{-N_{Q(i)}}_{Q(i)}\left(Y_D\right)_{ij}q^{-N_{d(j)}}_{d(j)}.
\end{align}
Here, there is no summation over $i,j=1,2,3$, while for each of the zero mode overlaps one needs to use the appropriate clockworking factor $q_{u(i)},~q_{d(i)},~q_{Q(i)}$ and chain lengths $N_{u(i)},~N_{d(i)},~N_{Q(i)}$ in Eqs. \eqref{eq:VRj0}, \eqref{eq:fpsi^0}.
The SM Yukawas give the SM quark mass matrices, as in the SM,
\beq\label{eq:SM:mass}
\left({\cal M}_u^{\rm SM}\right)_{ij}=\frac{v}{\sqrt2}\left(Y_u^{\rm SM}\right)_{ij}, \qquad 
\left({\cal M}_d^{\rm SM}\right)_{ij}=\frac{v}{\sqrt2}\left(Y_d^{\rm SM}\right)_{ij}.
\eeq
The ${\mathcal O}(v^2/M^2)$ corrections to the above expressions will be discussed below.

The hierarchy of quark masses is naturally obtained  if 
\begin{align}
q^{-N_{Q(1)}}_{Q(1)}\ll q^{-N_{Q(2)}}_{Q(2)}\ll q^{-N_{Q(3)}}_{Q(3)},\\
q^{-N_{u(1)}}_{u(1)}\ll q^{-N_{u(2)}}_{u(2)}\ll q^{-N_{u(3)}}_{u(3)},\\
q^{-N_{d(1)}}_{d(1)}\ll q^{-N_{d(2)}}_{d(2)}\ll q^{-N_{d(3)}}_{d(3)},
\end{align}
 so that there is the corresponding hierarchy between the zero mode overlaps. The above hierarchy is easy to achieve by choosing appropriately the $q_i$ and $N_i$ factors, while keeping $Y_U, Y_D$ still anarchic. Two limits are especially illuminating, when comparing to the other solutions of the SM flavor puzzle:
 \begin{itemize}
 \item
 {\it The universal $q$ limit} (or {\it the FN limit}) of clockwork is when all the clockwork factors are the same, $q_{Q(i)}=q_{u(i)}=q_{d(i)}\sim {\mathcal O}({\rm few})$, while 
 \beq\label{eq:FNlimit}
 N_{Q(1)}\gg N_{Q(2)} \gg N_{Q(3)},
 \eeq
  and similarly for up and down right-handed quarks. 
\item  
{\it The universal $N$ limit} (or {\it the RS limit}) is approached when
 \beq
 \label{eq:RSlimit}
 q_{Q(1)}\gg q_{Q(2)}\gg q_{Q(3)}, 
 \eeq
 and similarly for up and down right-handed quarks, while all the clockwork chains have the same length, $N_{Q(i)}=N_{u(i)}=N_{d(i)}\sim {\mathcal O}({\rm few})$. 
 \end{itemize}

\begin{figure}[t]
\begin{center}
\includegraphics[width=9.cm]{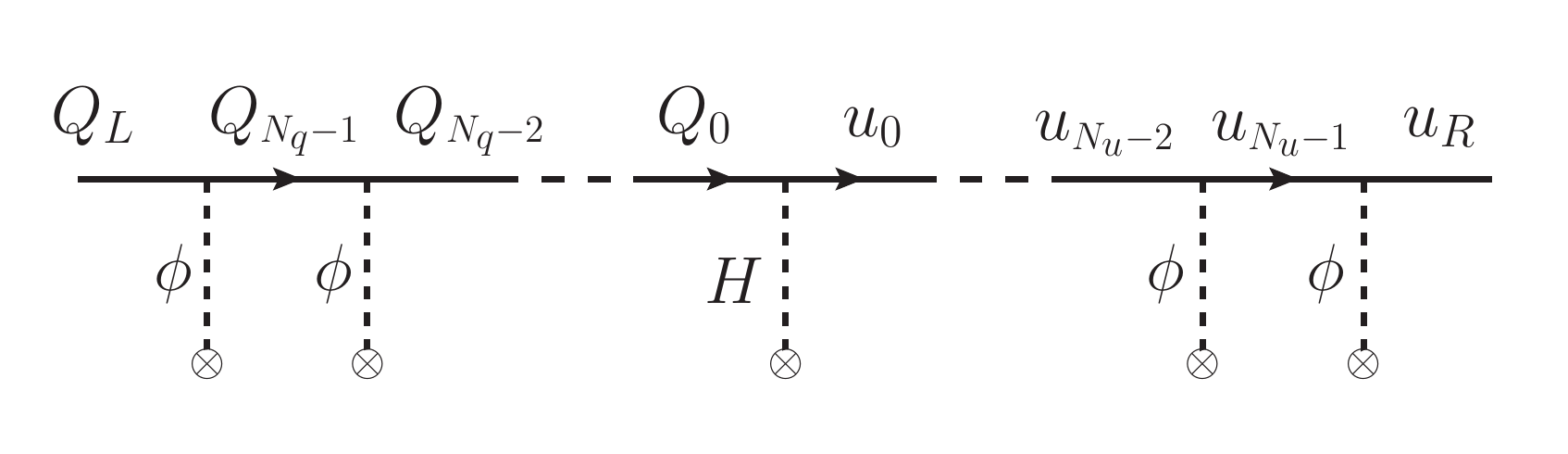}
\caption{\label{fig:FN:Feynman} An example of a Feynman diagram that generates the hierarchical quark masses in FN models. 
}
\end{center}
\end{figure}

 The two limits of the clockwork correspond, but are not entirely equivalent, to the two well known solutions of the SM flavor puzzle, the FN and the RS models of flavor, respectively. We discuss this further in Sec. \ref{subsec:UVflavor}.
 
  In both of the above limits we take $Y_U$ and $Y_D$ to be anarchic $3\times 3$ complex matrices. The SM quark mass matrices \eqref{eq:SM:mass} are diagonalized by bi-unitary transformations, $\diag(\overline{m}_u)=L_u\mathcal M_u^{\rm SM} R_u^\dagger$, $\diag(\overline{m}_d)=L_d \mathcal M_d^{\rm SM} R_d^\dagger$. The entries of the rotation matrices are given by the ratios of the zero mode profiles on the $0$-th node. 
  
  For the off-diagonal elements, $ i<j$, one has,
 \begin{subequations}
\label{eq:Rotshierarchy}
\begin{align}
|L_{u,d}|_{ij}&\sim |L_{u,d}|_{ji}
\sim \frac{f_{Q(i)}}{f_{Q(j)}}\sim \frac{\big(q_{Q(j)}\big)^{N_{Q(j)}}}{\big(q_{Q(i)}\big)^{N_{Q(i)}}}, 
\\
|R_u|_{ij}&\sim |R_u|_{ji}\sim \frac{f_{u(i)}}{f_{u(j)}}\sim \frac{\big(q_{u(j)}\big)^{N_{u(j)}}}{\big(q_{u(i)}\big)^{N_{u(i)}}}, 
\\
|R_d|_{ij}&\sim |R_d|_{ji}\sim \frac{f_{d(i)}}{f_{d(j)}}\sim \frac{\big(q_{d(j)}\big)^{N_{d(j)}}}{\big(q_{d(i)}\big)^{N_{d(i)}}},
\end{align}
\end{subequations}
while the diagonal elements are close to unity. Since the CKM matrix is given by
\beq
V_{\rm CKM}=L_u L_d^\dagger,
\eeq
this fixes the ratios
\beq
\frac{f_{Q(1)}}{f_{Q(2)}}\sim \lambda, \qquad \frac{f_{Q(2)}}{f_{Q(3)}} \sim \lambda^2,
\eeq
where $\sin\theta_C\simeq \lambda=|V_{us}|\simeq0.23$, with  $\theta_C$ the Cabibbo mixing angle. 

The SM quark mass eigenvalues are given by 
\beq
\begin{split}
	\overline{m}_{u(i)}&\sim v~f_{Q(i)}f_{u(i)}, 
\\
	\overline{m}_{d(i)}&\sim v~f_{Q(i)}f_{d(i)}.
\end{split}
\eeq
Taking as the parametric scaling of the quark masses (see also Sec.~\ref{sec:scans}),
\beq
\overline{m}_u\sim \lambda^7,\, \overline{m}_c\sim \lambda^3,\, \overline{m}_t\sim 1, \qquad 
\overline{m}_d\sim \lambda^7,\, \overline{m}_s\sim \lambda^5,\, \overline{m}_b\sim \lambda^2,
\eeq
the zero mode overlaps are required to be
\beq
\begin{split}
\label{eq:q:scaling}
	q_{Q(1)}^{-N_{Q(1)}}&\sim \lambda^3, \quad q_{Q(2)}^{-N_{Q(2)}}\sim \lambda^2, \quad~ q_{Q(3)}^{-N_{Q(3)}}\sim 1,
\\
	q_{u(1)}^{-N_{u(1)}}&\sim \lambda^4, \quad q_{u(2)}^{-N_{u(2)}}\sim \lambda, \quad\,\,\, q_{u(3)}^{-N_{u(3)}}\sim 1,
\\
	q_{d(1)}^{-N_{d(1)}}&\sim \lambda^4, \quad q_{d(2)}^{-N_{d(2)}}\sim \lambda^3, \quad~ q_{d(3)}^{-N_{d(3)}}\sim \lambda^2.
\end{split}
\eeq

Note that the above clockwork scenario can still provide a solution to the hierarchy problem, if we introduce an additional node chain for the graviton to induce a clockworking effect for the gravitational coupling.
In this case the SM and any clockwork extension of the fermion sector would be coupled to the $0$-th site of the clockwork-gravity model. {In the 5D picture, all the fermions would then have to be localized on a $N_{u(1)}+N_{u(2)}+\ldots + N_{Q(3)}+1$ stack of overlapping branes while only gravity propagates in the bulk.}

\subsection{Dynamical completions for clockwork models of flavor}
\label{subsec:UVflavor}
In this subsection we discuss the connection between the FN and RS models of flavor and the clockwork models in the two limits, Eqs. \eqref{eq:FNlimit} and \eqref{eq:RSlimit}.
In FN the flavor puzzle is solved by introducing a new $U(1)_H$ flavor symmetry, and a set of new fields, including a flavon scalar field, $\phi$. In the traditional FN models the chiral SM fermions carry integer generation-dependent $U(1)_H$ charges, $N_{u(i)}, N_{d(i)}$ and $-N_{Q(i)}$, whereas the flavon $\phi$ has charge $-1$. The $U(1)_H$ symmetry is broken spontaneously by the  flavon vev, $\langle \phi\rangle$, yet the original high energy symmetry preserving action leaves its imprint at low scales, dictating the form of the SM Yukawa couplings. For instance, the Yukawa couplings for the up-quarks are, using spurion analysis, 
\begin{align}
(Y^{\rm SM}_u)_{ij}\sim \left(\frac{\langle \phi\rangle }{\Lambda}\right)^{N_{Q(i)}+N_{u(j)}}\, \qquad\qquad\text{[traditional FN]},
\label{eq:FNY}
\end{align}
where $\Lambda$ is a heavy scale to be discussed momentarily and the analogy with the clockwork mechanism is evident with the association $\langle \phi\rangle/\Lambda =1/q$, cf. Eqs. \eqref{eq:YuSM}, \eqref{eq:YdSM}. In traditional FN we thus need $\langle \phi \rangle \ll \Lambda$, to generate flavor hierarchies. 

As shown below there exists also a different realization of FN models, which we refer to as the clockworked FN models, in which the spurion analysis still applies but it does so with inverse powers of the vev of $\phi$,
\begin{align}
(Y^{\rm SM}_u)_{ij}\sim \left(\frac{\Lambda}{\langle \phi^*\rangle }\right)^{N_{Q(i)}+N_{u(j)}}\,\qquad\qquad\text{[clockworked FN]},
\label{eq:FNY:cw}
\end{align}
with $\Lambda$ a dimensionful paremeter that has a different interpretation than in Eq.~\eqref{eq:FNY}. In this case the identification with the clockwork models of flavor is $\langle \phi\rangle/\Lambda =q$. In clockworked FN models therefore $\langle \phi\rangle \gg \Lambda$ generates the flavor hierarchies.

 \begin{figure}[t]
\begin{center}
\includegraphics[width=11.cm]{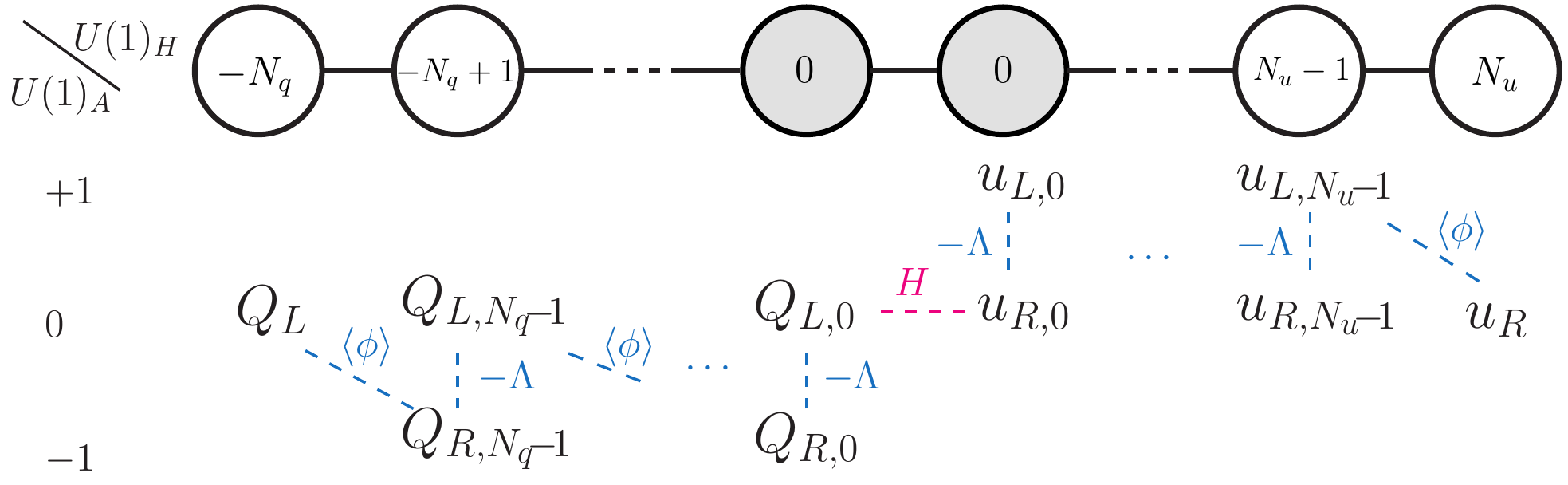}
\caption{\label{fig:FN:alter} The traditional FN chain with an additional axial $U(1)_A$ symmetry to prevent $\phi^*$ cross links.  The horizontal charges are denoted in the nodes -- the vector-like quarks on the two greyed out nodes, linked by the Higgs, do not carry a horizontal charge.
}
\end{center}
\end{figure}

To realize explicitly the two types of FN models we need to specify the full field content. We start with the traditional FN models, and then make the necessary modifications to arrive at the clockworked FN models.  
Each SM fermion field, $\psi^{(i)}$, is accompanied by a chain of $N_{\psi(i)}$ vector-like fermions of mass $\sim \Lambda$.
Taking for illustration the up-type quarks, there are $N_{u(i)}$ new Dirac fermions, $u^{(i)}_{k}$, added to the $i$-th generation SM quark, $u_R^{(i)}$. The vector-like Dirac fermions carry $U(1)_{H}$ charges from $0$ to $N_{u(i)}-1$, while the chiral fermion $u_R^{(i)}$ has a charge $N_{u(i)}$. With this matter content the {\it most general} mass and Yukawa interactions read,
\begin{align}
\label{eq:LFN}
\mathcal L_{\rm FN}\supset
\begin{pmatrix}
 \bar u^{(i)}_{L,0} & \bar u^{(i)}_{L,1}&\cdots& \bar u^{(i)}_{L,N_{u{(i)}}-1}
 \end{pmatrix} 
 \,
 \begin{pmatrix}
 -\Lambda & \langle\phi\rangle   & 0 &  \ldots & 0 \\ 
  \langle\phi^*\rangle & -\Lambda & \langle\phi\rangle &   \ldots &0 \\ 
 \vdots & \ddots&  \ddots & \ddots & 0 \\
 0 & \cdots &\langle\phi^*\rangle & -\Lambda &  \langle\phi\rangle\\ 
\end{pmatrix}
\begin{pmatrix}
u^{(i)}_{R,0}
 \\
\vdots
\\ u^{(i)}_{R,N_{u{(i)}}-1} 
\\ u^{(i)}_R
\end{pmatrix}\,,
\end{align}
where each entry has an ${\mathcal O}(1)$ dimensionless coefficient that we do not write out  for simplicity.
This mass matrix closely resembles that of the clockwork with $qm\sim \Lambda$ and $m\sim \langle\phi\rangle$, Eq.~\eqref{eq:matrix},
except for the $\phi^*$ terms. The analogy is complete if the theory is supersymmetric, so that 
such non-holomorphic terms are forbidden.  We choose a different possibility to forbid the $\phi^*$ terms
and introduce an axial $U(1)_A$ symmetry and a new scalar $S$ with charge $1$ under $U(1)_A$ and a vev $\langle S\rangle =\Lambda$, while the flavon $\phi$ has also $U(1)_A$ charge $1$ and those of the fermions are as in Fig.~\ref{fig:FN:alter}.\footnote{Since $U(1)_A$ is in general anomalous it would require additional structure were it to be gauged.} 

\begin{figure}[t]
\begin{center}
\includegraphics[width=11.cm]{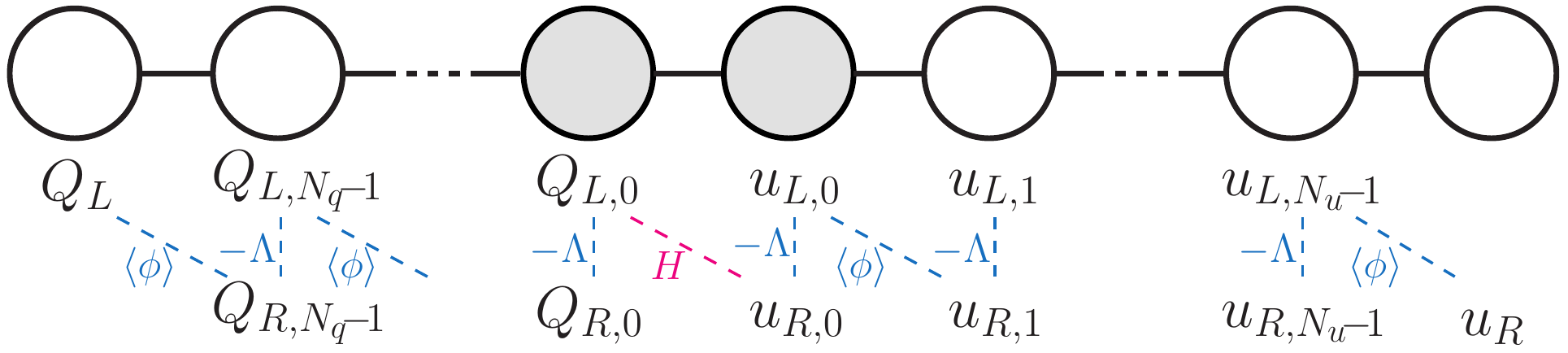}\\[-4mm]
\Huge{$\shortparallel$}\\[2mm]
\includegraphics[width=11.cm]{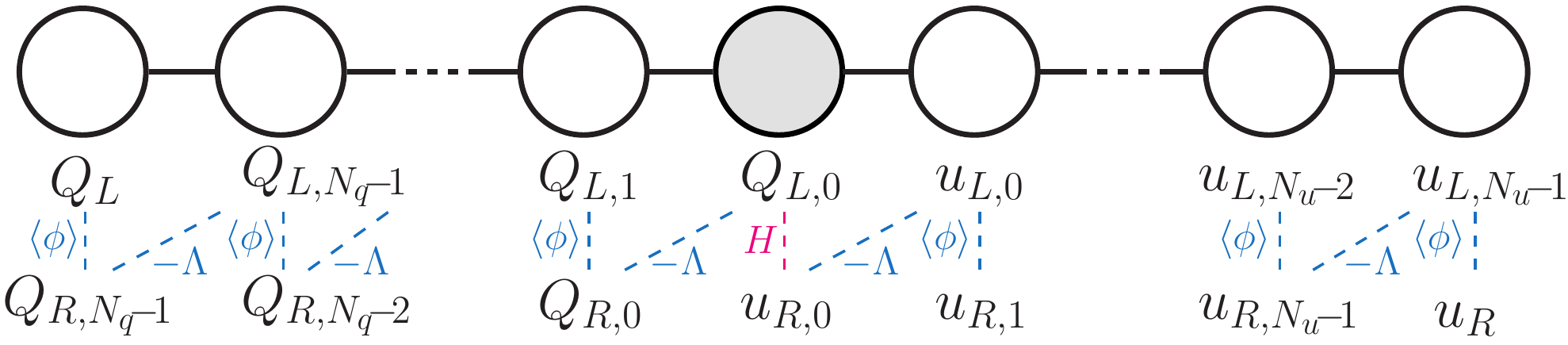}
\caption{\label{fig:FN:moose} Top: the traditional FN chain 
with the fields on the same node  carrying the same $U(1)_H$ horizontal (and electroweak) charges. The chiral fields on the outermost nodes are charged under $U(1)_H$.
Bottom: re-grouping into two clockwork chains connected through a Higgs Yukawa interaction on the middle node after flavon obtains a vev, $\langle \phi \rangle \ne 0$. 
}
\end{center}
\end{figure}

For the quark doublets there are, analogously, $N_{q(i)}$ new Dirac fermions, $Q^{(i)}_{k}$, added to the $i$-th generation SM quark, $Q_L^{(i)}$. The vector-like fermions $u_0^{(i)}$,  $Q_0^{(i)}$, both singlets under $U(1)_H$, then couple the two fermionic chains via the Higgs. Note that the $U(1)_A$ charge assignments allow only one chirality of the two vectorlike fermions to couple to the Higgs, see Fig.~\ref{fig:FN:alter}.  

It is now easy to see that, after $\phi$ obtains the vev, the traditional FN model with a $U(1)_A$ is equivalent to the clockwork model of flavor in the ``universal $q$'' limit, Eq. \eqref{eq:FNlimit}. All that is required is the identification, $\Lambda\to q m$,  $\langle \phi \rangle \to m$, setting all the ${\mathcal O}(1)$ factors in Eq.~\eqref{eq:LFN} to be exactly 1, and appropriately relabeling the fields, compare Fig.~\ref{fig:FN:moose}  with Fig.~\ref{fig:single:chain}. The traditional FN model and the clockwork model in the ``universal $q$'' limit, Eq.~\eqref{eq:FNlimit}, are therefore equivalent, if the degrees of freedom associated with the flavon $\phi$ are much heavier than the vectorlike fermions/gears, so that they can be integrated out. 
The FN expressions for the  SM quark masses, obtained using mass insertion approximation, Fig. \ref{fig:FN:Feynman}, then also offer 
an intuitive diagrammatic interpretation of the clockwork mechanism; we can identify $u_R$ ($Q_L$) at the end of the chain with the SM field. It then has to `go through' the rest of the chain to get to the Higgs, paying a ($\phi/\Lambda$) factor at every step.

  \begin{figure}[t]
\begin{center}
\includegraphics[width=11.cm]{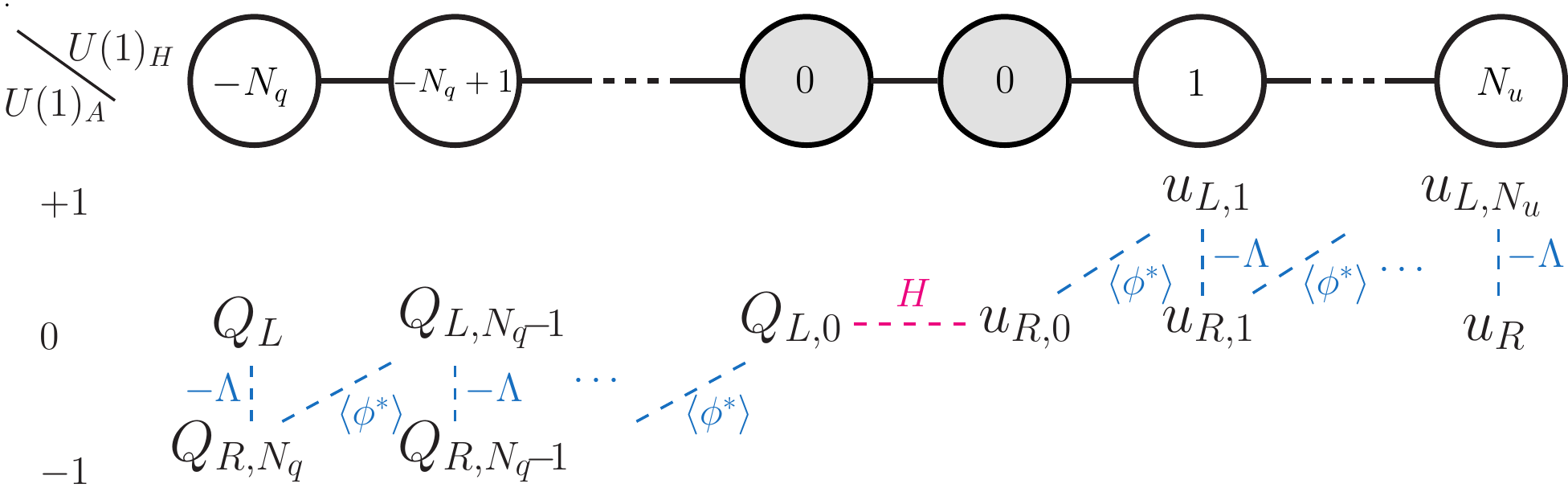}
\caption{\label{fig:FN:clockworked} The clockworked FN chain that naturally leads to the clockwork model of flavor.  The chiral fermions on the greyed-out Higgs node do not carry a horizontal charge.}
\end{center}
\end{figure}

We turn next to the clockworked FN models. In contrast to the traditional FN, there is only one $U(1)_H$-singlet fermion per chain, $u_{R,0}$ ($Q_{L,0}$), while the rest of the fermions come in pairs of opposite chirality but same charge, see Fig.~\ref{fig:FN:clockworked}.\footnote{Apart from relabeling, the choice is just whether one of the $u_{L,k}$, $Q_{R,k}$ chiral fermions carry vanishing horizontal charges (traditional FN), or charges $N_u$ and $-N_q$ (clockworked FN).} In the clockworked FN the $U(1)_H$ is anomaly free. This is in contrast to the traditional FN where additional field content is required to achieve anomaly free $U(1)_H$.
As before, we still use $U(1)_A$ to forbid the terms under the diagonal of the mass matrix (as in Eq. \eqref{eq:LFN}). However, the axial symmetry is no longer needed in order to arrive at just one Higgs term. Let us remark again that, even though $U(1)_A$ remains anomalous, it is auxiliary to the discussion, and can be avoided. 

The clockwork model of flavor in the ``universal $q$'' limit follows immediately from the field content in Fig.~\ref{fig:FN:clockworked} in the limit of a heavy flavon degrees of freedom, now identifying $\langle \phi^*\rangle\to q m$, $\Lambda\to m$, and relabeling just two fields, $Q_L, u_R\to Q_{L,N_q}, u_{R,N_u}$, compare Fig.~\ref{fig:single:chain}  with Fig.~\ref{fig:FN:clockworked}. 
Note that in this case $\Lambda$ does not correspond to the mass of any particle in the spectrum, but rather gives the mass band spread for vector-like fermions.\footnote{It does have a symmetry interpretation, though. Once $U(1)_H$ is broken by $\langle\phi\rangle$, one can define a new accidental horizontal $U(1)_\Lambda$ symmetry by shifting in Fig.~\ref{fig:FN:clockworked}  the $u_{L,i}$ to the left by one node,  the $Q_{R,i}$ to the right by one node, and assign the fields on the same node equal $U(1)_\Lambda$ charges. The $U(1)_\Lambda$ is broken by $\Lambda$, so that $\Lambda$ can be viewed as a spurion of this approximate symmetry.}
The hierarchies in masses and mixings can be understood by realizing that the zero modes are equal to the $Q_L, u_R$ fields, up to $\Lambda/\langle \phi \rangle$ corrections. The zero modes thus have effective horizontal charges that are to a good approximation equal to the ones of $Q_L, u_R$, i.e., they are $N_q,N_u$ respectively. This leads to the spurion expansion which is on {\it inverse} powers of $\langle \phi^*\rangle$. 
This is somewhat reminiscent of the models of gauged mininal flavor violation where 
the flavor symmetry is made anomaly free and 
the expansion is in inverse powers of flavon vevs~\cite{Grinstein:2010ve,Alonso:2016onw}.

 Given the above analogies between the FN models and the clockwork models of flavor in the ``universal $q$'' limit, the logical question is how to tell them apart. The clockwork models of flavor do not contain a dynamical flavon field, $\phi$. 
 The differences between the ``universal $q$'' clockwork models and the FN models, which do contain the flavon field $\phi$, will therefore depend on how heavy the radial and the angular modes of $\phi$ are (we denote their masses by $m_{|\phi|}, m_{\arg\phi}$), respectively. If both are parametrically heavier than the gears, they can be integrated out, and the two models are equivalent at the renormalizable level at the energy scales of the gears and below. The mass of the radial mode, $m_{|\phi|}$, depends on the details of the scalar potential, and is naturally at the scale $\langle \phi \rangle$. The angular mode, on the other hand, is the Goldstone boson of a spontaneously broken global $U(1)_H$ -- the axiflavon. For global $U(1)_H$ that is anomalous, as in the traditional FN models, the axiflavon can solve the strong CP problem~\cite{Calibbi:2016hwq,Ema:2016ops} or even act as a relaxion~\cite{Davidi:2018sii}. For non-anomalous global $U(1)_H$, as in the clockworked FN models, the mass of the axiflavon would have to come from explicit breaking. If this breaking is small, the axiflavon would appear in the spectrum, possibly pointing towards the dynamical symmetry origin of the clockwork.
 
 There is, however, a particular limit of the FN parameter space, where both the radial and the angular mode of $\phi$ are parametrically heavier than the gears. 
This is the case, if both the flavon-fermion couplings (we denote them by $Y'$) as well as the $U(1)_A$ breaking Dirac mass terms, are small, $Y'\ll 1$ and $\Lambda \ll \langle \phi\rangle$, respectively. We can then have a hierarchy $M\sim \max (Y'\langle \phi\rangle, \Lambda) \ll m_{\arg\phi}\ll \langle \phi \rangle$, such that the gear masses, $M$, are much smaller than $m_{\arg\phi}$, $m_{|\phi|}$, and still the spontaneous breaking of $U(1)_H$ dominates over the explicit one, $m_{\arg\phi} \ll \langle \phi \rangle$.  In this case the ``universal $q$'' clockwork models and FN models are exactly the same at the gear mass scale (at the renormalizable level).

The discussion changes, if the $U(1)_H$ is gauged (we denote the corresponding gauge coupling by $g_H$). In that case the angular mode of $\phi$ is absorbed by the $U(1)_H$ gauge boson after spontaneous symmetry breaking. If the mass of the gauge boson, $\sim g_H \langle \phi \rangle$ is parametrically bigger than  the gear masses, $M\sim \max (Y'\langle \phi\rangle, \Lambda)$, the gauge boson can be integrated out, and at the renormalizable level the ``universal $q$'' clockwork models and FN models are equivalent. 

Gauging $U(1)_H$ also has other consequences. FCNCs are generated from the tree level exchanges of the flavor gauge boson, so that 
the $\langle \phi \rangle$ mass scale is pushed well above the LHC.  The bounds from tree level exchanges of the radial flavon mode are typically weaker, see e.g., Ref.~\cite{Baldes:2016gaf,Calibbi:2012at} (these estimates are indicative, the actual limits depend on the details of the FN model). 
The new fermions, are also much more innocuous from the point of view of low energy constraints,  as we show in the next section. The low energy bounds only require them to be heavier than about a TeV.

 \begin{figure}[t]
\begin{center}
\includegraphics[width=9.5cm]{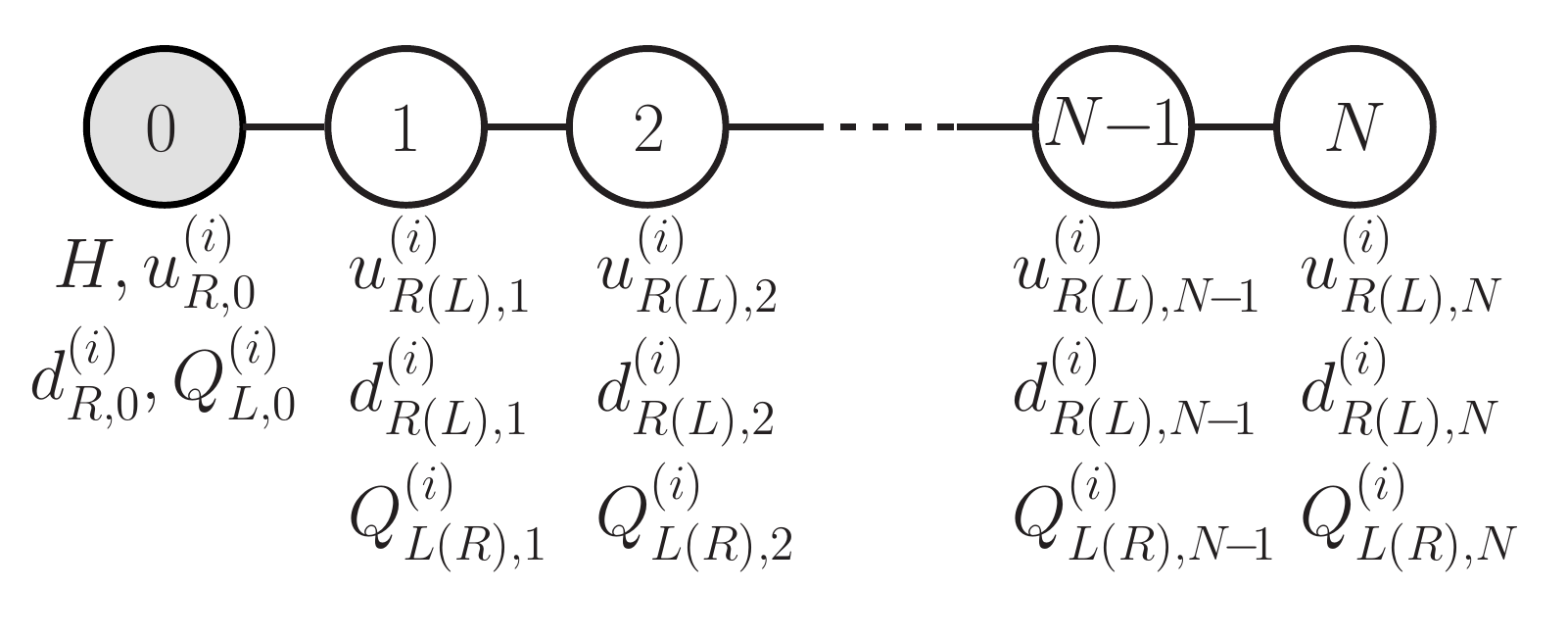}
\caption{\label{fig:clockworkRS} The ``universal $N$'' limit of the clockwork setup, where all the clockwork chains are of the same length, $N$,  and the hierarchy of the SM quark masses comes entirely from different values of the clockworking factors, $q$, see also Eq. \eqref{eq:RSlimit}.}
\end{center}
\end{figure}

Finally, we discuss the ``universal $N$'' limit of clockwork, Eq.~\eqref{eq:RSlimit}, which is reminiscent of the RS flavor models~\cite{Randall:1999ee, Grossman:1999ra,Agashe:2004cp,Agashe:2004ay}.  In this case the clockwork chains are of equal length, so that the set-up in Fig.~\ref{fig:clockwork:diagram} can be projected to a single chain with $N$ nodes, shown in Fig.~\ref{fig:clockworkRS}.
It is tempting to think of the $N$ nodes as a partial realization of the deconstructed extra dimension. However, a crucial difference with a properly deconstructed extra dimension is that in the clockwork model there is only a single gauge group, the SM one, which spans all the nodes, while in the deconstructed extra dimensional models there is one copy of the gauge group for each of the nodes. Taking the $N\to \infty$ limit thus does not correspond to a continuum limit. In the continuum limit the SM gauge fields would correspond to the zero modes of the 5D gauge fields, but there are no corresponding KK states (a proper extension of clockwork to 5D is possible, but leads to exponentially suppressed gauge couplings, see App.~\ref{sec:continuum}).

The behavior of the zero modes in clockwork and the RS is very similar, while the differences arise at the mass scale of the gears. For instance, the form of the SM Yukawa matrices in terms of zero mode overlaps, Eqs. \eqref{eq:YuSM}, \eqref{eq:YdSM}, is exactly the same as the well known form in the RS models of flavor \cite{Agashe:2004cp,Agashe:2004ay,Csaki:2008eh,Blanke:2008zb}. In the RS the $f_{Q(i)}, f_{u(i)}, f_{d(i)}$ are the zero mode overlaps with the IR brane that contains the Higgs. In our case these are the values of the zero modes on the $0$-th node of the clockwork chain, which is the node that couples to the Higgs.

On the other hand, the massive modes are quite different in the clockwork and the RS. The RS contains KK states of both gauge bosons and the SM fermions, while in clockwork the SM is supplemented only by the fermionic gears. Furthermore, the typical mass gap between neighbouring clockworking gears is much smaller than the mass gap between the gears and the zero mode, while the RS KK states have mass splittings that are all ${\mathcal O}(1)$. We discuss the implications of this for flavor and high $p_T$ observables in Secs.~\ref{sec:flavor} and~\ref{sec:collider}.

\subsection{The QCD Landau pole}

The addition of  new degrees of freedom charged under the SM gauge group modifies the renormalization group evolution (RGE) of the SM gauge couplings above the scale $\mu=M$. The most pronounced effect is in the QCD coupling $\alpha_s$ potentially destroying asymptotic freedom~\cite{Gross:1973ju}. At one loop, the RGE of $\alpha_s$ is given by
\begin{align}\label{eq:QCD}
\frac{d\alpha_s}{d{\rm \ln\mu}}=-2\beta_0\frac{\alpha_s^2}{4\pi},~~~\beta_0=\frac{11N_c-2N_f}{3}, 
\end{align}
where $N_c=3$ is the number of colors and $N_f$ is the number of fermions in the fundamental representation of $SU(N_c)$, i..e, all the SM quarks and gears lighter than scale $\mu$. 
The sign of the beta function
depends on $N_f$, with $N_f=16$ the maximum value for which QCD is asymptotically free  at one loop. This corresponds to 6 SM flavors plus $N_{\rm gears}= 10$ gears.  For $N_{\rm gears}>10$ the theory develops a UV Landau pole at the scale
\begin{align}\label{eq:QCDLandau}
\Lambda_{\rm Landau}=M~e^{\frac{-2\pi}{\alpha_s(M)\beta_0}}, 
\end{align}
where $M$ is the scale at which the gears are integrated out (so roughly the average gear mass). In the setup of Eq.~(\ref{eq:q:scaling}) $N_{\rm gears}=26$ which gives $\Lambda_{\rm Landau}\simeq 2\times10^4$ TeV for $M=5$ TeV. The value of $\Lambda_{\rm Landau}$ can be increased through trivial modifications of the setup. For instance, increasing $q$ while reducing the length of the clockwork chains results in fewer new colored states contributing to the QCD $\beta$ function. 

\begin{figure}[t]
\begin{center}
\includegraphics[width=8.5cm]{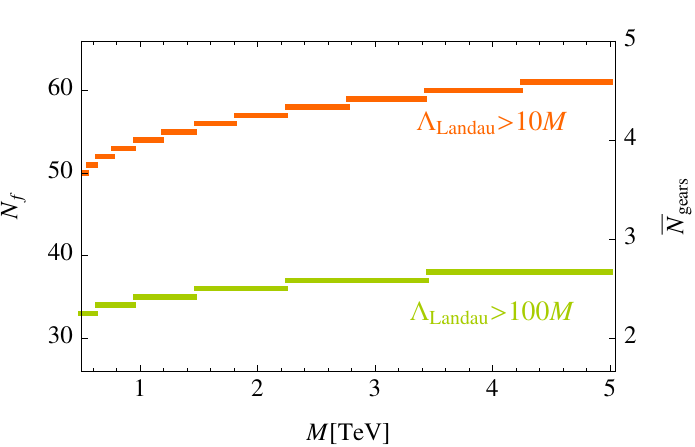}
\caption{\label{fig:LandauPole} 
Example upper bounds on the total number of colored Dirac fermions ($N_f = 6 + N_{\rm gears} $) or, equivalently, on the  effective number of gears per quark flavor ($\bar N_{\rm gears} \equiv N_{\rm gears} / 12$) as a function of the assumed common gear mass, $M$. Requiring there is no Landau pole in $\alpha_s$ below $\Lambda_{\rm Landau}=10 (100) M$ gives the bounds shown as orange (lower, green) lines,  when rounding $N_f$ to the closest integer. 
}
\end{center}
\end{figure}

In Fig.~\ref{fig:LandauPole} we show upper bounds on $N_f$  and on the effective number of gears per quark flavor, $\bar N_{\rm gears} \equiv N_{\rm gears} / 12$, as a function of the common gear mass, $M$.  We require that the Landau pole is not reached below $10M (100M)$, with the bounds on $N_f, \bar  N_{\rm gears}$ shown in red (green). The bounds were computed using the three-loop $\alpha_s$ RGE~\cite{Chetyrkin:2000yt} with $\alpha_s(m_Z) = 0.118$~\cite{Patrignani:2016xqp}.     
We observe that in order for the Landau pole to be parametrically above the gear masses the discrete clockwork chains cannot be arbitrarily long. The maximum number of colored gears is $N_{\rm gears} \sim {\mathcal O}(30-60)$ for $M$ in the (few) TeV region.

\subsection{Perturbativity and stability of the Higgs potential}

\noindent Quark loops also provide an important negative contribution in the one-loop beta function for the quartic self-coupling of the Higgs~\cite{Rodejohann:2012px,Xiao:2014kba}. In the SM this causes the quartic coupling to run to negative values at $\sim 10^{10}$ GeV \cite{Degrassi:2012ry,Buttazzo:2013uya}. This is pushed lower, when gears are added to the SM field content. 
The contributions from the SM and the clockwork vector-like quarks to the Higgs quartic, $V\supset\lambda|H|^4/2$, can be written as 
\beq
\beta_\lambda\supset 12\text{Tr}\left(Y_U^\dagger Y_U+Y_
D^\dagger Y_D\right)\lambda-12\text{Tr}\left(Y_U^\dagger Y_U Y_U^\dagger Y_U+Y_
D^\dagger Y_DY_
D^\dagger Y_D\right),
\eeq
where ${d\lambda}/{d\ln\mu}=\beta_\lambda/{16\pi^2}$ and making use of the basis-invariance of the result we performed the computation in the interaction basis of eq.~(\ref{eq:LagSMCK0}) where only the $0$-th site couples to the Higgs.

The scale at which the Higgs quartic becomes negative is, at leading logarithmic approximation,
\begin{align}
\label{eq:Lambdalambda}
\Lambda_{\rm Decay}=M~e^{-\frac{16\pi^2\lambda_0}{\beta_\lambda}},
\end{align}
where $\lambda_0\simeq 0.258$. Requiring that the beta function remains perturbative,	
\beq
\text{Tr}\left( Y_U^\dagger  Y_U Y_U^\dagger \tilde Y_U\right)+\text{Tr}\left(\tilde Y_D^\dagger \tilde Y_D \tilde Y_D^\dagger \tilde Y_D\right)\ll \frac{4\pi^2}{3},
\eeq
puts a self-consistency constraint on the clockwork flavor models --  while entries in the Yukawa matrices can be $\mathcal{O}(1)$ they should be mostly smaller than 1.

The problem of vacuum stability near the mass scale of vector-like quark states is a common problem in many NP models.
A well known solution to increase the scale at which perturbativity or vacuum stability is lost (albeit at the cost of some tuning), is to add additional scalars. 
A coupling between a new singlet $\phi$ and the Higgs of the form $\mathcal{L}\supset \lambda_{S}H^\dagger H\phi^2$, gives a positive contribution to the beta function, $\delta\beta_\lambda=2\lambda_S^2$, thus potentially raising the scale $\Lambda_{\rm Decay}$.

\section{Flavor constraints}
\label{sec:flavor}
\subsection{General considerations}

We start the discussion by considering the clockwork model before electroweak symmetry breaking. The clockwork Lagrangian \eqref{eq:LagSMCK0}, after mass diagonalization of the clockwork chains using the unitary transformation in Eq. \eqref{eq:singleVR0}, is given by
\beq
\begin{split}
\label{eq:LagSMCK1}
	\mathcal L=&\mathcal L_{\rm kin}-\sum_{i,j}\left[f_{Q(i)} f_{d(j)}(Y_D)_{ij}
    \bar Q_L^{(i)} H d_{R}^{(j)}+f_{Q(i)} f_{u(j)}(Y_U)_{ij}
    \bar Q_L^{(i)}\tilde H u_{R}^{(j)}\right]
	\\
	&-\sum_{k}\sum_{i,j}\left[f_{Q(i)} f_{d(j)}^{k}(Y_D)_{ij}\bar Q_{L}^{(i)} H d_{R,k}^{(j)}+f_{Q(i)} f_{u(j)}^{k}(Y_U)_{ij}\bar Q_{L}^{(i)} \tilde H u_{R,k}^{(j)}\right.
	\\
	&+\left.f_{Q(i)}^k f_{d(j)}(Y_D)_{ij}\bar Q_{L,k}^{(i)} H d_{R}^{(j)}+f_{Q(i)}^k f_{u(j)}(Y_U)_{ij}\bar Q_{L,k}^{(i)} \tilde H u_{R}^{(j)}\right]
	\\
	&-\sum_{k,k^\prime}\sum_{i,j}\left[f_{Q(i)}^k f_{d(j)}^{k^\prime}(Y_D)_{ij}\bar Q_{L,k}^{(i)} H d_{R,k^\prime}^{(j)}+f_{Q(i)}^k f_{u(j)}^{k^\prime}(Y_U)_{ij}\bar Q_{L,k}^{(i)} \tilde H u_{R,k^\prime}^{(j)}\right]+\text{h.c.}.
\end{split}
\eeq
To shorten the notation above we denoted the zero modes by $Q_{L}^{(i)}$, $d_{R}^{(i)},u_{R}^{(i)}$, and dropped the primes on gear mass eigenstates, $Q_{L,k}^{(i)}$, $d_{R,k}^{(i)}$ and $u_{R,k}^{(i)}$, where $k=1,\ldots,N_{\psi(i)}$. The zero modes are identified with the SM fields, which obtain their mass only after electroweak symmetry breaking.

The Yukawa couplings between the SM fields and the gears, shown in the second and third lines of Eq. \eqref{eq:LagSMCK1}, induce new contributions to the Flavor Changing Neutral Currents (FCNCs)~\cite{delAguila:2000rc,Ishiwata:2015cga,Bobeth:2016llm}. The relevant tree level contributions are shown in Fig.~\ref{fig:flavormix}, and for the low energy observables we can work in the limit where the gears are integrated out. After electroweak symmetry breaking the SM quarks become massive, with the leading contribution being the first line of Eq. \eqref{eq:LagSMCK1}, corrected by $v^2/M^2$ suppressed terms from couplings to the gears. The latter also induce FCNC couplings of the SM quarks to the $Z$-boson and the Higgs, and produce additional flavor breaking contributions to the charged currents. The $B_{d,s}^0-\bar B_{d,s}^0$,  $D^0-\bar D^0$ and $K^0-\bar K^0$ mixing amplitudes therefore receive NP corrections from the tree level exchanges of the $Z$ and the Higgs. 
In addition, neutral-meson mixings also receive phenomenologically relevant one loop corrections due to the exchanges of the gears, shown in Fig.~\ref{fig:flavormixloop}.

For generic flavor violating couplings, heavy fermions would need to have PeV-scale masses, in order to avoid experimental constraints on FCNCs. In contrast, in clockwork flavor models the FCNCs are suppressed by the overlaps of the zero-modes that also lead to the hierarchy of SM quark masses. The bounds on the gear masses are therefore only in the TeV mass range, as we show below. This protection against FCNCs from zero-mode overlaps, the clockwork GIM (CW-GIM), is well known in the RS models of flavor, where it was dubbed RS-GIM \cite{Agashe:2004ay,Agashe:2004cp}, and is a general feature of sequestered models, including the FN models \cite{Davidson:2007si,Gilad:comm}. 

\subsection{Flavor mixing in the EFT}
\label{eq:flavor:EFT}

\begin{table}[t]
\centering
\caption{Dimension six $SU(3)_c\times SU(2)_L\times U(1)_Y$-invariant operators that receive contributions from the clockwork gears. The notation for operators follows Ref.~\cite{Grzadkowski:2010es}.\label{tab:SMEFTops}}
\vspace{0.2cm}
\begin{tabular}{cccc}
\hline\hline
Name & Operator & ~~~Name & Operator\\
\hline
$\mathcal{O}_{HQ}^{(1)}$&$\left(H^\dagger i\overleftrightarrow{D}_\mu H\right)\bar Q_L\gamma^\mu Q_L$&$\mathcal{O}_{QQ}^{(1)}$&$(\bar Q_L\gamma^\mu Q_L)(\bar Q_L\gamma_\mu Q_L)$\\
$\mathcal{O}_{HQ}^{(3)}$&$\left(H^\dagger i\overleftrightarrow{D}^I_\mu H\right)\bar Q_L\gamma^\mu \tau^I Q_L$&$\mathcal{O}_{QQ}^{(3)}$&$(\bar Q_L\gamma^\mu\tau^I Q_L)(\bar Q_L\gamma_\mu\tau^I Q_L)$\\
$\mathcal{O}_{Hu}$&$\left(H^\dagger i\overleftrightarrow{D}_\mu H\right)\bar u_R\gamma^\mu u_R$&$\mathcal{O}_{Qd}^{(1)}$&$(\bar Q_L\gamma^\mu Q_L)(\bar d_R\gamma_\mu d_R)$\\
$\mathcal{O}_{Hd}$&$\left(H^\dagger i\overleftrightarrow{D}_\mu H\right)\bar d_R\gamma^\mu d_R$&$\mathcal{O}_{Qu}^{(1)}$&$(\bar Q_L\gamma^\mu Q_L)(\bar u_R\gamma_\mu u_R)$\\
$\mathcal{O}_{Hud}$&$\left(\tilde H^\dagger i D_\mu H\right)\bar u_R\gamma^\mu d_R$&$\mathcal{O}_{dd}$&$(\bar d_R\gamma^\mu d_R)(\bar d_R\gamma_\mu d_R)$\\
$\mathcal{O}_{uH}$&$\left(H^\dagger H\right)\bar Q_L \tilde H u_R$&$\mathcal{O}_{uu}$&$(\bar u_R\gamma^\mu u_R)(\bar u_R\gamma_\mu u_R)$\\
$\mathcal{O}_{dH}$&$\left(H^\dagger H\right)\bar Q_L H d_R$&&\\
\hline\hline
\end{tabular}
\end{table}

\begin{figure}[t]
\begin{center}
\includegraphics[width=14cm]{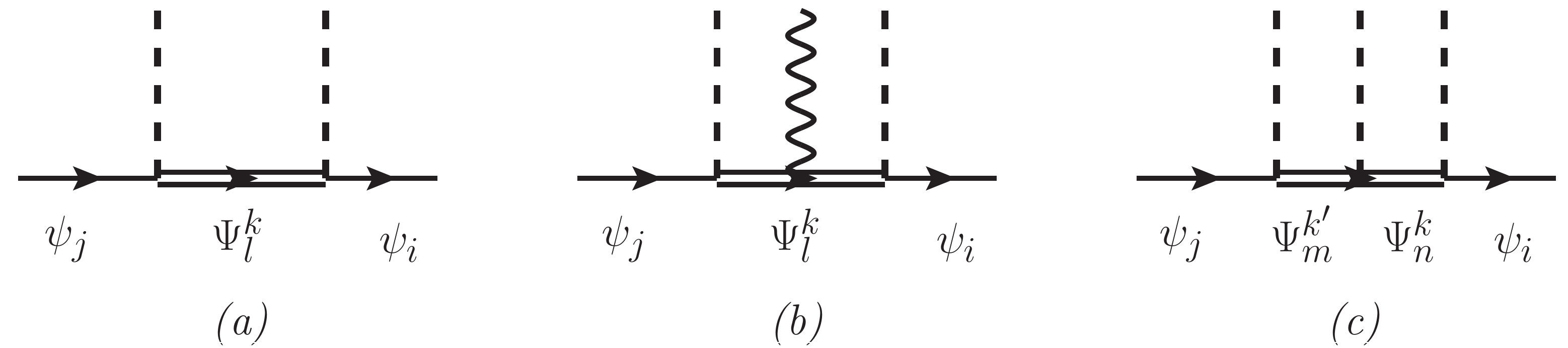}
\caption{\label{fig:flavormix} Three types of diagrams generating NP induced ${\mathcal   O}(v^2/M^2)$ flavor violating transitions among the SM fermions (single lines) from tree level exchanges of gears (double lines). 
The dashed lines denote the SM Higgs, while the wiggly line in diagram \textit{(b)} denotes either a  $W$ or $Z$ boson. 
}
\end{center}
\end{figure}

\begin{figure}[t]
\begin{center}
\includegraphics[width=13cm]{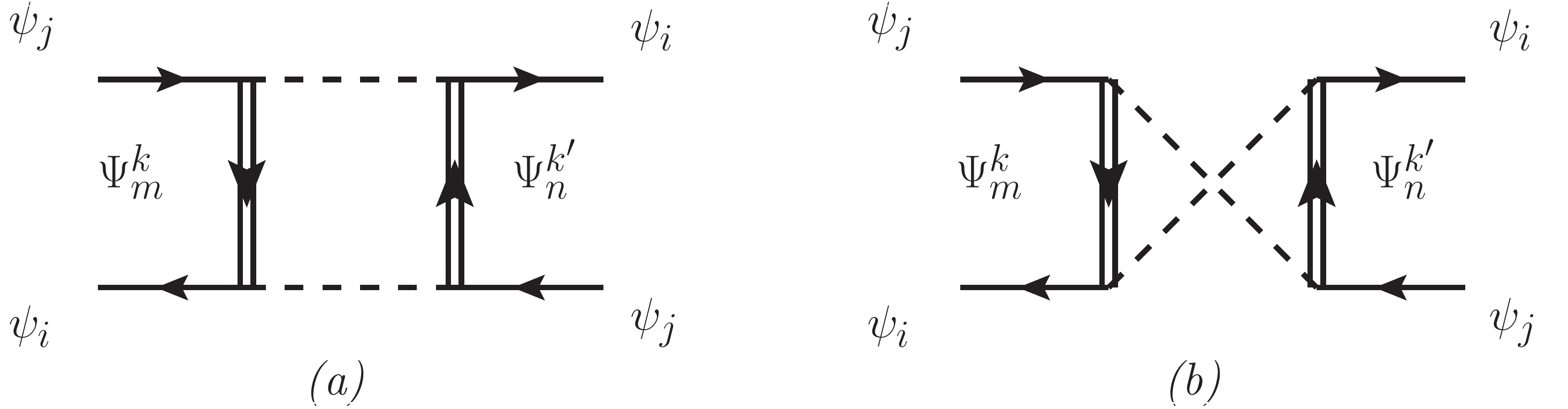}
\caption{\label{fig:flavormixloop} One loop diagrams contributing to neutral-meson mixing induced at ${\mathcal   O}(v^2/M^2)$ by Higgs-mediated interactions (dashed lines) with the gears (double lines).
}
\end{center}
\end{figure}

In this subsection we first prepare the necessary formalism that allows for a systematic comparison with the experimental constraints on low-energy observables, to be used in the subsequent subsections.
 In the first step we integrate out the gears at the scale $\mu\sim M$, 
matching the diagrams in Figs.~\ref{fig:flavormix} and~\ref{fig:flavormixloop} to dimension six operators of the SM effective field theory (SMEFT)~\cite{Buchmuller:1985jz,Grzadkowski:2010es} 
\begin{align}
\mathcal L_{\rm SMEFT}=\sum w_i\,\mathcal{O}_i \label{eq:SMEFT}.
\end{align}
The Wilson coefficients scale as $w_i\sim1/M^2$, where $M$ is the typical mass of the gears. 
Matching to SMEFT amounts to working in the mass-insertion approximation, i.e., we keep in the analysis terms that are leading in the $Y_{U,D} v/M$ expansion.

 The dimension six operators generated in the matching are listed in Tab.~\ref{tab:SMEFTops}.
The tree-level diagrams in Fig.~\ref{fig:flavormix} match onto the Higgs-current fermion-current operators ($\psi^2H^2 D$) shown  in the left column of Tab.~\ref{tab:SMEFTops}. The \textit{(a)} and \textit{(b)} diagrams  in  Fig.~\ref{fig:flavormix} match onto the operators $\mathcal{O}_{HQ}^{(1,3)}$ for the $SU(2)$-singlet gears, $u_{R,k}^{(i)}$, $d_{R,k}^{(i)}$, and onto the operators $\mathcal{O}_{Hu}$, $\mathcal{O}_{Hd}$ and $\mathcal{O}_{Hud}$ for the doublet gears, $Q_{L,k}^{(i)}$. These diagrams also contribute to the chirallity-flipping operators $\mathcal{O}_{uH}$ and $\mathcal{O}_{dH}$ via equations of motion~\cite{delAguila:2000rc}, modifying the effective SM Yukawa couplings~\cite{Harnik:2012pb}. Diagram \textit{(c)} requires the mixing of the gears in doublet and singlet representations and also contributes to the latter operators. 

For example, the contribution to the operator $\mathcal{O}_{HQ}^{(1)}$ reads,
\beq
\begin{split}
	[w_{HQ}^{(1)}]_{ij}&=\frac{1}{4}f_{Q(i)}f_{Q(j)}\sum_k\sum_r\left\{-[Y_D]_{ir}\frac{\big(f_{d(r)}^k\big)^2}{\big(M_{d(r)}^k\big)^2}[Y_D^\dagger]_{rj}+[Y_U]_{ir}\frac{\big(f_{u(r)}^k\big)^2}{\big(M_{u(r)}^k\big)^2}[Y_U^\dagger]_{rj}\right\}
	\\
	&\simeq \frac{1}{4}\left[F_Q\left(Y_U M_{u}^{-2} Y_U^\dagger-Y_D M_{d}^{-2}Y_D^\dagger\right)F_Q\right]_{ij}~,\label{eq:matchingTLsample}
\end{split}
\eeq
with 
\beq
M_{\psi}={\rm diag}[q m_{\psi(1)},q m_{\psi(2)},q m_{\psi(3)}].
\eeq
 The contributions to the other Wilson coefficients can be found in App.~\ref{sec:matching}. In  Eq.~\eqref{eq:matchingTLsample} the zero-mode overlaps are written in a matrix notation, $F_Q=\text{diag}[f_{Q(1)},f_{Q(2)},f_{Q(3)}]$, $F_u=\text{diag}[f_{u(1)},f_{u(2)},f_{u(3)}]$, $F_d=\text{diag}[f_{d(1)},f_{d(2)},f_{d(3)}]$.  The equality in the second line applies when  $q$ is universal with $q\gg1$, so that all the gears in a given clockwork chain are degenerate and one can use the unitarity relation of Eq.~\eqref{eq:unitarity}. For quarks $Q_{L(3)}$ and $u_{R(3)}$, that are not clockworked, \textit{cf.} Eq.~\eqref{eq:q:scaling}, this needs to be replaced with $f_{\psi}^k=0$ in the first line of Eq.~\eqref{eq:matchingTLsample} and with vanishing contributions to the appropriate Wilson coefficients in the second line.

A double insertion of a flavor-violating $\psi^2 H^2 D$ operator gives a tree-level  $Z$ and $H$ exchange contribution to the four-fermion operators relevant for neutral-meson mixing. These contributions are of order ${\mathcal O}(v^4/M^4)$. 
The loop diagrams in Fig.~\ref{fig:flavormixloop},  on the other hand, contribute at $\mathcal O(v^2/M^2)$ and can thus give the leading contribution to the meson-mixing~\cite{Bobeth:2016llm} for gear masses in the few-TeV range. The relevant four-fermion operators ($\psi^4$) are listed in the right row of Tab.~\ref{tab:SMEFTops}. Diagram \textit{(a)} in Fig.~\ref{fig:flavormixloop} with gears $u_{R,k}^{(i)}$ or $d_{R,k}^{(i)}$ (with gears $Q_{L,k}^{(i)}$) on the interal lines  matches onto $\mathcal{O}_{QQ}^{(1,3)}$  (onto $\mathcal{O}_{dd}$ and $\mathcal{O}_{uu}$). Diagram \textit{(b)} in Fig.~\ref{fig:flavormixloop} matches onto $\mathcal{O}^{(1)}_{Qd}$ (onto $\mathcal{O}^{(1)}_{Qu}$) when on one of the internal lines the gear is $Q_{L,k}^{(i)}$, while on the other it is $d_{R,k}^{(i)}$ $(u_{R,k}^{(i)})$. Contributions to $\mathcal{O}_{QQ}^{(1,3)}$ are generated by diagram \textit{(b)} with $u_{R,k}^{(i)}$ and $d_{R,k}^{(i)}$ gears on the internal lines, although these do cancel out for meson mixing. The results of this matching at one loop are shown in App.~\ref{sec:matching}.

In order to connect the Wilson coefficients evaluated at the high-energy scale $\mu=M$ to the experimental data at the electroweak or low-energy scales we need to add other loop-corrections to $\psi^4$~\footnote{In case that several clockwork chains contribute to the same Wilson coefficient we assume that the scale $M$ is equal to the mass of the lightest gear, neglecting the RG running above this threshold. 
}. We include EW contributions to the mixing of $\psi^2 H^2 D$ into the four-fermion operators~\cite{Jenkins:2013zja,Jenkins:2013wua,Alonso:2013hga}, running from $\mu=M$ down to $\mu=m_W$ in the leading logarithm (LL) approximation. Pure gauge interactions (\textit{viz.} in the symmetric electroweak phase) cannot produce $\Delta F=2$ contributions in the loop corrections to $\psi^2 H^2 D$ operators~\cite{Alonso:2013hga}, while the Yukawa corrections can potentially be more important than the loop contributions computed in Fig.~\ref{fig:flavormixloop} because some are proportional to the top Yukawa~\cite{Jenkins:2013wua} and are logarithmically enhanced by $\log(M/m_W)$. The matching between four-fermion operators of the SMEFT and those of the low-energy EFT (LEEFT) at $\mu=m_W$ generates extra finite pieces at one loop~\cite{Aebischer:2015fzz,Bobeth:2017xry}. The QCD corrections also induce  important rescalings and mixings among the four-fermion operators that we include in the running to LL accuracy~\cite{Gilman:1982ap,Ciuchini:1997bw,Buras:2000if,Buras:2001ra} despite being formally a two-loop contribution. As for the $\psi^2 H^2 D$, they are not renormalized by QCD interactions and we neglect the corresponding EW loop corrections which are small compared to the tree-level contributions in Fig.~\ref{fig:flavormix}.     

Finally, in matching the SMEFT and LEEFT at the electroweak scale, we need to transform the fermions in flavor space from the interaction basis to the mass basis by
\begin{align}
(u_L,~d_L,~u_R,~d_R)\longrightarrow (L_u^\dagger u_L,~ L_d^\dagger d_L,~R_u^\dagger u_R,~R_d^\dagger d_R),\label{eq:tomassbasis}
\end{align}
where $L_u$, $L_d$, $R_u$, $R_d$ are the unitary transformation matrices and where the relation $V_{\rm CKM}=L_u L_d^\dagger$ is understood.

\subsection{Low-energy constraints} 
\label{sec:lowEconstraints}
A necessary condition for the self-consistency of the clockwork model of flavor is that the presence of the gears do not parametrically change the mass hierarchies of the zero modes. The quark masses, including the dimension-6 corrections, are 
\begin{align}
\left[{\cal M}_u \right]_{ij}=\frac{v}{\sqrt2}\left[Y_u^{\rm SM}\right]_{ij}= \frac{v}{\sqrt2}\Big(\left[F_Q Y_U F_u\right]_{ij}-\frac{v^2}{2}\left[w_{uH}\right]_{ij}\Big), 
\end{align}
where we have written $\big(Y_u^{\rm SM}\big)_{ij}$ in Eq.~\eqref{eq:YuSM} up to second order and the down-type mass matrix is obtained with the obvious substitutions.
The explicit results in App.~\ref{sec:matching} show that the ${\mathcal O}(v^3)$ contributions from the gears have the same suppression factors $f_{Q(i)}~f_{u(i)}$ and $f_{Q(i)}~f_{d(i)}$, and thus do not change the flavor patterns.

The suppression of corrections by the zero-mode overlaps is a general feature of the contributions of the clockwork gears to processes 
involving SM fermions. For instance, 
the operators $\mathcal O_{uH}$ and $\mathcal O_{dH}$ missalign the masses of the quarks and their Higgs couplings
\begin{align}
\mathcal L_Y= -
\left(\bar u_{L}\left[\frac{Y_u^{\rm SM}}{\sqrt 2}+\delta y_u\right]u_{R}+\bar d_{L}\left[\frac{Y_{d}^{\rm SM}}{\sqrt 2}+\delta y_d\right]d_{R}\right)h+{\rm h.c.},
\end{align}
where the corrections to the SM relation are, in the quark mass basis, given by 
\begin{align}
\begin{split}
\label{eq:deltayuYukawas}
	\left[\delta y_u\right]_{ij}&=-\frac{v^2}{\sqrt{2}}\left[L_u w_{uH}R_u^\dagger \right]_{ij}=-\frac{v^2}{2\sqrt{2}}\left[L_u\,F_Q\left(Y_U M_u^{-2} Y_U^\dagger F_Q^2 Y_U\right.\right.
\\
	&\left.\left.+Y_U F_u^2 Y_U^{\dagger} M_{Q}^{-2} Y_U-2Y_U M_u^{-1} Y_U^\dagger M_Q^{-1} Y_U\right)F_u\,R_u^\dagger\right]_{ij},
\end{split}
\\
\begin{split}
\label{eq:deltaydYukawas}
	~~~~\left[\delta y_d\right]_{ij}&=-\frac{v^2}{\sqrt{2}}\left[L_d w_{dH} R_d^\dagger\right]_{ij}=-\frac{v^2}{2\sqrt{2}}\left[L_d\,F_Q\left(Y_D M_d^{-2} Y_D^\dagger F_Q^2 Y_D\right.\right.
\\
	&\left.\left.+Y_D F_d^2 Y_D^{\dagger} M_{Q}^{-2} Y_D-2Y_D M_d^{-1} Y_D^\dagger M_Q^{-1} Y_D\right)F_d\,R_d^\dagger\right]_{ij}.
\end{split}
\end{align}
Note that despite the unitary rotations, these couplings still receive a clockwork suppression of the quarks involved in the process, $F_Q$ and $F_{u,d}$, because the flavor hierarchies in Eq.~(\ref{eq:q:scaling}) are inherited by the rotation matrices, cf. Eq.~(\ref{eq:Rotshierarchy}).

This is a general feature of all the contributions of the gears to low-energy observables.
For instance, for processes involving only $d_i$ quarks, the dominant contribution is generally given by the operators with external quark doublets. For example, the contribution of $[\mathcal O_{Hd}]_{33}$ to $Z\to b b$ is suppressed by a clockwork factor $\lambda^4$, to be compared to the one given by $[\mathcal O_{HQ}^{(1,3)}]_{33}$, which  is unsuppressed because $Q_{L(3)}$ is not clockworked. Analogously, the contribution of $[\mathcal O_{QQ}^{(1,3)}]_{12}$ to $K$-$\bar K$ mixing is $\mathcal O(\lambda^{10})$ and of the same order as the top-box diagram in the SM, while the one from $[\mathcal O_{Qd}^{(1)}]_{12}$ and $[\mathcal O_{dd}]_{12}$ are further suppressed by factors $\lambda^2$ and $\lambda^4$, respectively. In the case of neutral processes involving the $u_i$ quarks the singlet-field contribution is again suppressed with respect to the doublet for the first family, whereas it is the opposite for the second family and there is no relative suppression for the third. For example, in $D$-$\bar D$ mixing the operators $[\mathcal O_{QQ}]_{12}$, $[\mathcal O_{Qu}^{(1)}]_{12}$ and $[\mathcal O_{uu}]_{12}$ are all suppressed by the same factor $\lambda^{10}$.

In the following we discuss the stronger bounds that can be derived from low-energy observables on the parameters of the clockwork model. 

\subsubsection{Weak boson decays}
\label{sec:weakbosondecays}

The couplings of weak gauge bosons to the SM quarks can be appreciably affected by the gears. Rates and angular asymmetries of the weak boson hadronic decays have been measured with a relative precision below the permille level in $e^+e^-$ collisions~\cite{ALEPH:2005ab}, imposing strong bounds on the couplings and masses of the clockwork chains. The $Z$ interactions with the quarks can be generally parametrized as
\begin{align}
\label{eq:L_Z}
\mathcal L_Z \supset&\frac{g}{\cos\theta_w}\sum_{ij}\left[\left(\frac{\delta_{ij}}{2}+\left[\delta g_{L}\right]_{ij}^{Z_u}\right)\bar u_{i}\gamma^\mu P_Lu_{j}+\left(-\frac{\delta_{ij}}{2}+\left[\delta g_{L}\right]_{ij}^{Z_d}\right)\bar d_{i}\gamma^\mu P_Ld_{j}\right.\nonumber\\
&\left.+\left[\delta g_{R}\right]_{ij}^{Z_u}\bar u_{i}\gamma^\mu P_R u_j+\left[\delta g_{R}\right]_{ij}^{Z_d}\bar d_{i}\gamma^\mu P_R d_j \right]Z_\mu,
\end{align}
where $\delta g_{L,R}^{Z_{(d,u)}}$ encode the corrections due to NP.  In our case these are given by,
\begin{subequations}
\label{eq:gZgears}
\begin{align}
\label{eq:gLZdgears}
	\left[\delta g_{L}\right]_{ij}^{Z_d}&=-\frac{v^2}{2}\,\left[L_d\left(w_{HQ}^{(1)}+w_{HQ}^{(3)}\right)L_d^\dagger\right]_{ij}=\frac{v^2}{4}\left[L_d F_Q Y_D M_d^{-2}Y_D^\dagger F_Q L_d^\dagger\right]_{ij},
 \\
 \label{eq:gLZugears}
	\left[\delta g_{L}\right]_{ij}^{Z_u}&=-\frac{v^2}{2}\,\left[L_u\left(w_{HQ}^{(1)}-w_{HQ}^{(3)}\right)L_u^\dagger\right]_{ij}=-\frac{v^2}{4} \left[L_u F_Q Y_U M_u^{-2} Y_U^\dagger F_Q L_u^\dagger\right]_{ij},
 \\ 
 \label{eq:gRZdgears}
	\left[\delta g_{R}\right]_{ij}^{Z_d}&=-\frac{v^2}{2}\,\left[R_d\,w_{Hd}\,R_d^\dagger\right]_{ij}=-\frac{v^2}{4} \left[R_d F_d Y_D^\dagger M_Q^{-2}Y_D F_d R_d^\dagger\right]_{ij},
 \\
 \label{eq:gRZugears}
	\left[\delta g_{R}\right]_{ij}^{Z_u}&=-\frac{v^2}{2}\,\left[ R_u\,w_{Hu}\,R_u^\dagger\right]_{ij}=\frac{v^2}{4} \left[R_u F_u Y_U^\dagger M_Q^{-2} Y_U F_u R_u^\dagger\right]_{ij}.
\end{align}
\end{subequations}
The hermiticity of ${\cal L}_Z$, Eq.~\eqref{eq:L_Z}, implies that the above anomalous couplings are real for $i=j$.

Experimental data on $Z$ couplings to the left-handed SM fermions translate to constraints on the clockwork chains of the singlet fermions. Similarly, experimental ranges on $Z$ couplings to the right-handed SM fermions translate to constraints on the clockwork chains of the doublet fermions.  Of special interest  is the coupling $\left[\delta g_{L}\right]_{33}^{Z_d}$, measured in $Z\to b\bar b$ decays. Since it does not receive suppression from the zero-mode overlap (which needs to be large to give large enough top mass), it provides a strong bound on the $M_{d(i)}$ masses. Couplings giving access to $M_{Q(i)}$ and  $M_{u(i)}$ are $\left[\delta g_{R}\right]_{22}^{Z_u}$ and $\left[\delta g_{L}\right]_{33}^{Z_u}$, respectively. The latter is restricted indirectly using $SU(2)_L$ symmetry from the $t\to b W$ decay~\cite{Efrati:2015eaa}. A global fit to the electroweak and low-energy data gives ~\cite{Falkowski:2017pss},
\begin{subequations}
\label{eq:dgZbounds}
\begin{align}
\left[\delta g_{L}\right]_{33}^{Z_d}&=(-0.3\pm0.7)\times10^{-3},
\\
\left[\delta g_{L}\right]_{33}^{Z_u}&=(0.7\pm3.8)\times10^{-2},
\\ 
\left[\delta g_{R}\right]_{22}^{Z_u}&=(0.8\pm2.3)\times10^{-3}. 
\end{align}
\end{subequations}
To obtain the above ranges we used the numerical likelihood provided in Ref.~\cite{Falkowski:2017pss}. The measurements in Eq.~\eqref{eq:dgZbounds} correspond to lower bounds 
\begin{align}
	M_{d(i)}\gtrsim 3.8~\text{TeV},~~~M_{u(i)}\gtrsim 0.5~\text{TeV},~~~M_{Q(i)}\gtrsim 0.5~\text{TeV}, 
\end{align}
at $90\%$ C.L., assuming Yukawa couplings of order ${\mathcal O}(1)$.~\footnote{The SMEFT breaks down for such small gear masses. Nonetheless, as discussed in Sec.~\ref{sec:collider}, direct searches set lower limits of gear masses above 1 TeV, for which our treatment of low-energy observables remains valid.}

The gears also contribute to the charged-current interactions, modifying the $W$ couplings to quarks, 
\begin{align}\label{eq:dgWgears}
\mathcal L_W \supset&\frac{g_L}{\sqrt{2}}\left(V_{ij}+\left[\delta g_L\right]_{ij}^{W_q}\right)\bar u_i\gamma^\mu P_L d_j W_\mu^++\frac{g_L}{\sqrt{2}}\left[\delta g_R\right]_{ij}^{W_q}\bar u_i\gamma^\mu P_R d_j W_\mu^++{\rm h.c.},
\end{align}
where,	
\begin{align}
	\left[\delta g_L\right]_{ij}^{W_q}&=v^2\left[L_u w_{Hq}^{(3)} L_d^\dagger\right]_{ij}=-\frac{v^2}{4}\left[L_u F_Q \left(Y_U M_u^{-2} Y_U^\dagger+Y_D M_d^{-2} Y_D^\dagger\right) F_Q L_d^\dagger\right]_{ij},\nonumber\\
	\left[\delta g_R\right]_{ij}^{W_q}&=\frac{v^2}{2}\left[R_u w_{Hud} R_d^\dagger\right]_{ij}=\frac{v^2}{2}\left[R_u F_u Y_U^\dagger M_{Q}^{-2} Y_D F_d R_d^\dagger\right]_{ij}.
\end{align}
For tree-level processes, these contributions can be thought of as being absorbed in the definition of the measured CKM matrix elements, but with a CKM matrix that is in principle not unitary.  In the numerical analysis we take for the SM inputs the CKM elements that are determined from tree-level processes only, but nevertheless employ CKM unitarity to recover the full CKM to be used in computing FCNC observables. As we will see below, the most stringent constraints on the clockwork parameters come from both tree-level and one-loop induced FCNCs. They restrict the phenomenologically viable clockwork parameter space to only small deviations in the CKM unitarity tests, such as  $|V_{ud}|^2+|V_{us}|^2+|V_{ub}|^2=1$ (see e.g. Ref.~\cite{Gonzalez-Alonso:2016etj}).  In particular, the deviations we obtain within the clockwork model are below current experimental precision of such measurements, making our approach self-consistent and the use of CKM unitarity justified.

\subsubsection{Rare meson decays}
\label{sec:raredecays}

Important constraints on new physics generally arise from rare meson decays triggered by the $b\to s\ell\ell$, $b\to s\nu\bar\nu$ and $s\to d\nu\bar\nu$ transitions. 
The contributions of the gears to these processes enjoy a flavor suppression from the overlaps of the external
fields equivalent to the GIM mechanism in the SM, but they occur at tree level from the diagrams in Fig.~\ref{fig:flavormix} where the external $Z$ or Higgs boson connects to a dilepton pair. 
The effective Hamiltonian used to describe the ``short-distance'' contributions to these decays at low energies is,
\beq
\label{eq:HeffDF1}
\begin{split}
\mathcal H_{\rm ew}\supset&-\frac{\alpha}{2\pi v^2}\lambda_{ij}^{(t)}\left[C_9\left(\bar d_i\gamma^\mu d_{L j}\right)\left(\bar\ell\gamma_\mu\ell\right)+C_{10}\left(\bar d_i\gamma^\mu d_{L j}\right)\left(\bar\ell\gamma_\mu\gamma_5\ell\right)\right.
\\
&\left.+C_{\nu}\left(\bar d_i\gamma^\mu d_{L j}\right)\left(\bar\nu\gamma_\mu(1-\gamma_5)\nu\right)\right],
\end{split}
\eeq
where $\alpha$ is the electromagnetic structure constant, $\lambda^{(t)}_{ij}=V_{ti}^*V_{tj}$, and $C_9^{\rm SM}(m_b)=4.32$, $C_{10}^{\rm SM}=-4.41$~\cite{Bobeth:2003at} and $C_\nu^{\rm SM}=-6.35$~\cite{Brod:2010hi} are the Wilson coefficients  of the LEEFT. The scale-dependence of $C_9$ is due to the substantial RGE effects in QCD. Other terms in $\mathcal H_{\rm ew}$, not displayed in Eq.~(\ref{eq:HeffDF1}), do not receive important contributions from the gears. For instance, the ``primed operators'' corresponding to the operators in Eq.~(\ref{eq:HeffDF1}) with $d_{L i}\to d_{R i}$ replacement are  further suppressed by the zero-mode overlaps. 

The contributions of the gears to the $C_{9,10,\nu}$ Wilson coefficients are due to the tree level $Z$ exchange and are, for a given $d_j\to d_i$ process,
\begin{align}\label{eq:C910gears}
[\delta C_{10}]_{ij}&=[\delta C_\nu]_{ij}=-\frac{1}{1-4s^2_W}[\delta C_9]_{ij}=-\frac{2\pi}{\alpha \lambda_{ij}^{(t)}}\left[\delta g_{L}\right]_{ij}^{Z_d},
\end{align}
with the $Z d_i d_j$ coupling given in Eq.~\eqref{eq:gLZdgears}.
 These rare decays then set constraints on the tower of gears of the $d_{R(i)}$ fields.

Branching fractions and angular distributions of different $b\to s\ell\ell$ decays were measured by a number of experiments.   For instance, the $B_s\to\mu\mu$ branching fraction has been measured by LHCb, CMS and ATLAS~\cite{Aaij:2013aka,Chatrchyan:2013bka,Aaboud:2016ire}. Normalizing the experimental measurements to the  theoretical predictions~\cite{Bobeth:2013uxa} gives,
\begin{align}\label{eq:R}
 R=\frac{\overline{\rm BR}(B_s\to\mu\mu)}{\overline{\rm BR}(B_s\to\mu\mu)^{\rm SM}}=\left|\frac{C_{10}
 }{C_{10}^{\rm SM}}\right|^2=0.83(16),
\end{align}
which implies $\left[\delta C_{10}\right]_{23}=0.39\pm0.37$.  This is consistent with a tower of gears of mass $M_{d(i)}\simeq5.9$ TeV, and excludes $M_{d(i)}\gtrsim 3.5$ TeV, at 90\% C.L., barring cancellations and taking Yukawa couplings to be ${\mathcal O}(1)$ in the sums in Eq.~(\ref{eq:C910gears}). Taking $M_{d(i)}\simeq5.9$ TeV then also leads to  $\left[\delta C_9\right]_{23}\simeq-0.04\pm0.04$, which is beyond the current sensitivity of ongoing experiments.  Similar bounds can also be extracted from a different flavor entry, $\left[\delta C_{10}\right]_{13}$, using the current limits on $B_d\to\mu\mu$.

For $q_i\to q_j \nu\bar\nu$ transitions the experimental upper bounds have been set on the $B\to K^{(*)}\nu\bar\nu$ branching ratio, while there is a measurement of the $K^+\to \pi^+\nu\bar\nu$ rate~\cite{Artamonov:2008qb}. Comparing the latter  to the SM prediction~\cite{Brod:2010hi} gives $\left[\delta C_\nu\right]_{12}=2.7^{+2.2}_{-3.2}$, which corresponds to a generic bound $M_{d(i)}\gtrsim 1.7$ TeV. While this is worse than the bound we obtained from $\left[\delta C_{10}\right]_{23}$, it corresponds to a different combination of Yukawa couplings in Eq.~(\ref{eq:C910gears}). The decay $B\to K^{(*)}\nu\bar\nu$, on the other hand, probes exactly the same combination of couplings already constrained by $B_s\to\mu\mu$. Combining the experimental bounds with the SM predictions in Ref.~\cite{Buras:2014fpa}, we obtain $\left[C_\nu\right]_{23}\in[-6.8,\,19.5]$ at 90\% C.L., which translates to a bound $M_{d(i)}\gtrsim 0.8$ TeV.      

	Rare decays involving up-type quarks suffer severe GIM suppression of the penguin and box diagrams in the SM. The corresponding top decays are extremely rare, with branching ratios in the range ${\mathcal O}(10^{-12}$-$10^{-15})$~\cite{Eilam:1990zc,Mele:1998ag,AguilarSaavedra:2002ns} in the SM, while the rare $D$-meson decays are dominated by the ``long-distance'' contributions that are difficult to quantify~\cite{Burdman:2001tf}. Current experimental bounds on, e.g., BR$(t\to u\,Z)\lesssim0.022\%$ and BR$(t\to c\,Z)\lesssim0.049\%$ (at 95\% C.L.)~\cite{Sirunyan:2017kkr} are not strong enough yet to significantly constrain new physics with mass scales larger than 1 TeV~\cite{Durieux:2014xla}. Experimental bounds on rare $D$ meson decays, on the other hand, have reached a considerable sensitivity in the $D\to\mu\mu$ decay, ${\rm BR}(D\to\mu\mu)^{\rm exp.}<6.2\times 10^{-9}$ at 90\% C.L.~\cite{Aaij:2013cza}, while the SM prediction is estimated to lie below $ 10^{-10}$~\cite{Fajfer:2015mia}. The contributions of the clockwork chains to the decay rates are suppressed by at least $\lambda^{10}$, leading to only a very weak bound, with data still compatible with  $M\gtrsim0.1$ TeV. Bounds from the  other charm quark decays such as $D\to\pi \mu\mu$ lead to even weaker limits~\cite{Fajfer:2015mia}.

Finally, we comment on the experimental hints of non-standard contributions in $b\to s\ell\ell$ transitions that could be solved by lepton-non-universal new-physics effects in  $C_{9,10}$~\cite{Aaij:2014ora,Aaij:2017vbb,Geng:2017svp,Altmannshofer:2017yso,Capdevila:2017bsm,Ciuchini:2017mik,DAmico:2017mtc}. Although the gears 
can significantly contribute to these Wilson coefficients, the couplings to the leptons are governed in our set-up  by the SM couplings to the $Z$, which are lepton flavor universal. This, for instance, explains the accidental suppression by $1-4s^2_W\sim0.1$ of the vectorial coupling to the charged leptons in Eq.~(\ref{eq:C910gears})  compared to the axial one. In our framework the lepton non-universal contributions to $b\to s\ell\ell$ only arises from the exchange of a Higgs in the diagram \textit{(c)} in Fig.~\ref{fig:flavormix}. Although this is a (scalar) tree-level contribution it is further suppressed by the SM lepton Yukawas. Extending the clockwork mechanism to the lepton sector could, in principle, explain the simultaneous non-standard violations of quark and lepton flavor suggested by the data.

\subsubsection{Neutral-meson mixing}
\label{sec:mesonmixing}

The LEEFT for neutral meson mixing can be described by 
\beq
\mathcal H_{\rm ew}\supset \sum C_\alpha \mathcal O_\alpha^{ij},
\eeq
 where the operators in the $d$-quark sector and in the so-called ``chiral basis'' are~\cite{Buras:2001ra},
 \begin{subequations}
 \label{eq:Omixing}
\begin{align}
\mathcal O_{\rm VLL}^{ij}&=(\bar d_i\gamma^\mu P_L d_j)(\bar d_i\gamma_\mu P_L d_j), &&
\\
\mathcal O_{{\rm LR},1}^{ij}&=(\bar d_i\gamma^\mu P_L d_j)(\bar d_i\gamma_\mu P_R d_j),
&\mathcal O_{{\rm LR},2}^{ij}&=(\bar d_i P_L d_j)(\bar d_i P_R d_j),
\\
\mathcal O_{{\rm SLL},1}^{ij}&=(\bar d_i P_L d_j)(\bar d_i P_L d_j),
&\mathcal O_{{\rm SLL},2}^{ij}&=(\bar d_i \sigma_{\mu\nu}P_L d_j)(\bar d_i \sigma^{\mu\nu}P_L d_j),
\end{align}
\end{subequations}
and in addition the three operators $\mathcal O_{VRR}^{ij}$ and $\mathcal O_{SRR,1(2)}^{ij}$ obtained through the $P_L\to P_R$ replacement from $\mathcal O_{VLL}^{ij}$ and $\mathcal O_{SLL,1(2)}^{ij}$, respectively. The SM interactions only generate $\mathcal O_{VLL}^{ij}$, while other operators can mix under RG running in QCD.
\begin{figure}[t]
\begin{center}
\includegraphics[width=12cm]{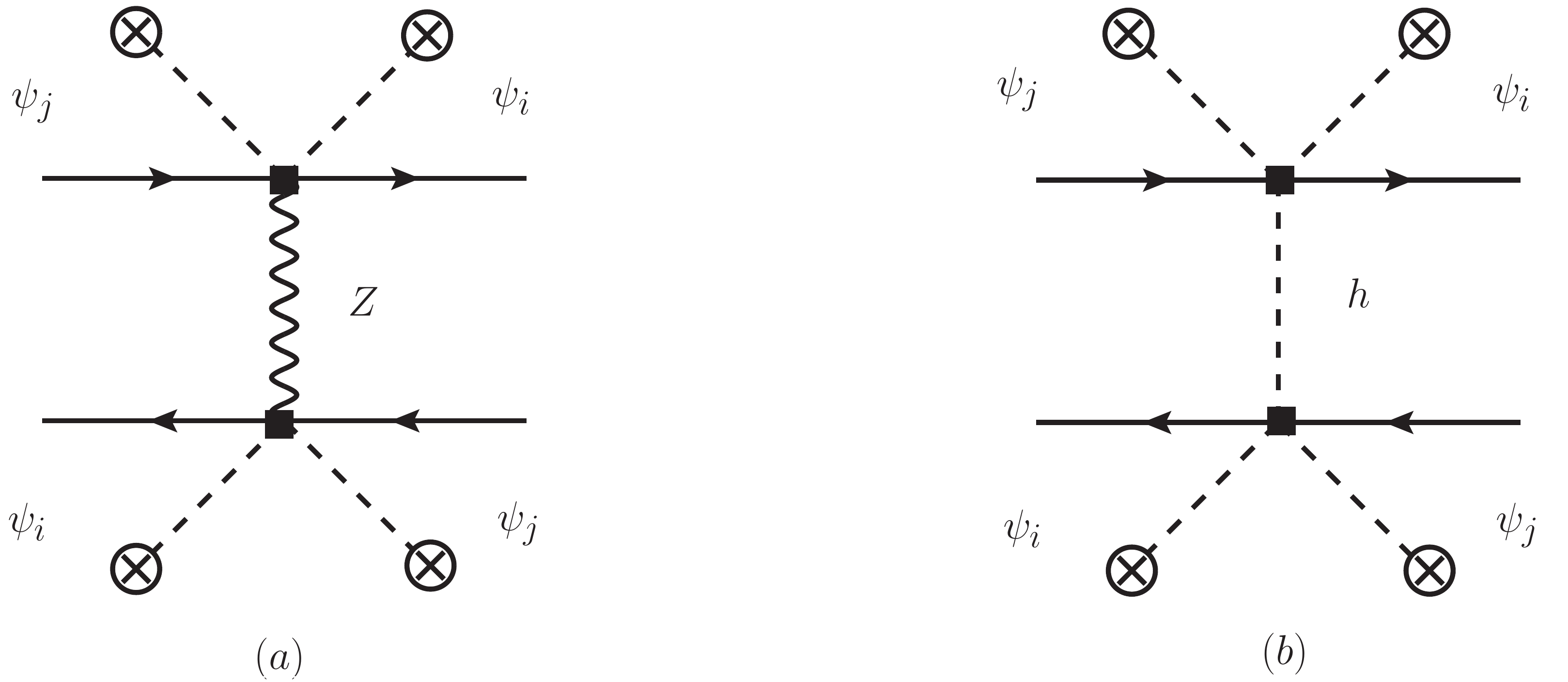}
\caption{\label{fig:flavormixTL2} Diagrams contributing to neutral-meson mixing at tree-level and ${\mathcal   O}(v^4/M^4)$ induced by two insertions of $\psi^2H^2D$ operators. 
}
\end{center}
\end{figure}

For $B_q-\bar B_q$ mixing meson oscilations, the leading contributions from clockwork gears are to $\mathcal O_{\rm VLL}$, while contributions to the other operators in Eq.~\eqref{eq:Omixing} are suppressed by further powers of $\lambda$.
Integrating out the gears, the gauge boson and the top quark at the electroweak scale $\mu=m_W$, gives
\beq
\begin{split}
\label{eq:mixing1}
\left[C_{\rm VLL}\right]_{ij}(m_W)&=\frac{2}{v^2}\left(\left[\delta g_L\right]_{ij}^{Zd}\right)^2-\eta_6^{2/7}\left[(L_d L_d)\cdot\left(w_{QQ}^{(1)}(M)+w_{QQ}^{(3)}(M)\right)\cdot (L_d L_d)^\dagger\right]_{ij}
\\
&-\frac{y_t^2\,\lambda_{ij}^{(t)}}{8\pi^2v^2}\left[V\,\delta g_L^{Zu}\,V^\dagger\right]_{ij}\log\frac{M}{m_W}+\left[\Delta C_{\rm VLL}(m_W)\right]_{ij}, 
\end{split}
\eeq
where we introduced the following notation
\begin{align}
\hspace{-0.2cm}\left[(L_d L_d)\cdot\left(w_{QQ}^{(1)}+w_{QQ}^{(3)}\right)\cdot (L_d L_d)^\dagger\right]_{ij}=\sum_{rstu}[L_d]_{ir}[L_d]_{it}[L_d^\dagger]_{sj}[L_d^\dagger]_{uj}[w_{QQ}^{(1)}+w_{QQ}^{(3)}]_{rstu},
\end{align}
for the contraction of the indices of the unitary rotations with those of the Wilson coefficients. The SMEFT result in \eqref{eq:mixing1} was evolved from the heavy scale $m\approx M_{\psi(i)}$. The first term in this equation is the $\mathcal{O}(v^4/M^4)$ tree-level contribution from the $Z$-boson exchange  in Fig.~\ref{fig:flavormixTL2} \textit{(a)} with a double insertion of the anomalous $Z\,d_id_j$ left-handed coupling, Eq.~\eqref{eq:gLZdgears}. This term is thus sensitive to the same combination of masses and couplings of the $d_R$-type gears as the $d_i \to d_j \ell \ell$ decays, discussed in Section \ref{sec:raredecays}. The second term encompasses the finite pieces of order $(4\pi)^{-2}\times \mathcal O(v^2/M^2)$ obtained from the box diagrams in Fig.~\ref{fig:flavormixloop} at $\mu=M$. The dots denote matrix multiplication in the generation indices, and the SMEFT Wilson coefficients $w_{QQ}^{(1,3)}$ are given in Eq. \eqref{eq:wQQ13}.
The prefactor is the result of QCD running of the $\mathcal O_{\rm VLL}$ operator, performed with six dynamical flavors, $\eta_{6}=\alpha^{(6)}_s(M)/\alpha^{(6)}_s(m_W)$. This contribution is sensitive to the clockwork chains of both the $d_{R(i)}$ and $u_{R(i)}$ quarks, \textit{cf.} Eq.~\eqref{eq:wQQ13}. 

The first term in the second line of Eq. \eqref{eq:mixing1} is due to the mixing of the $\mathcal O_{HQ}^{(1,3)}$ operators into $\mathcal O_{QQ}^{(1,3)}$~\cite{Jenkins:2013zja,Jenkins:2013wua,Alonso:2013hga}. It is of order $(4\pi)^{-2}\times\mathcal O(v^2/M^2)$ but logarithmically enhanced by $\log(M/m_W)$.
It depends on the anomalous $Z\,u_iu_j$ left-handed coupling, Eq.~\eqref{eq:gLZugears}, and is sensitive only to the gears of the $u_{R(i)}$ quarks. Finally, the last piece in Eq.~(\ref{eq:mixing1}) is a finite contribution that is due to the matching of SMEFT onto LEEFT at $\mu=m_W$, whose explicit expression was obtained in Ref.~\cite{Bobeth:2017xry} and that we reproduce  in App.~\ref{sec:matching}, Eq.~\eqref{eq:CVLLfinite}, adapted to our normalization of the operators. This contribution is sensitive to the clockwork chains of both the $d_{R(i)}$ and $u_{R(i)}$ quarks.

For neutral-kaon mixing both the contributions to $\mathcal O_{\rm VLL}^{ij}$, Eq. \eqref{eq:mixing1}, and to the other operators in \eqref{eq:Omixing} are important. The additional $\mathcal{O}(\lambda^2)$ suppression 
of gear contributions to the operators  $\mathcal O_{{\rm SRR},1}$ and $\mathcal O_{{\rm LR},1(2)}$ is compensated by their large enhancements that is a combination of QCD running  and the chiral enhancement of the corresponding hadronic matrix elements. The matching conditions for these operators at  $\mu=m_W$ are
\begin{align}
[C_{\rm VRR}]_{ij}(m_W)&=\frac{1}{v^2}([\delta g_R^{Z_d}]_{ij})^2-\eta_6^{2/7}\left[R_d R_d\cdot\,w_{dd}(M)\cdot (R_d R_d)^\dagger\right]_{ij},\label{eq:mixingVRR}\\
\begin{split}
\label{eq:mixingCLR1}
[C_{{\rm LR},1}]_{ij}(m_W)&=\frac{1}{v^2}[\delta g_L^{Z_d}]_{ij}[\delta g_R^{Z_d}]_{ij}-\eta_6^{1/7}\left[L_d R_d\cdot\,w_{Qd}^{(1)}(M)\cdot (L_d R_d)^\dagger\right]_{ij} 
\\
&\qquad-\frac{y_t^2\,\lambda_{ij}^{(t)}}{8\pi^2v^2}\left[\delta g_{R}^{Z_d}\right]_{ij}\log\frac{M}{m_W}+\left[\Delta C_{{\rm LR},1}(m_W)\right]_{ij},
\end{split}
\\
\begin{split}
\label{eq:mixingCLR2}
[C_{{\rm LR},2}]_{ij}(m_W)&=-\frac{2}{3}(\eta_6^{1/7}-\eta_6^{-8/7})\left[L_d R_d\cdot\,w_{Qd}^{(1)}(M)\cdot (L_d R_d)^\dagger\right]_{ij}
\\
&\qquad-\frac{1}{m_h^2}[\delta y_d]_{ij}[\delta y_d]_{ji}^*,
\end{split}
\\
[C_{{\rm SLL},1}]_{ij}(m_W)&=-\frac{([\delta y_d]^\dagger_{ij})^2}{2m_h^2},
\\
[C_{{\rm SRR},1}]_{ij}(m_W)&=-\frac{([\delta y_d]_{ij})^2}{2m_h^2}.
\end{align}
Contributions to $\mathcal O_{{\rm SLL},2}$ and $\mathcal O_{{\rm SRR},2}$ are absent but generated by the renormalization of $\mathcal O_{{\rm SLL},1}$ and $\mathcal O_{{\rm SRR},1}$ in QCD and the RG running~\cite{Buras:2001ra}.
The operator $\mathcal O_{\rm VRR}$ receives contributions from the tree-level exchange of the $Z$-boson in Fig.~\ref{fig:flavormixTL2} \textit{(a)} and from the loop contribution in Fig.~\ref{fig:flavormixloop} evaluated at $\mu=M$. On the other hand, the structure of the contributions to $C_{{\rm LR},1}$ is similar to the one of $C_{\rm VLL}$ in Eq.~(\ref{eq:mixing1}). The first term in Eq.~\eqref{eq:mixingCLR1} is due to the tree-level $Z$ exchange, the second is the finite part from the loops, the third is the contribution induced by electroweak mixing from the $\psi^2 H^2D$ operators, while the last term is the finite part from the matching between SMEFT and LEEFT at $\mu=m_W$, Eq. \eqref{eq:CVLRfinite}. These contributions depend on $\delta g_R^{Z_d}$ and $w_{Qd}^{(1)}$ and are due to clockwork towers of the doublet quarks. 

\begin{table}[t]
\centering
\begin{tabular}{cccccc}
\hline\hline
Process & $U$ & $D$ & $Q$ & $UQ$ & $DQ$\\  
\hline
$B_s$-$\bar B_s$ &{\color{TealBlue} $\lambda^4$, \textcurrency$^*$ }& {\color{TealBlue}$\lambda^4$, $\times$ and \textcurrency} & $\lambda^7$, \textcurrency$^*$ & $\lambda^7$, \textcurrency  & $\lambda^{6}$, $\times$ \\
$B$-$\bar B$ & {\color{TealBlue}$\lambda^6$, \textcurrency$^*$ }& {\color{TealBlue}$\lambda^6$, $\times$ and \textcurrency }& $\lambda^9$, \textcurrency$^*$ &  $\lambda^9$, \textcurrency & $\lambda^{8}$, $\times$\\
$K$-$\bar K$ & {\color{TealBlue}$\lambda^{10}$, \textcurrency$^*$}&  {\color{TealBlue}$\lambda^{10}$, $\times$ and \textcurrency} & {\color{TealBlue}$\lambda^{12}$, \textcurrency$^{*\dagger}$} &  {\color{TealBlue}$\lambda^{12}$, \textcurrency$^\dagger$ }& {\color{TealBlue}$\lambda^{12}$, $\times^{\dagger}$ and \textcurrency$^\dagger$ } \\
$D$-$\bar D$ & $\lambda^{10}$, $\times$ and \textcurrency & $\lambda^{10}$, \textcurrency & $\lambda^{10}$, $\times$ and \textcurrency &  {\color{TealBlue} $\lambda^{8}$, $\times^\dagger$} & $\lambda^{10}$, \textcurrency$^\dagger$\\
\hline\hline
\end{tabular}
\caption{Structure of the contributions of the clockwork model to neutral-meson mixing, where the gears of a given type can appear alone (listed as $U$, $D$ and $Q$) or in pairs (listed as $UQ$ and $DQ$). In each entry we show the order in $\lambda$ at which the contribution starts. We also indicate by the symbol ``$\times$'' if the contributions is an $\mathcal O(v^4/M^4)$ tree-level contribution and by the symbol ``\textcurrency'' if it is a $\mathcal O(v^2/M^2)$ loop contribution. A ``~$\dagger$~'' superscript in either of the two symbols further indicates that the given contribution receives a chiral enancement, and a ``~*~'' superscript in the loop symbol indicates that the contribution is logarithmically enhanced. Entries in teal blue are the leading contributions taken into account in the numerical analyses. \label{tab:mixingsummary}}
\end{table}

In $C_{{\rm LR},2}$ the first  term is the one that is produced by the box diagrams, while the second term is the $\mathcal{O}(v^4/M^4)$ contributions from tree-level diagram in Fig.~\ref{fig:flavormixTL2} \textit{(b)} with a Higgs-boson exchanged and a double insertion of the anomalous Yukawa couplings in Eq.~(\ref{eq:deltaydYukawas}). Finally, the contributions to $C_{{\rm LR},1}$, $C_{{\rm SRR},1}$ and $C_{{\rm SLL},1}$ are due to the tree level diagrams in Fig.~\ref{fig:flavormixTL2} \textit{(b)}. These chirally-enhanced operators contribute to kaon mixing with a suppression $\mathcal{O} (\lambda^{12})$ and are sensitive mainly to the $Q_{L(i)}$-gears and to a combination of $Q_{L(i)}$- and $d_{R(i)}$-gears.

For  $D^0-\bar D^0$ mixing the same operators as in Eq.~(\ref{eq:Omixing}) are relevant, but with $d_i$- and $d_j$-quarks replaced by $u$- and $c$-quarks. One obtains similar expressions for the diagrams in Fig.~\ref{fig:flavormixTL2}, now with modified couplings to the up-quarks, and for the contributions from the box diagrams in Fig.~\ref{fig:flavormixloop}, replacing the rotation matrices and using $w_{uu}$ and $w_{Qu}^{(1)}$ in Eqs.~\eqref{eq:mixingVRR} and (\ref{eq:mixingCLR1},~\ref{eq:mixingCLR2}), respectively. In this case the contributions of the electroweak mixing from $\psi^2H^2D$ operators and the finite contributions in the matching between the SMEFT and LEEFT are negligible because the flavor mixing comes suppressed by the square of SM down-type quark Yukawas. The structure of the clockwork suppression is now different, with the leading contributions due to $\mathcal O_{\rm SRR,1}$ at $\mathcal O(\lambda^{8})$ that translates to bounds on a combination of gear masses of type $Q_{L(i)}$ and $u_{R(i)}$. Contributions to $\mathcal O_{{\rm LR},1}$, $O_{{\rm LR},2}$ and $\mathcal O_{{\rm VRR}}$ are further suppressed by $\lambda^{10}$.

In Tab.~\ref{tab:mixingsummary} we summarize concisely the sensitivity of neutral meson-mixings to different gear species, specifiying the structure of the dominant contributions, which are taken into account in the numerical analysis. 

The experimental data we use are the observables $C_{B_q}$ and $\phi_{B_q}$ for $B_q$ - $\bar B_q$, $C_{\epsilon_K}$ and $C_{\Delta m_K}$ for  $K$ - $\bar K$ and $|M_{12}|$ and $\Phi_{12}$ for $D$ - $\bar D$ mixing, as defined in Ref.~\cite{Bona:2007vi} and reported in the latest results of the UTfit collaboration~\cite{Bona:2017kam,Alpigiani:2017lpj}. For the predictions we use the masses and CKM mixing parameters in Tab.~\ref{tab:numinputs} as the fundamental input parameters, whereas for the hadronic matrix-elements (bag parameters) we use results of lattice QCD calculations. In particular, for the neutral $K$-meson system we use the calculation of the SWME collaboration~\cite{Jang:2015sla} 
(for other similar calculations see Refs.~\cite{Garron:2016mva,Carrasco:2015pra}). For $D$ - $\bar D$ mixing, we use the calculation of the ETMC~\cite{Carrasco:2015pra}, while for the $B_q$ neutral meson system the leading gear contributions to $B_q$ - $\bar B_q$ mixing have the same structure as the SM one (see Sec.~\ref{sec:mesonmixing}), and the bag-parameter factors out from $C_{B_q}$ and $\phi_{B_q}$. Finally, to connect the contributions to the operators at the electroweak scale to the ones at the low-energy scales, we use the master formulas for the RG running in QCD up to LL accuracy as given in Ref.~\cite{Buras:2001ra}.

\subsection{Numerical scan}
\label{sec:scans}

In order to investigate the impact of different low-energy constraints we perform a numerical scan
over Yukawa matrices and gear masses. We start by inverting the relations between diagonal quark mass matrices and Yukawa couplings $Y_{U,D}$, whereby we factor out the flavor suppression induced by the clockwork zero-mode overlaps, 
\begin{align}
\label{eq:scan0} 
	Y_U=F_Q^{-1}\,L_u^\dagger\,\diag(\overline{m}_u)\,R_u\,F_u^{-1},~~~~~Y_D=F_Q^{-1}\,L_d^\dagger\,\diag(\overline{m}_d)\,R_d\,F_d^{-1}.
\end{align}
For the overlaps we take
\begin{align}
	\label{eq:overlaps}
	F_Q=\text{diag}(\lambda^3,~\lambda^2,~1),~~F_u=\text{diag}(\lambda^4,~\lambda,~1),~~F_d=\text{diag}(\lambda^4,~\lambda^3,~\lambda^2),
\end{align}
where in the scan the exact relations are used. The numerical scan is set-up in such a way that all the generated $Y_{U,D}$ matrices give the  central values of the measured CKM matrix elements and the light quark masses. 

Provided the entries of $Y_U$ and $Y_D$ are anarchical, the flavor hierarchies in $Y_U^{\rm SM}$ and $Y_D^{\rm SM}$ manifest itself in the structure of the rotation matrices,
\begin{subequations}
\label{eq:scan1}
\begin{align}
L_u&\approx L_d \approx V_{\rm CKM}\approx \left( \begin{array}{ccc}
1 & \lambda & \lambda^3 \\
\lambda & 1 & \lambda^2 \\
\lambda^3 & \lambda^2 & 1 \end{array} \right),
\\
R_u&\approx
 \left( \begin{array}{ccc}
1 & \lambda^3 & \lambda^4 \\
\lambda^3 & 1 & \lambda \\
\lambda^4 & \lambda & 1 \end{array} 
\right),
\hspace{2cm} R_d\approx 
\left( \begin{array}{ccc}
1 & \lambda & \lambda^2 \\
\lambda & 1 & \lambda \\
\lambda^2 & \lambda & 1 \end{array} 
\right).
\end{align}
\end{subequations}

\begin{figure}[t]
\begin{center}
\includegraphics[width=10cm]{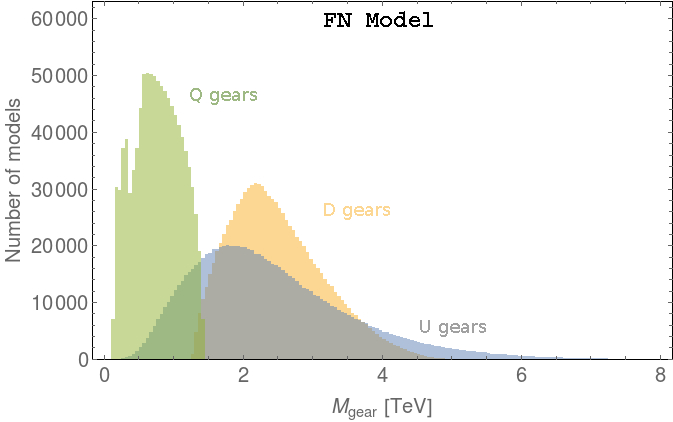}
	\caption{\label{fig:scan1} Distribution of the lower-mass bounds (in TeV) for a million pairs of randomly generated anarchic Yukawa matrices, $Y_{U,D}$, such that the SM quark masses and CKM matrix elements are at their experimental values. The lower bounds on the $M_{u,d,Q}$ masses follow from low-energy observables assuming that only one type of gear is active at a time.  
}
\end{center}
\end{figure}

\begin{table}[t]
\centering
\begin{tabular}{cccccccccc}
\hline\hline
	$\overline{m}_u$ & $\overline{m}_c$ & $\overline{m}_t$ & $\overline{m}_d$ & $\overline{m}_s$ & $\overline{m}_b$ & $|V_{us}|$ & $|V_{ub}|$ & $|V_{cb}|$ & $\gamma$ \\
\hline
0.0010& 0.47 & 135 & 0.0021 & 0.043 & 2.3 & 0.22508& $3.73\cdot 10^{-3}$ & $4.17\cdot 10^{-2}$ & $72.1^o$\\
\hline\hline
\end{tabular}
\caption{Inputs to the numerical scan, with masses given in GeV. The $\overline{\text{MS}}$ masses from Ref.~\cite{Patrignani:2016xqp} are evolved up to $\mu=2$ TeV using 3 loop QCD RGE. The CKM entries and the angle $\gamma$ are taken from measurements of tree-level processes as averaged by the PDG.\label{tab:numinputs}}
\end{table}

In the scan we randomly generate the three unitary matrices $R_u$, $R_d$ and $L_d$, while preserving their parametric structure, Eq.~\eqref{eq:scan1}. The $L_u$  matrix is then obtained from $L_u=V_{\rm CKM}\,L_d$. Each of the three matrices, $R_{u,d}$ and $L_d$, is parametrized by three angles and six complex phases (see e.g. Ref.~\cite{Bronzan:1988wa}). The scans are over flat priors in angles and weak phases. The angles are restricted to be in the region $[\pi/10,~\pi/2]\cdot\lambda^n$ with the appropriate power of $n$, while the phases are left unconstrained. 

The values of the CKM matrix elements and the quark masses used in the scan are given in Tab.~\ref{tab:numinputs}. 
The $\overline{\text{MS}}$ quark masses were RG  evolved to the typical gear mass scale,  which we take to be $\mu=2$ TeV.  We ensure the unitarity of CKM matrix through the use of Wolfenstein parametrization including up to $\mathcal{O}(\lambda^5)$ corrections. The four required experimental inputs are taken exclusively from tree-level processes. The resulting CKM matrix elements are polluted by the gear contributions but not at a level that is detectable at the current precision of the CKM-unitarity tests, as discussed in Sec.~\ref{sec:weakbosondecays}. 

To the generated Yukawa matrices we apply all the low-energy constrains described in Sec.~\ref{sec:lowEconstraints}, giving the lower bound on the masses of the gears.  For simplicity, we take these to be generation independent and neglect $\mathcal{O}(1/q)$ and $\mathcal O (v^2/M^2)$ terms, so that, 
\beq
	M_u=M^k_{u(i)}\,, \quad
	M_d=M^k_{d(i)}\,,  \quad
	M_Q=M^k_{Q(i)}\,.
\eeq

Since the neutral-mixing amplitudes receive several different contributions, as summarized in Tab.~\ref{tab:mixingsummary}, we perform the scan in two steps. 
We start with an exploratory scan in which we assume that only one of the gear species is active at a time, either $Q$-, $u$- or $d$- type gears, and then decouple the other two. We generate a million pairs of anarchic Yukawa matrices, $Y_{U,D}$. For each such pair we find the most stringent bound on the masses $M_u$, $M_d$ and $M_Q$ by demanding that the observables are within the $1\sigma$ experimental band~\footnote{To be more precise, we center the SM prediction on the experimental central value and determine the bound on the given mass by saturating the $1\sigma$ range by the clockwork contribution.}. The results are plotted in Fig.~\ref{fig:scan1}. The strongest bound is on the mass of $d$-type gears, $M_d$, and is due to the experimental bound on NP corrections in $Z\to b_L \bar{b}_L$ decay. The strongest bounds on $M_u$ are either due to $B_s$-$\bar B_s$, $B_d$-$\bar B_d$ mixing observables or due to $\epsilon_K$, depending on the actual values of the Yukawa couplings, while the constraint on the deviations of the $Z$ couplings typically still allow gear masses below $1$ TeV. Finally, the least constrained are the $Q$-type gears, with lower bounds on the masses never stronger than $\sim 1.5$ TeV,  given by $\epsilon_K$. 

\begin{figure}[t]
\begin{center}
\includegraphics[width=9cm]{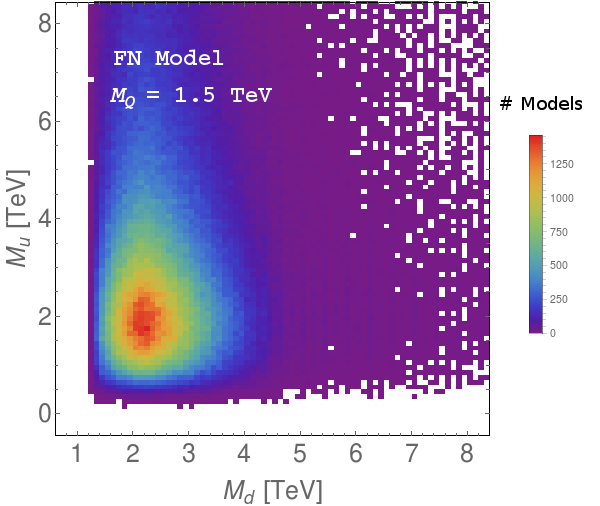}
	\caption{\label{fig:scan2} Distribution of the combined lower bounds on the masses of the lightest gears of up and down type, $M_d,~M_u,$ due to low-energy observables
	for a million pairs of anarchic Yukawa matrices, 
	holding $M_Q$ fixed to 1.5 TeV.
}
\end{center}
\end{figure}

The exploratory scan neglects the potentially important Higgs-mediated contributions to $K$-$\bar K$ and to $D$-$\bar D$ oscillations that involve two types of gears, $d$- and $q$-type or $u$- and $q$-type, respectively. These contributions are parametrically the dominant ones for the two mixing amplitudes. In order to perform a more realistic study we start by fixing $M_Q=1.5$ TeV, which is in the range of direct searches for vector-like quarks at ATLAS and CMS. We repeat the scan including the Higgs-mediated contributions and obtain bounds on $M_u$ and $M_d$ for each pair of randomly generated $Y_{U,D}$ matrices, demanding again a $1\sigma$ consistency with the data. Given the additive nature of these contributions to the observables, our procedure guarantees that our results will be, at worse, within the 2$\sigma$ range of the measured values. The resulting distribution of the lower bounds on $M_d$, $M_u$  are shown in Fig.~\ref{fig:scan2}, which shows that most of the viable points is consistent with $u_R$- and $d_R$-gear masses below 5 TeV.   

Among the viable points we choose two representative examples on which we perform the collider study in the next section.
In both benchmark models we take
\begin{align}
\label{eq:benchmark}
Y_U &=\left( \begin{array}{ccc}
0.275-0.215\,i & 0.167+0.179~i & 0.282-0.076~i \\
0.081-0.154~i &  0.287-0.133~i& -0.105-0.057~i \\
-0.016-0.123~i&-0.417-0.333~i & 0.046+0.765~i\end{array} \right),\nonumber\\
Y_D &=\left( \begin{array}{ccc}
-0.094-0.010\,i & 0.193-0.051~i & 0.140-0.125~i \\
0.426+0.288~i &  0.271+0.247~i& 0.094-0.141~i \\
0.118-0.004~i&0.035-0.073~i & 0.258-0.040~i\end{array} \right).
\end{align}
For the above $Y_U$ and $Y_D$ Yukawa matrices we obtain $\beta_\lambda/(4\pi)^2=-0.0692$, that leads to
$\Lambda_\textgoth{Hell}\equiv\Lambda_{\rm Decay}=41.6\times M_{\rm gear}$. Setting $M_Q=1.5$ TeV, the bounds from low-energy observables require
\begin{align}
\label{eq:le_FN}
	M_u\gtrsim 1.21~\text{TeV},~~~~~M_d\gtrsim 1.38~\text{TeV}.
\end{align}
The values of $N_{\psi(i)}$ are those leading to the overlaps in Eq.~\eqref{eq:overlaps} for $q=1/\lambda$, i.e., 
\beq
\begin{split}
\label{eq:gears}
	N_{Q(1)}&=3,\quad N_{Q(2)}=2, \quad N_{Q(3)}=0,\\
	N_{u(1)}&=4,\quad N_{u(2)}=1, \quad\, N_{u(3)}=0, \\
	N_{d(1)}&=4,\quad N_{d(2)}=3, \quad\,\, N_{d(3)}=2.
\end{split}
\eeq
The two benchmark models used in the collider study are defined as follows:
\begin{description}
	\item[Benchmark 1]
		\quad We take 
		\begin{align}
			m_{Q}&\equiv m_{Q(1)}=m_{Q(2)}=400~\rm{GeV},\nonumber\\
			m_{u}&\equiv m_{u(1)}=m_{u(2)}=367~\rm{GeV},\\
			m_{d}&\equiv m_{d(1)}=\,m_{d(2)}=\,m_{d(3)}\gg m_{Q},m_u,\nonumber
		\end{align}
which corresponds, e.g., to $M_{Q,1}^{(1)}= 1.50~\rm{TeV}$ and $M_{u,1}^{(1)}= 1.33~\rm{TeV}$ and decoupled $d$-type gears.
	\item[Benchmark 2]
		\quad In this case we choose \begin{align}
			m_{Q}&\equiv m_{Q(1)}=m_{Q(2)}=400~\rm{GeV},\nonumber\\
			m_{u}&\equiv m_{u(1)}=m_{u(2)}=367~\rm{GeV},\\
			m_{d}&\equiv m_{d(1)}=\,m_{d(2)}=\,m_{d(3)}=418~\rm{GeV},\nonumber
		\end{align}
		which corresponds. e.g., to $M_{Q,1}^{(1)}= 1.50~{\rm TeV},~ M_{u,1}^{(1)}= 1.33~\rm{TeV}$ and $M_{d,1}^{(1)}=1.52~\rm{TeV}.$

\end{description}

Note that we have chosen our scalings in Eq.~(\ref{eq:q:scaling}) such that the entries of the proto-Yukawas are anarchic but of $\mathcal{O}(\lambda)$ yet one of the eigenvalues is $\mathcal O(1)$. 
Were we to change the scalings for some of the singlet-fields, $q_{u(1)}^{-N_{u(1)}}=q_{d(1)}^{-N_{d(1)}}=\lambda^5$, $q_{u(2)}^{-N_{u(2)}}=\lambda^2$ and $q_{d(3)}^{-N_{d(3)}}=\lambda^3$, the SM fermion spectrum would then be reproduced with all Yukawa entries of $\mathcal{O}(1)$. However, this would also increase drastically the loop contributions to the Higgs self-coupling $\beta$-function making $\Lambda_\text{Decay}$ effectively of the same order as the masses of the gears. Even with our choice of scalings, only a subset of models in the scan (about $600$ out of a million) give a contribution to the Higgs quartic such that $\Lambda_\text{Decay}>10~M_{\rm gear}$, since $Y_U$ must contain a large eigenvalue corresponding to the mass of the top. This turns out to be an especially accute problem for the ``universal $N$'' limit of clockwork flavor model. The latter requires clockworking the third generation doublet and top-singlet fields, which in turn implies that each of the two corresponding zero modes will receive at least an overlap suppression of $\sim 1/2$ from the $q\to 1$ limit, \textit{cf.} Eq.~\eqref{eq:Omixing}. The corresponding increase of some of the proto-Yukawa entries needed to reproduce $m_t$ unavoidably leads to the problem with the running of the Higgs quartic. The only way to realize the ``universal $N$'' limit scenario, seems to require identifying the UV-cut off of the theory with the mass of the heavier gears in the setup.

\section{Collider phenomenology}
\label{sec:collider}

\subsection{Gear spectrum and decay patterns}

\begin{figure}
\begin{center}
\includegraphics[width=7cm]{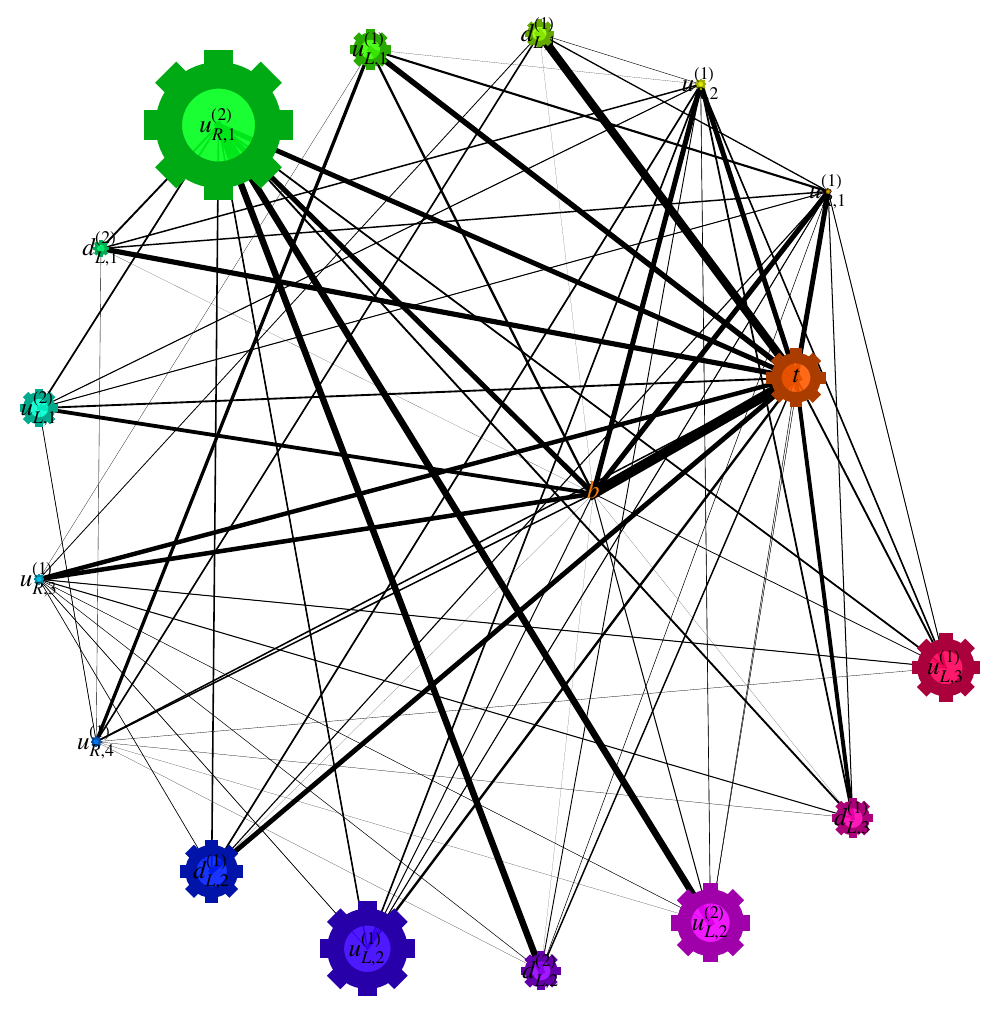}
~~
\includegraphics[width=7cm]{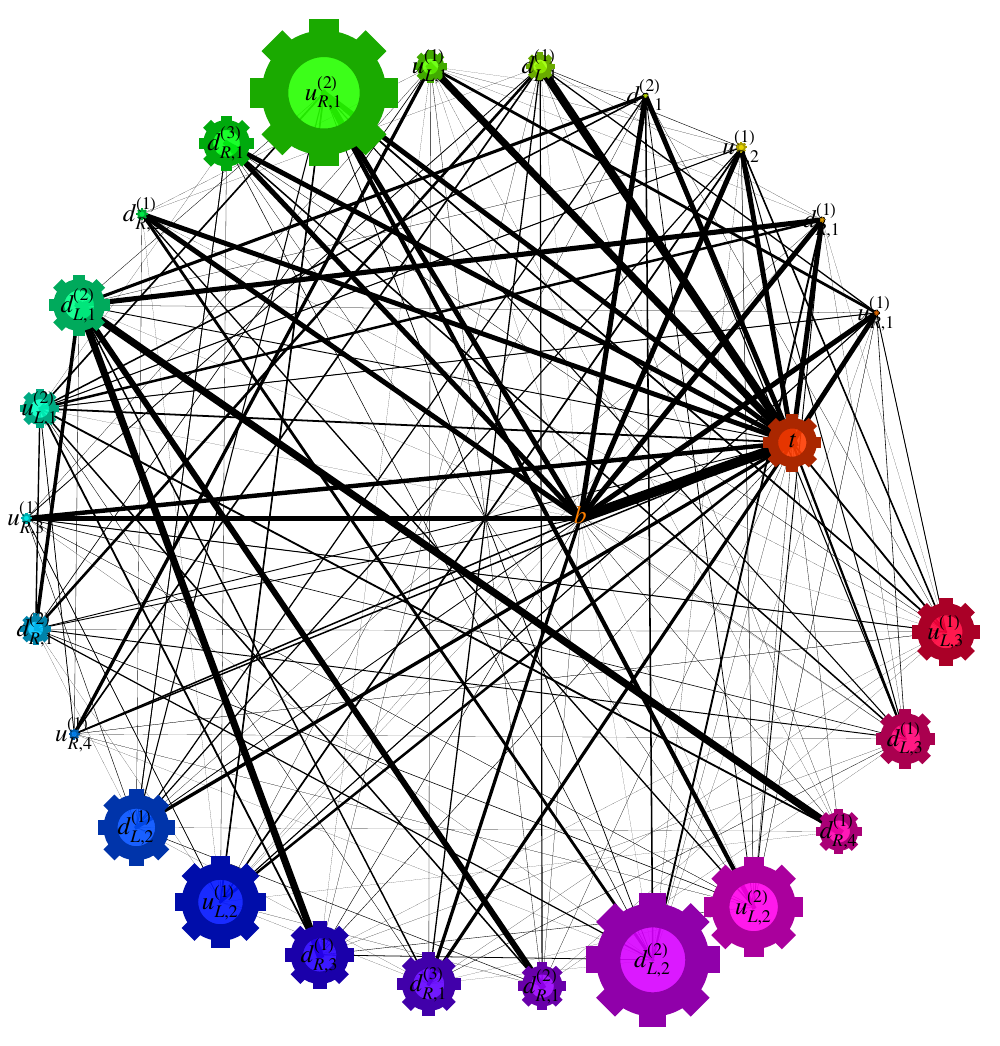}
\caption{\label{fig:spectrum} Spectrum and decay patterns of gears for benchmarks 1 (left plot) and 2 (right plot). The collider accessible gears, along with the SM $b$ and $t$ quarks, are drawn in a mass ordered counter-clockwise spiral,  where $Q_{L,k}^{(i)}=(u_{L,k}^{(i)}, d_{L,k}^{(i)})$ are the mass eigenstates of the vector-like quark $SU(2)_L$ doublet chain, and $u_{R,k}^{(i)}, d_{R,k}^{(i)}$ for the singlet chains. The distance from the center is logarithmically proportional to particle's mass, $\propto \log(M/1 \,{\rm GeV})$, while the symbol's diameter is proportional to particle's width, $\propto \log(1+\Gamma/1{\rm GeV})$. The main decay channels are denoted by black lines, with the width of the lines proportional to the respective branching ratios (the decays have in addition $Z,W$, or $ H$ in the final state, not shown above, and for neutral current transition we sum over the final states with $Z$ and $H$).}
\end{center}
\end{figure}

Gears with TeV scale masses, as allowed by the present low energy constraints, can be searched for at the LHC and at future high energy colliders. The main production channel for the gears is the QCD pair production with the corresponding cross sections precisely calculable~\cite{Czakon:2017wor}. The collider signatures, on the other hand, do depend on the details of the gear decay patterns. The gears decay predominantly through their coupling to the Higgs doublet, cf. Eq.~\eqref{eq:LagSMCK1}, into gears from a different-chirality chain.
The lightest gears decay directly to SM fermions via the emission of $W,Z$ or $h$ as do heavier gears for which these are the only kinematicaly allowed channels. Given the overlap suppression the decays are predominantly to $t$ and $b$.
To illustrate these features we represent in Fig.~\ref{fig:spectrum} the spectrum and decay patterns of the gears for the two benchmarks introduced in the last section.

\subsection{Existing collider constraints}

The main existing collider constraints on clockwork flavor models are  expected to arise from searches for pair production of vector-like quarks, in final states involving third generation SM quarks. In particular, we find the searches for down-like gears decaying to the $tW$ channel, as well as searches for up-like gears decaying to the $tH$ and $tZ$ final states, to be most sensitive. We tested our benchmarks with {\tt CheckMate 2.0.26}~\cite{Dercks:2016npn} and found that they are consistent with all 13TeV searches implemented therein.

To perform a more detailed analysis, we recast the recent ATLAS search for vector-like quarks decaying into $tW$ final states~\cite{Aaboud:2018uek} as well as the analogous search employing the $tZ$ and $tH$ final states~\cite{Aaboud:2018xuw}, both using 35 fb${}^{-1}$ of LHC data at 13TeV. 
To this end we implemented the benchmarks into a {\tt Feynrules}~\cite{Alloul:2013bka} model and simulated gear production and decays using {\tt aMC@NLO}~\cite{Alwall:2014hca}. The resulting cross-sections obtained at LO in QCD were rescaled to match the full NNLO+NNLL results using K-factors computed with {\tt top++}~\cite{Czakon:2011xx}.

\begin{figure}
\begin{center}
\includegraphics[width=7cm]{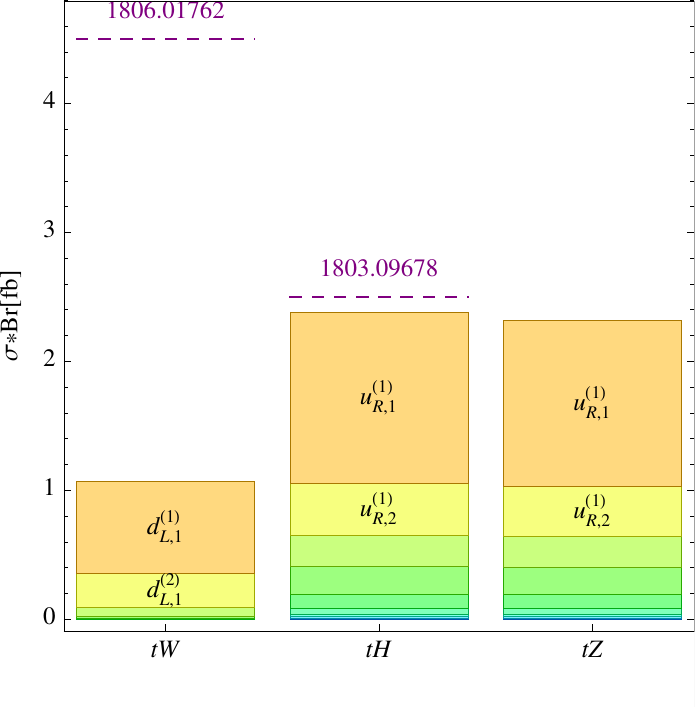}
~~
\includegraphics[width=7cm]{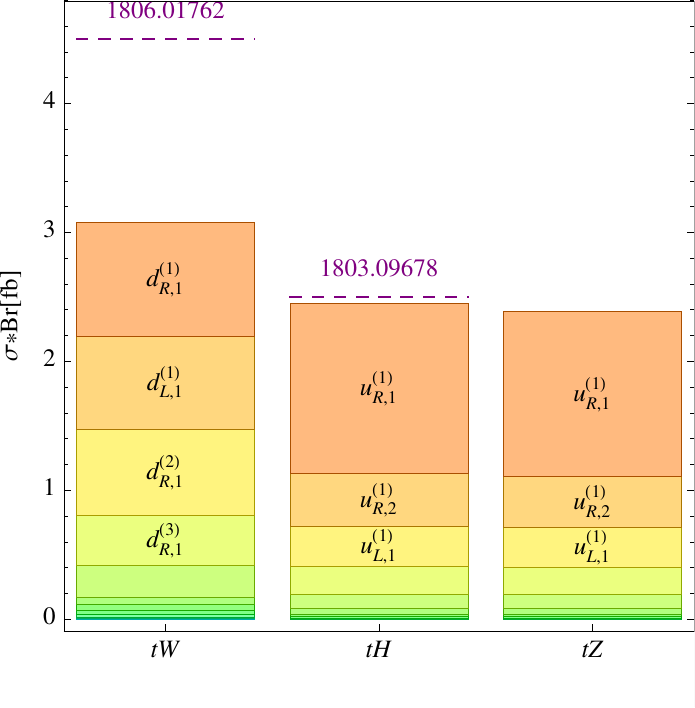}
\caption{\label{fig:xsection} The total gear pair production cross-sections in the final states $tW+X$, $tH+X$ and $tZ+X$ for benchmarks 1 (left plot) and 2 (right plot). The contributions of individual gears are shown stacked and ordered top down by their increasing mass (decreasing cross-section). The currently most stringent upper bounds, obtained by recasting of searches for vector-like quarks in the $tW+X$~\cite{Aaboud:2018uek} and $tH+X$ (the 1-lepton channel)~\cite{Aaboud:2018xuw} final states, are denoted with dashed lines. The corresponding bound on the $tZ+X$~\cite{Aaboud:2018xuw} final state is much weaker and is not shown (it is outside the figure bounds).}
\end{center}
\end{figure}

While the experimental searches~\cite{Aaboud:2018uek,Aaboud:2018xuw} target pair production of a single vector-like quark state, in the two clockwork benchmarks several of the lightest gears contribute significantly to the signal cross section. Fortunately, the sensitivities of both searches, Refs.~\cite{Aaboud:2018uek,Aaboud:2018xuw}, have a plato in the interesting mass range $M \in [1.2,1.8]\,$TeV.  To obtain the predicted signal we are thus able to simply sum the individual contributions of the lowest lying gears, which fall into this mass range, and compare the resulting total signal cross-section, $\sigma \cdot Br$, with the reported upper bounds.
 In the case of the combined $tZ$ and $tH$ search, we only use the so-called $1$-lepton channel which (partly due to an apparent downwards fluctuation in the background) exhibits the best sensitivity overall and targets specifically the $tH$ channel. Since all gears have comparable branching ratios into $tH$ and $tZ$ final states, we do not consider the significantly weaker limits on the later mode. The results for both benchmarks are shown in Fig.~\ref{fig:xsection}.

Since several lightest gears give significant contributions to the total signal cross-sections, the lightest gears need to have masses that are appreciably above the reported mass limits on the individual vector-like quarks with the same quantum numbers and decay channels. For instance, Ref.~\cite{Aaboud:2018xuw} puts a lower bound on an up-like $SU(2)_L$ singlet vector-like quark mass of $1.2$\,TeV. Both of the benchmarks almost saturate the corresponding upper bound for the $tH$ final state cross-section even though the lightest gear, $u_{R,1}^{(1)}$, in both cases has a mass of $1.33$\,TeV.

\subsection{Reconstructing the gear spectrum at colliders}

The dense spectrum of gears and the potentially complex pattern of gear decays poses a challenge also in the case a signal is discovered. In the conventional vector-like quark searches the clockwork signal will appear as an excess of events with high transverse energies or $H_T$, but without a dominant single peak in the invariant mass of any particular final state, such as $tH$ or $tW$. 

In the following we propose a novel reconstruction strategy targeting pair production of heavy quarks with a-priori unknown but potentially long decay chains resulting in a single heavy flavored quark, $t$ or $b$, plus any number of massive weak or Higgs bosons per decay chain. Our procedure is based on the so-called hemisphere clustering algorithm, defined in Section 13.4 of Ref.~\cite{Ball:2007zza}, and already used by several existing experimental analyses in the context of searches for production of new particles at the LHC (see e.g. Refs.~\cite{Khachatryan:2015vra, CMS:2017kmd}). All the visible objects, i.e., jets, as well as isolated leptons and photons, are clustered into exactly two pseudojets, where the clustering is performed by minimizing the Lund distance measure~\cite{Sjostrand:1982am}. The original hemisphere algorithm is seeded by the two objects with the largest combined invariant mass. Since each gear decay chain results in exactly one heavy flavored quark ($t$ or $b$) we instead seed our algorithm with $t$- and $b$-tagged jets. The idea is that, at least for the moderately boosted pair produced gears, the two pseudojets will predominatly capture the decay products of the individual gears. Finally, we select events, where the invariant masses of the two pseudojets are comparable. 

\begin{figure}
\begin{center}
\includegraphics[width=7cm]{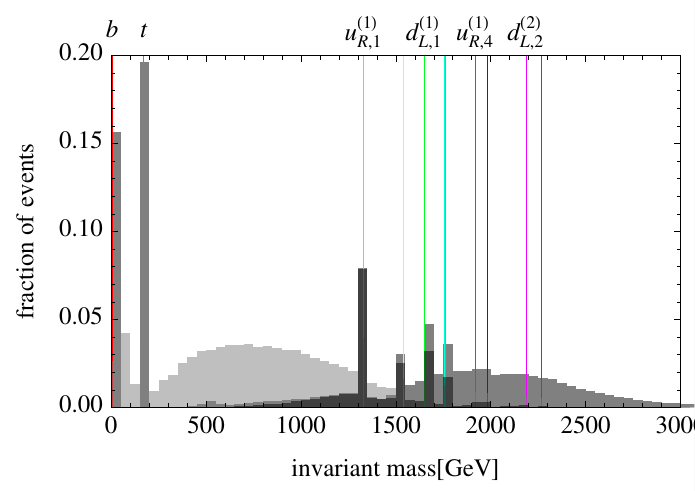}
~~
\includegraphics[width=7cm]{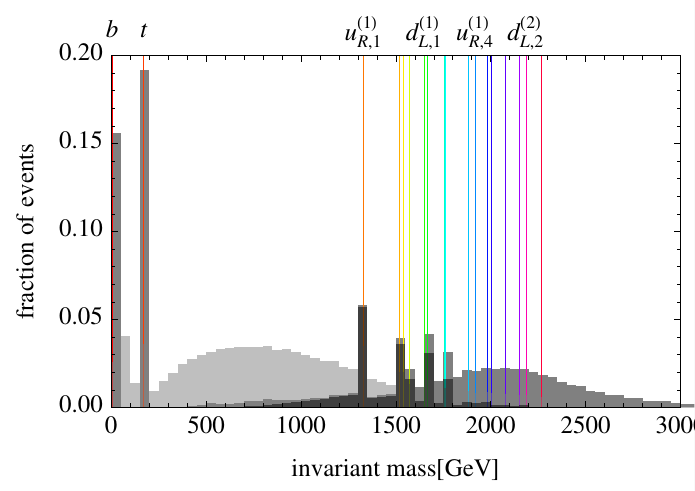}
\caption{\label{fig:hemisphere} Invariant mass spectrum of individual pseudojets clustered using the hemisphere algorithm applied to partonic gear pair production and decay at the 13 TeV LHC  for model benchmarks 1 (left plot) and 2 (right plot). The original hemisphere clustering results, using the two highest invariant mass objects as seeds, are shown in light gray. Our modified hemisphere clustering results, which use two heavy flavored quarks as seeds, are shown in mid-gray. Finally, the modified hemisphere clustering results, where in addition the masses of the two pseudojets are required to differ by less than 30\%,  are shown in dark gray. The spectral lines corresponding to the actual gear masses are ovelaid in the same colors as in Figs.~\ref{fig:spectrum} and~\ref{fig:xsection} with a subset of gear labels printed on the top. See main text for details.}
\end{center}
\end{figure}

We demonstrate the usefulness of our procedure by simulating gear production and decay for the two model benchmarks at parton level, using the same inputs as in the previous subsection. We do not decay tops, $b$-quarks, $W$, $Z$ and the Higgs, but rather use these directly as objects in our clustering procedure.
The results are shown in Fig.~\ref{fig:hemisphere}, where we plot the invariant mass distributions of individual pseudojets and overlay the spectral lines of the gears in the two benchmark models. 

The original hemisphere clustering results, obtained by using the two highest invariant mass objects as seeds, are shown in light gray in Fig.~\ref{fig:hemisphere}. The resulting spectrum does not exhibit any sharp features, with the bulk of the invariant mass distribution lying well below the mass of the lightest gear. The results of modified hemisphere clustering, obtained using two heavy flavored quarks as seeds, but for putting no restrictions on the pseudojet masses, are shown in mid-gray in Fig.~\ref{fig:hemisphere}. The invariant mass distribution already exhibits clear spectral line features. The pseudojets with masses of the top and $b$-quarks are abundantly identified, but also those of a few lowest lying gears. Finally, we show in dark gray the results of modified hemisphere clustering, but keeping only the events for which the masses of the two pseudojets differ by less than 30\%. The low invariant mass peaks corresponding to the pseudojets containing only a single top or $b$-quark  are rejected by this requirement. In addition, the gear peaks are even more pronounced after this cut, with little loss in the number of signal events in the peaks. It would be interesting to see how many of these features survives a more realistic analysis using the $b$-, top-, or mass-drop tagged jets, either narrow or wide, as well as the leptons, as the relevant objects in the clustering procedure. We defer such a more detailed study, which would also need to include backgrounds and detector effects, to a future publication.

\subsection{Indirect probes at colliders}

The large multiplicity of colored particles expected in these scenarios motivates also their indirect probes at the LHC. In particular, Higgs physics provide two well known examples of loop induced processes exhibiting a large sensitivity to the virtual exchange of  new resonances: gluon fusion  and $h\to \gamma \gamma$. In the case of new fermionic resonances, the discussion is almost the same for both observables, so we will focus in the following on $gg\to h$ for the sake of simplicity. 

Assuming real Yukawa couplings, the leading order partonic gluon-fusion cross section  is given by \cite{Ellis:1975ap}
\begin{align}
	\sigma(gg\to H)=\frac{\alpha_s^2 m_h^2}{576 \pi}\left|\sum_f \frac{y_f}{m_f}A_{1/2}(\tau_f) \right|^2 \delta(\hat{s}-m_H^2)
\end{align}
where $\tau_f=4(m_f)^2/m_h^2$,
\begin{equation}
   A_{\frac12}(\tau) = \frac{3\tau}{2} \left[ 1 + (1-\tau)\,f(\tau) \right] , \qquad
\end{equation}
and
\begin{equation}\label{ftau}
   f(\tau) = \begin{cases}
    & \left( \arcsin\frac{1}{\sqrt\tau} \right)^2 ; \quad \text{for } \tau\ge 1 \,, \\[3mm]
    &-\frac14 \left( \ln\left (\frac{1+\sqrt{1-\tau}}{1-\sqrt{1-\tau}}\right )
     - i\pi \right)^2 ; \quad \text{for } \tau<1 \,.
    \end{cases}
\end{equation}
In the above equations, $m_f$ and $y_f$ are defined in the physical mass basis after EWSB, i.e. $\mathcal{L}\supset -\sum_f m_f\bar{\psi}_f\psi_f-\sum_f y_f\bar{\psi}_f\psi_f h=-\sum_f m_f(1+y_f/m_f h)\bar{\psi}_f \psi_f.$ 

Generically, the contribution of the heavy gears to the above cross-section is twofold. First of all, they modify the top contribution to the process, which is the leading one in the SM, by changing the top Yukawa coupling via dimension 6 operators, see Eq. (\ref{eq:deltayuYukawas}). On the other hand,  they also provide an extra contribution through their virtual exchange in the loop giving rise to the process. In the limit $\tau_f\gg 1$, $A_{1/2}(\tau_f)\to 1$, which simplifies the discussion  greatly. Indeed, this is a good approximation even for the top quark, leading to an amplitude \cite{Azatov:2011qy}, 
\begin{align}
	\mathcal{A}(gg\to h)&\propto \sum_f \frac{y_f}{m_f} A_{1/2}(\tau_f)\approx  \sum_f \frac{y_f}{m_f}=\mathrm{Tr}\left[Y\cdot \mathcal{M}^{-1}  \right]=\mathrm{Tr}\left[\frac{\partial \mathcal{M}}{\partial v}\cdot \mathcal{M}^{-1}  \right]\nonumber\\
	&=(\det \mathcal{M})^{-1} \frac{\partial \det(\mathcal{M})}{\partial v}=\frac{\partial }{\partial v} \log(\det \mathcal{M})=\frac{1}{2}\frac{\partial}{\partial v}\log(\det (\mathcal{M}^{\dagger}\mathcal{M})),
	\label{eq:amp}
\end{align}
where $\mathcal{M}(v)$ is just the mass matrix containing all the mass terms for the heavy gears and the top quark. This well-known result is just the manifestation of the Higgs low-energy theorem (LET)~\cite{Ellis:1975ap, Shifman:1979eb,Kniehl:1995tn} which tells us that in order to compute $g g\to h^n$  in zero-momentum limit, we just need to consider the Higgs as a background field and take the Higgs-dependent mass of each field as threshold for the running of $\alpha_s$. For the case of fermionic degrees of freedom, this leads to 
\beq
\mathcal{L}_{h^ngg}=\frac{\alpha_s}{24\pi}G^{a,\mu\nu}G^a_{\mu\nu}\sum_f \log m_f^2(h)=\frac{\alpha_s}{24\pi}G^{a,\mu\nu}G^a_{\mu\nu} \log\left[\det\left(\mathcal{M}^{\dagger}(h)\mathcal{M}(h)\right)\right].
\eeq
Expanding in powers of $h$ around $v$, one gets
\beq
\mathcal{L}_{h^ngg}=\frac{\alpha_s}{24\pi}G^{a,\mu\nu}G^a_{\mu\nu}\left(A_1h+\frac{1}{2}A_2h^2+\ldots\right),
\eeq
where 
\beq
\label{An}
A_n=\frac{\partial^n}{\partial v^n}\log \left[\det \left(\mathcal{M}^\dagger(v)\mathcal{M}(v)\right)\right].
\eeq
We can see that for the case of $A_1$ we recover the result of Eq. (\ref{eq:amp}), as expected.

As an example of how this applies to clockwork we consider first the basic scenario of one clockworked generation of light doublet $Q$ and singlet $u$ quarks, with the Higgs field coupled to the zeroth site mixing the two chains.
We write the masses and $q$ factors for each chain as $q_{Q,U}$ and $m_{Q,U}$ and for demonstration purposes we assume that each chain has length $N=3$.
The masses for the up-type quarks and gears are then written as $\bar{\Psi}_L\mathcal{M}\Psi_R$ with
\begin{equation}
	\mathcal{M}(v)=
\begin{pmatrix}
yv&0&0&0&-q_Um_U&0&0 \\
-q_Qm_Q&m_Q&0&0&0&0&0\\
0&-qm_Q&m_Q&0&0&0&0\\
0&0&-qm_Q&m_Q&0&0&0\\
0&0&0&0&m_U&-q_Um_U&0\\
0&0&0&0&0&m_U&-q_Um_U\\
0&0&0&0&0&0&m_U
\end{pmatrix}
\end{equation}
and
\begin{align}
\Psi_L&=(Q_{L,0},Q_{L,1},\ldots,Q_{L,N},U_{L,1},\ldots,U_{L,N})^T \nonumber\\
\Psi_R&=(U_{R,0},Q_{R,1}\ldots,Q_{R,N},U_{R,1},\ldots,U_{R,N})^T.
\end{align}
Note that the coefficients in Eq.~(\ref{An}) are invariant under unitary rotations of $\mathcal{M}$ so  we can write the mass matrix in any basis we like.
Due to the structure of the mass matrix and its dependence on $v$ the determinant scales quadratically with $v$, and thus we find $A_1=\tfrac{2}{v}$.
This is indeed the factor that arises due to the presence of chiral quarks with masses generated solely from the Higgs mechanism, i.e. the zero modes of the clockwork chains.
Therefore the presence of the gears does not generate corrections to the effective Higgs-gluon vertex at one-loop.
Extending this to two or thee generations does not change the conclusions.
Say, for example, we add another generation with a doublet and an up-type right-handed quark, both of which are clockworked and have a Yukawa mixing via the Higgs at the zeroth site, the determinant of $\mathcal{M}^\dagger \mathcal{M}$ is $\sim v^4$ and $A_1=\tfrac{4}{v}$.
We obtain a factor of $\tfrac{2}{v}$ for each pair of clockwork chains coupled via a Yukawa coupling to Higgs that we integrate out, which is the same as the result when we integrate out two pairs of chiral quarks which get their mass from the Higgs mechanism.
Therefore even with more than one generation the presence of the gears does not generate corrections to the effective Higgs-gluon vertex.
For three generations the same conclusions hold, and since other $A_n$ couplings are simply derivatives of $A_1$, these also do not receive contributions from the presence of gears.
The same applies for the fermionic contribution to $h\to \gamma \gamma$ in such a way that both loop induced processes reduce to the SM expectation in the model at hand. 

A similar cancellation to the one present here,  between the modification of the top Yukawa coupling  -- induced from the dimension 6 operators generated after integrating out the heavy gears -- and the direct contribution of the gears, was  observed in the context of composite Higgs models, see e.g. Refs.~\cite{Falkowski:2007hz, Low:2009di, Low:2010mr, Azatov:2011qy}, since quite generically  
\beq
\det (\mathcal{M}^\dagger(v) \mathcal{M}(v))=F(v)\times \xi(...)
\eeq
where $F(v)$ is some generic function carrying all the dependance of the determinant on the Higgs vev, and the dots inside $\xi$ refer to other parameters of the particular model. One possible way  of breaking this degeneracy that was put forward  consisted in looking rather to  $gg\to hg$ \cite{Harlander:2013oja, Banfi:2013yoa, Azatov:2013xha, Grojean:2013nya}, since for large $p_T$, the large virtuality of the additional gluon will allow to probe much shorter distances than the original process. One could do something similar in this case, to probe for the presence of the heavy gears indirectly. In particular, the $gg\to hg$ cross-section is expected to be resonantly enhanced at gear pair production thresholds. Two potentially interesting observables sensitive to this behavior are the Higgs-jet invariant mass as well as their $p_T$. A detailed study of this is however beyond the scope of this paper.

\section{Conclusions}
\label{sec:conclusions}
In this paper we explored the possibility that the clockwork mechanism solves the SM flavor puzzle.  In clockwork models of flavor the mass hierarchies arise from SM chiral fermions coupling to chains of vector-like fermions. There are several important parameters that determine the phenomenology of the clockwork models: the lengths of the individual clockwork chains, $N_{\psi(i)}$, the clockworking factors $q_{\psi(i)}$, and the mass  scale for the gears, $q_{\psi(i)}m$.  The clockwork models are reminiscent of the two other common ways of generating the quark mass hierarchies, the FN and the RS models, but retain only the bare minimum of ingredients needed to generate the flavor hierarchies.

The FN models most easily match onto clockwork models in the limit where the FN flavons are much heavier than the gears, taking $q_{\psi(i)}$ to be the same for all fermions, while $N_{\psi(i)}$ are generation and flavor dependent. Then the most natural realization of clockwork is when the chiral fermions in the FN models do not carry the horizontal charge, which is not the usual choice that has been made in the FN models. The traditional FN models, where the SM chiral fermions do carry horizontal charges and/or when the FN flavons are lighter than the gears, also match onto clockwork at the low energies. The two realizations differ above the flavon mass scale and in the fact that the clockworked FN has an anomaly-free horizontal symmetry. 

The connection between the RS models of flavor and a particular limit of clockwork -- flavor universal $N_{\psi(i)}$ and generation and flavor dependent $q_{\psi(i)}$ -- is just approximate. Typically the gears will form a band roughly $m$ above the zero mode, while RS has well separated fermionic KK modes. The clockwork also does not contain excitations of the gauge bosons. Furthermore, while the solutions to flavor puzzle and hierarchy problem are intertwined in RS, they are orthogonal in clockwork. All the SM fields and the clockwork chains that solve the flavor puzzle would have to have gravity clockworked in the same way.  Finally, while the ``universal $N$'' limit of clockwork would appear to be a natural candidate to be UV completed in the framework of the linear dilaton model, this is not the case -- the continuous 5D limit leads to phenomenologically unacceptable exponentially small gauge couplings. 

In this paper we also studied in detail the phenomenological consequences of the clockwork flavor models. The lengths of clockwork chains are constrained by the impact they have on the running of QCD and the Higgs quartic, which both have important implications for the viability of the models.  We settled on a representative clockwork model with 19 gears  \eqref{eq:gears}.
Integrating out all the new heavy particles -- the gears -- we first matched onto the SM effective field theory, which we used to analyze the constraints from low-energy experiments: from weak boson decays, rare meson decays and neutral meson mixings. Similarly to what happens in the RS models \cite{Agashe:2004ay,Agashe:2004cp}, the clockwork models of flavor are endowed with a powerful flavor protection against flavor-changing neutral currents (FCNCs).  The FCNCs with light quarks on the external legs are suppressed by the small overlaps of the zero-modes, which is the same suppression that gives rise to hierarchies between the SM quark masses. This CW-GIM  mechanism, along with the requirements arising from the stability of the Higgs potential, suffices to alleviate the flavor constraints to the level that TeV scale gear masses are compatible with experimental bounds.

 We performed a complete numerical study of the low-energy constraints, using which we singled out two benchmark models. Their phenomenology at the LHC was then studied in detail using Monte-Carlo simulations, recasting existing searches for vector-like quarks at the LHC. Due to the rich spectrum present in these setups, with several gears contributing simultaneously  to the $tH, tZ$ and $tW$ final states, the bounds on the gear masses are somewhat stronger than on the individual vector-like quarks, in the $1.2$ TeV and $1.4$ TeV regime for up-quark and down-type quark gears, respectively. 
 Using a modified hemisphere clustering algorithm, that we propose, there are good prospects of discrimination between contributions from different gears, in case an excess is observed in one of these searches. Finding such a multiple peak structure would be a smoking gun for flavor clockwork.

The analysis we performed is not the most general one. We assumed that each of the SM chiral fermions is clockworked separately.  In principle, the clockworking itself could mix different generations, by either having the clockworking factors $q$ or the gear mass terms $m$ promoted to $3\times3$ matrices. The connection with FN suggests a way to prevent this from happening and keeping the gears from different generations separate -- introducing a horizontal symmetry for each generation $(U(1)_H, \phi)\to ((U(1)_{H})^3,\phi_i)$. On the other hand it would be interesting to 
explore the implications of flavor non-diagonal clockworking for the natural generation of hierarchy in the SM Yukawa couplings and the CW-GIM suppression of new physics effects.

{\bf Acknowledgements.} We thank Matthew McCullough, Kfir Blum and Gilad Perez for many useful discussions. We thank Svjetlana Fajfer for encouragements to finish the project on time. AC, BD, JFK and JZ would like to thank the CERN Theoretical Physics Department for its hospitality and support while finalizing this work. JZ acknowledges support in part by the DOE grant DE-SC0011784. The research of AC has been supported by the Cluster of Excellence {\it Precision Physics, Fundamental Interactions and Structure of Matter} (PRISMA-EXC1098) and grant 05H12UME of the German Federal Ministry for Education and Research (BMBF). BD acknowledges funding from Grant No. EP/P005217/1. JFK acknowledges the financial support from the Slovenian Research Agency (research core funding No. P1-0035 and J1-8137). JMC acknowledges support from the Spanish MINECO through the Ram\'on y Cajal program RYC-2016-20672.  
 
\appendix

\section{A continuum description?}
\label{sec:continuum}
In this appendix we give the details of the continuum limit of the clockwork and discuss 
the difficulties in using it 
to generate hierarchies among Yukawa couplings for the SM fermions.
The difficulty arises because having fermions living in the 5D bulk necessarily means that the SM gauge bosons must also live in the bulk, which in turn means that the 4D gauge coupling is exponentially suppressed with respect to the 5D gauge coupling.
Generating a realistic 4D gauge coupling from a perturbative 5D theory is therefore not possible.
The suppression occurs because having a bulk field in the continuum limit is equivalent to having that field clockworked in the 4D model.

We now give the derivations that lead us to the above conclusion. We start with the ansatz for the  metric  \cite{Giudice:2016yja}, 
\beq\label{eq:metric}
ds^2=e^{2\sigma } \big(\eta_{\mu\nu}d x^\mu dx^\nu -e^{-6 \ell \sigma} dy^2\big),
\eeq
where $\sigma(y)=2k|y|/3$. This metric can interpolate between the clockwork case, $\ell=0$;  the RS models of flavor, $\ell=1/3, k=3/(2 R)=3 \hat k/2$, where $R$ is the compactification radius; and the flat space, $k=0$. Note that we are using the metric $\eta_{\mu\nu}=\diag(+,-,-,-)$, which is opposite to the one in Ref. \cite{Giudice:2016yja}. The coordinate in the 5th dimension is chosen such that $y=0$ corresponds to the IR brane, and $y=\pi R$ to the UV brane. In order to solve the hierarchy problem, one has \cite{Giudice:2016yja}
\beq
k R \simeq 10, \qquad \text{ or, equivalently,}\qquad \sigma' R\simeq 7.
\eeq

\subsection{Fermions}
We assume that the clockwork stabilization mechanism is provided by the dilaton, $S$ \cite{Giudice:2016yja}. 
The action for a fermion $\Psi$ in a warped space is given in the Jordan frame  by (see, e.g., Refs.~\cite{Grossman:1999ra,Bertlmann:1996xk,Cox:2012ee,Antoniadis:2011qw})
\beq\label{eq:action:Jordan}
{\cal S}_J=\int d^4 x \int_{-\pi R}^{\pi R} dy \sqrt{G} e^S \Big\{\frac{i}{2}E_a^A  \left[ \bar\Psi \gamma^a(\partial_A +\omega_A)\Psi -  \bar\Psi (\stackrel{\leftarrow}{\partial}_A -\omega_A) \gamma^a\Psi\right]- m_\Psi \bar \Psi \Psi\Big\},
\eeq
where  $G=\det(G_{AB})$ is the determinant of the metric, and the two derivatives are written in such a way that they act only on fermions. We use $AB$ for indices in the curved space, $a,b$ for the indices in the tangent space, $E_a^A$ is the inverse vielbein, $\omega_{A}$ the spin connection, while the gamma matrices are $\gamma^a=(\gamma^\mu, i\gamma_5)$. The mass term $m_\Psi$ is understood to be odd under the orbifolding $Z_2$, $y\to -y$. The fermions can be either even or odd under the orbifolding, 
$\Psi_\pm(-y)=\pm \Psi_\pm(y)$. The $\Psi_{+(-)}$ will lead to right-handed (left-handed) zero modes.

Going to the Einstein frame is achieved through the metric transformation, $g_{MN}\to \exp(-2S/3) g_{MN}$, giving 
\beq\label{eq:action}
{\cal S}_E=\int d^4 x \int_{-\pi R}^{\pi R} dy \sqrt{G} e^{-S/3} \Big\{\frac{i}{2}E_a^A  \left[ \bar\Psi \gamma^a(\partial_A +\omega_A)\Psi -  \bar\Psi (\stackrel{\leftarrow}{\partial}_A -\omega_A) \gamma^a\Psi\right]- m_\Psi e^{-S/3} \bar \Psi \Psi\Big\}.
\eeq
The metric in the Einstein frame is given by Eq.~\eqref{eq:metric}. The resulting spin connection  is $\omega_{A}=\frac{1}{2}e^{3\ell \sigma}\sigma' (i\gamma_\mu\gamma_5,0)$, while the inverse  vielbein
is given by $E_a^A=\diag(e^{-\sigma},e^{-\sigma}, e^{-\sigma},
e^{-\sigma},e^{(3\ell-1)\sigma})$. We are not interested in the dilaton dynamics,
so that we can set it to its background profile, $S=3\sigma=2 k|y|$ \cite{Giudice:2016yja}. After integrating by parts we get
\beq\label{eq:action:perparts}
S=\int d^4 x \int dy\,  e^{3(1-\ell)\sigma} \bar\Psi \Big\{i\cancel{\partial} -\gamma_5 e^{3\ell \sigma}\left(\partial_y+\frac32\sigma^{\prime}\right)- m_\Psi \Big\}\Psi,
\eeq
where, in particular, the spin connection has cancelled  out.  Defining $\Psi_{L,R}=\frac{1}{2}(1\mp\gamma_5) \Psi$ and using 
\beq
\Psi_L(x,y)=\frac{e^{-\frac{3}{2}\sigma}}{\sqrt{2\pi R}}\sum_n f_L^{(n)}(y)\psi_L^{(n)}(x), \qquad 
\Psi_R(x,y)=\frac{e^{-\frac{3}{2}\sigma}}{\sqrt{2\pi R}}\sum_n f_R^{(n)}(y)\psi_R^{(n)}(x),
\eeq
together with the  four dimensional equations of motion 
\beq
i\cancel{\partial}\psi_L^{(n)}(x)-m_n\psi_R^{(n)}(x)=0, \qquad 
i\cancel{\partial}\psi_R^{(n)}(x)-m_n\psi_L^{(n)}(x)=0,
\eeq 
we get 
\beq\label{eq:zeromode:eom}
 \left(\pm  e^{3\ell \sigma}\partial_y+m_\Psi \right)f_{R,L}^{(n)}=m_nf_{L,R}^{(n)}.
\eeq
Iterating the two equations we obtain 
\beq
\left[-e^{3\ell \sigma}\partial_ye^{3\ell \sigma}\partial_y\mp e^{3\ell \sigma}m_{\Psi}^{\prime}+m_{\Psi}^2-m_n^2\right]f_{R,L}^{(n)}=0.
\label{eq:bulk}
\eeq

The Kaluza-Klein modes $f_{L,R}^{(n)}(y)$ obey separate orthonormal conditions 
\beq
\frac{1}{\pi R} \int_0^{\pi R} d y e^{-3 \ell \sigma} f_{L,R}^{(n)}(y) f_{L,R}^{(m)}(y) = \delta_{nm}\,.
\eeq
They also need to satisfy proper boundary conditions at $y=0,\pi R$. For odd solutions the Dirichlet boundary conditions apply,
\beq\label{eq:Dirichlet}
\left.f_{R,L}^{(n)}(y)\right|_{y=0,\pi R} = 0\,.
\eeq
Even solutions, including zero modes, are subject to a different boundary condition, 
\beq\label{eq:even:bc}
\left.\left(  \partial_y  \pm m_\Psi e^{-3\ell\sigma}   \right) f^{(n)}_{R,L} (y) \right|_{y=0,\pi R} = 0\,.
\eeq
For clockwork, $\ell=0$, the zero mode profile is given by 
\beq\label{eq:zeromode}
f_{R,L}^{(0)}(y)=N_0 \exp\left(-\frac{c_{R,L}\, y}{\pi R}\right),
\eeq
where $c_{R,L}=\pm  m_\Psi \pi R$. For $c>0$ the zero mode is localized near the IR brane, $y=0$, while for $c<0$ the zero mode is localized near the UV brane, $y=\pi R$. The overlaps with the IR brane are given by the normalization factors, 
\begin{align}
f_{R,L}^{(0)}(0)=&N_0=\frac{\sqrt{2 c}}{(1-e^{-2 c})^{1/2}}\simeq\sqrt{2 c}, \qquad c>0,
\\
	f_{R,L}^{(0)}(0)=&N_0=\frac{\sqrt{2 c}}{(1-e^{-2 c})^{1/2}}\simeq\sqrt{2|c|}\, e^{-|c|}, \quad c<0,
\end{align}
where we suppressed the $L,R$ indices on the coefficients $c_{R,L}$, as well as generation indices, in order to shorten the expressions.  The last approximate equalities on the right hand side are valid for $|c|\gtrsim 1$. 

For a Higgs boson localized on the IR brane (at $y=0$), the effective 4D Yukawas of SM fermions are proportional to $f^{(0)}_R (0)f^{(0)}_L (0)$, giving the hierarchy among the SM quark masses. The suppression of the light quarks comes from the exponential suppression of the zero mode overlaps with the IR brane, very similar to the RS.  There are also a number of difference with respect to the RS. The inclusion of the dilaton was essential in clockwork in order to obtain the necessary zero mode profiles. In particular, the $y$ dependence due to the dilaton multiplying the mass term, $m_\Psi \bar \Psi \Psi$,  in  Eq.~\eqref{eq:action}, exactly matches the $y$ dependence in the four-dimensional part of the kinetic term. This is the reason that the $m_\Psi \bar \Psi \Psi$ and $\bar \Psi i\slashed \partial \Psi$ in Eq.~\eqref{eq:action:perparts} come with the same prefactor. For $\ell=0$ the mass term and the derivative then have no additional $y$ dependence in Eq.~\eqref{eq:zeromode:eom} (for constant $m_\Psi$), leading to exponential zero mode profiles. Without the dilaton the zero mode profiles would be given by double exponentials, $f_{R,L}^{(0)}(y)\propto \exp\left[\mp 3 m_{\Psi}e^{(1-3\ell)\sigma}/2k(1-3\ell)\right]$, giving,  for $m_\Psi/k\sim {\mathcal O}(1)$, phenomenologically unacceptable quark masses. The RS case, $\ell=1/3$, represents a special choice which does lead to simple exponential profiles (and vice versa, introducing the 5D dilation in the RS would lead to double exponential zero mode profiles).

We discuss next the KK excitations. Specializing to the case of clockwork, $\ell=0$, Eq. (\ref{eq:bulk}) reads
\beq
f_{R,L}^{(n)\prime\prime}(y)+(m_n^2-m_\Psi^2)f_{R,L}^{(n)}(y)=0.
\eeq
The general solutions to the above equations, assuming $m_n> |m_{\Psi}|$, are given by,\footnote{For $m_{n}\le |m_{\Psi}|$ there is only one possible solution corresponding to the zero mode, $m_n=0$.}
\beq
f_{R,L}(y)=A_n\cos\left(\sqrt{m_n^2-m_\Psi^2}y\right)+B_n\sin\left(\sqrt{m_n^2-m_{\Psi}^2}y \right).
\eeq
The boundary conditions \eqref{eq:Dirichlet}, \eqref{eq:even:bc}, then give for odd and even KK modes 
\begin{align}
	f_{R,L}^{\rm odd}(y)&=\sqrt{2} \sin\left(\frac{ny}{R} \right),\\
	f_{R,L}^{\rm even}(y)&=\sqrt{2}\frac{ m_\Psi}{m_n} \left[\sin\left(\frac{ny}{R} \right) \mp\frac{n}{m_{\Psi} R}\cos\left(\frac{ny}{R} \right)\right],
\end{align}
with $n\in\mathbb{N}=\{1,2,\ldots\}$. The mass of the $n$-th KK mode is given by 
\beq
m_n^2=m_\Psi^2+\frac{n^2}{R^2}.
\eeq

\subsection{Gauge bosons}
For simplicity we consider the example of an abelian $U(1)$ gauge group. The action in the Jordan frame is
\beq
{\cal S}_J=\int d^4x \int_{-\pi R}^{\pi R}dy\, \sqrt{G}e^S \left(-\frac{1}{4}F_{MN}F_{PQ}G^{MP}G^{NQ}\right)+{\cal S}_{J,0}+{\cal S}_{J,1},
\eeq
where, 
\begin{align}
	{\cal S}_{J,k}&= 2\int d^4 x \int_{-\pi R}^{\pi R} d y \sqrt{|G_{k}|}e^S \biggr[-\frac{\theta_k}{4}RF_{\mu\nu}F_{\rho\sigma}G^{\mu\rho}G^{\nu\sigma} \biggr]\delta(y-y_k),\quad k=0,1,
\end{align}
with $y_0=0, y_1=\pi R$, and $G_{0,1}=\mathrm{det}G_{\mu\nu}|_{y=0,\pi R}$ the determinants of the two induced 4D  metrics required to assure 5D general covariance.  The dimensionless parameters $\theta_0R$ and $\theta_1 R$ control the size of the localized gauge kinetic terms, that we include for generality.

In the Einstein frame the action is given by
\begin{align}
	{\cal S}_E&=\int d^4x \int_{-\pi R}^{\pi R} dy\, \sqrt{G}e^{\frac{2}{3}S} \left(-\frac{1}{4}F_{MN}F_{PQ}G^{MP}G^{NQ}\right)+{\cal S}_{E,0}+{\cal S}_{E,1},
\\
	{\cal S}_{E,k}&= 2\int d^4 x \int_{-\pi R}^{\pi R} dy \sqrt{|G_{k}|}e^{S} \biggr[-\frac{\theta_k}{4} R F_{\mu\nu}F_{\rho\sigma} G^{\mu\rho} G^{\nu\sigma} \biggr]\delta(y-y_k),
\end{align}
which, after integrating by parts and setting $S$ to its background value, $S=3\sigma $, leads to
\begin{align}
{\cal S}_E=&-\frac{1}{4}\int d^4x \int dy\,  e^{3(1-\ell)\sigma}\Big(F_{\mu\nu}F^{\mu\nu} - 2 e^{6\ell\sigma}\partial_\mu A_5 \partial^\mu A_5  \Big)+\sum_k{\cal S}_{E,k}
\\
	& -\int d^4 x \int dy\, e^{3(1-\ell)\sigma}\Big[\partial_\mu A^\mu e^{-3(1-\ell)\sigma}\partial_y \big(e^{3(1+\ell)\sigma} A_5\big)+\frac{1}{2} A_\mu e^{-3(1-\ell)\sigma}\partial_y\big( e^{3(1+\ell)\sigma}\partial_y A^\mu\big)\Big]\nonumber\\
	{\cal S}_{E,k}&=2 \int d^4 x\int dy\, e^{3\sigma} \biggr[-\frac{\theta_k}{4} R F_{\mu\nu}F^{\mu\nu}\biggr]\delta(y-y_k),
\end{align}
where lowering and raising of the 4D indices are, here and below, performed using Minkowski metric, so that, e.g.,  $A^{\mu}=A_{\nu}\eta^{\mu\nu},\, \partial^{\mu}=\partial_{\mu}\eta^{\mu\nu},\, \ldots$. 
 To cancel the mixing between $A_{\mu}$ and the  scalar $A_5$ we add the gauge-fixing term
\begin{align}
	{\cal S}_{GF}=&-\frac{1}{2\xi}\int d^4 x \int_{-\pi R}^{\pi R} d y\, e^{3(1-\ell)\sigma}\left[\partial_{\mu}A^{\mu}-\xi e^{-3(1-\ell)\sigma}\partial_y (e^{3(1+\ell)\sigma} A_5)\right]^2
\end{align}
which, after some algebra, leads to the following bulk equation of motion
\begin{align}
	&\left[\left(\partial^2\left(1+2\theta_0 R e^{3\ell\sigma}\delta(y)+2\theta_1 Re^{3\ell\sigma}\delta(y-\pi R)\right)-e^{-3(1-\ell)\sigma}\partial_y e^{3(1+\ell)\sigma}\partial_y\right)\eta^{\mu\nu}\right.\nonumber\\
	&\left.-\partial^{\mu}\partial^{\nu}\left(1+2\theta_0 R e^{3\ell\sigma}\delta(y)+2\theta_1 Re^{3\ell\sigma}\delta(y-\pi R)-\frac{1}{\xi}\right)\right]A_{\nu}=0.
\end{align}
Expanding in KK modes for the clockwork case $\ell=0$
\begin{equation}
	A_{\mu}(x,y)=\frac{1}{\sqrt{2 \pi R}}e^{-\frac{3}{2}\sigma}\sum_{k=0}^{\infty}f_A^{(n)}(y)A_{\mu}^{(n)}(x),
\end{equation}
and using the 4D equations of motion in the unitary gauge $\xi\to\infty$
\beq
\left[\left(\partial^2+m_n^2\right)\eta^{\mu\nu}-\partial^{\mu}\partial^{\nu}\right]A_{\nu}^{(n)}(x)=0,
\eeq 
gives the following differential equations for the 5D profiles 
\beq\label{eq:5Dwave}
\left(\partial_y^2 +m_n^2-\frac{9}{2}\sigma^{\prime 2}\right)f_{A}^{(n)}(y)=0.
\eeq
as well as  boundary conditions 
\begin{align}
	\label{eq:vector:bc}
	\left.f_A^{(n)}(y)\right|_{y=y_k}&=0\qquad {\rm (odd)}.\\
	\left.\left[ (-1)^{k+1}\partial_y-\theta_k R m_n^2\right]e^{-3\sigma/2}f_A^{(n)}(y)\right|_{y=y_k}&=0\qquad {\rm (even)}.
\end{align}
In addition, the vector KK modes satisfy orthonormality conditions
\beq
\frac{1}{2\pi R}\int_{-\pi R}^{\pi R}dy \left(1+2\theta_0 R\delta(y-y_0)+2\theta_1 R\delta(y-y_1)\right)f_A^{(n)}(y)f_A^{(m)}(y)=\delta_{nm}.
\eeq 
The solution to Eqs. \eqref{eq:5Dwave} and \eqref{eq:vector:bc} always contains a zero mode, $m_0=0$, with 
\begin{align}
	f_A^{(0)}(y)=e^{\frac{3}{2} y \sigma^{\prime}}\sqrt{3\pi \sigma^{\prime} R}\left[e^{3\sigma^{\prime}\pi R}(1+3\theta_1\sigma^{\prime} R)-1+3\theta_0 \sigma^{\prime}R\right]^{-1/2}.
\end{align} 
The higher KK modes are in general given by 
\begin{align}
	f_n(y)=A_n \cos\left(y\sqrt{m^2_n-9 \sigma^{\prime 2}/4}\right) + B_n  \sin\left(y\sqrt{m^2_n-9\sigma^{\prime 2}/4}\right).
\end{align}
This leads, in agreement with the results of Ref.~\cite{Ahmed:2016viu}, to	 (for simplicity we set $\theta_0=0$),
\begin{align}
	f_n(y)&=\sqrt{2} \sin\left(\frac{ny}{R}\right)\qquad {\rm (odd)},\\
	f_n(y)&=B_n\left[\lambda_n\cos\left(\lambda_n y\right)+\frac{3}{2}\sigma^{\prime}\sin\left(\lambda_n y\right)\right]\qquad {\rm (even)},
\end{align}
where the masses of the clockwork gears are given by
\beq
m_n^2=\lambda_n^2+9\frac{\sigma^{\prime 2}}{4}=\frac{n^2}{R^2}\left(1+\Delta_n\right)^2+9\frac{\sigma^{\prime 2}}{4}, \quad n=1,2,\ldots\,, 
\eeq
with
\beq
\Delta_n\approx \frac{1/\pi+1/\theta_1+3/2 \sigma^{\prime}R - \sqrt{(1/\pi +1/\theta_1 +3/2
    \sigma^{\prime} R)^2+2 n^2}}{ n^2}.
\eeq
and
\beq
B_n=\frac{\sqrt{2}R}{\sqrt{n^2\left(1+\frac{2\theta_1}{\pi}\right)+\frac{9}{4}\sigma^{\prime 2} R^2}}+\mathcal{O}(\Delta_n).
\eeq

The coupling of any fermion zero-mode to a massless gauge boson, such as the photon, will be given by 
\beq
g_4=\frac{g_5}{\sqrt{R}}\sqrt{3/2\sigma^{\prime} R}\left[e^{3\sigma^{\prime}\pi R}(1+3\theta_1\sigma^{\prime} R)-1+3\theta_0 \sigma^{\prime}R\right]^{-1/2},
\eeq
which leads to
\beq
g_4\approx \frac{g_5\sqrt{3/2\sigma^{\prime}R}}{\sqrt{R(1+3\theta_1 \pi \sigma^{\prime} R)}}e^{-3/2\sigma^{\prime} \pi R},
\eeq
unless there is a truly enormous tuning, $1+3\theta_1\sigma^{\prime} R =\mathcal{O}(e^{-3\sigma^{\prime}\pi R})\approx 10^{-29}$. Since the dimensionless quantity $g_5/\sqrt{R}$ can not be taken to be arbitrary large, it becomes impossible to have a $\mathcal{O}(1)$ 4D gauge coupling and solve the hierarchy problem at the same time.  Another possibility would be to make the warp factor irrelevant by choosing values of $\sigma^{\prime} R\sim \mathcal{O}(1)$ or slightly smaller. In this case one could still have naturally  $k\ll M_5\approx M_{\rm Planck}$, since $k=0$ is technically natural by a dilaton shift symmetry. However, this limit would just correspond to the well known flat extra-dimensional case.

\section{Matching onto SMEFT}
\label{sec:matching}
In this appendix we perform the matching at $\mu\simeq M$ from the clockwork flavor model to SMEFT, integrating out the gears.  The tree-level exchanges of the gears, shown in Fig.~\ref{fig:flavormix}, give the following contributions to the SMEFT operators in Tab.~\ref{tab:SMEFTops},
\begin{subequations}
\begin{align}
	[w_{HQ}^{(1)}]_{ij}&=\frac{f_{Q(i)}f_{Q(j)}}{4}\left[Y_U M_{u}^{-2} Y_U^\dagger-Y_D M_{d}^{-2}Y_D^\dagger\right]_{ij}~,
\\
[w_{HQ}^{(3)}]_{ij}&=-\frac{f_{Q(i)}f_{Q(j)}}{4}\left[Y_U M_u^{-2} Y_U^\dagger+Y_D M_d^{-2} Y_D^\dagger\right]_{ij},
\\
[w_{Hud}]_{ij}&=f_{u(i)}f_{d(j)}\left[Y_U^\dagger M_{Q}^{-2} Y_D\right]_{ij},
\\
[w_{Hu}]_{ij}&=-\frac{f_{u(i)}f_{u(j)}}{2}\left[Y_U^\dagger M_Q^{-2} Y_U\right]_{ij}~,
\\
[w_{Hd}]_{ij}&=\frac{f_{d(i)}f_{d(j)}}{2}\left[Y_D^\dagger  M_Q^{-2} Y_D\right]_{ij}~,
\\
\begin{split}
[w_{uH}]_{ij}&=\frac{f_{Q(i)}}{2}\sum_r\left[Y_U M_u^{-2} Y_U^\dagger\right]_{ir}f_{Q(r)}\left[Y_u^{\rm SM}\right]_{rj}
\\
&\qquad+\frac{f_{u(j)}}{2}\sum_r\left[Y_u^{\rm SM}\right]_{ir}f_{u(r)}\left[Y_U^{\dagger} M_{Q}^{-2} Y_U\right]_{rj}
\\
&\qquad -f_{Q(i)}f_{u(j)}\left[Y_U M_u^{-1} Y_U^\dagger M_Q^{-1} Y_U\right]_{ij},
\end{split}
\\
\begin{split}
\label{eq:matchTL}
\left[w_{dH}\right]_{ij}&=\frac{f_{Q(i)}}{2}\sum_r\left[Y_D M_d^{-2} Y_D^\dagger\right]_{ir}f_{Q(r)}\left[Y_d^{\rm SM}\right]_{rj}
\\
&\qquad+\frac{f_{d(j)}}{2}\sum_r\left[Y_d^{\rm SM}\right]_{ir} f_{d(r)}\left[Y_D^{\dagger} M_Q^{-2}Y_D\right]_{rj}
\\
	&\qquad-f_{Q(i)}f_{d(j)}\left[Y_D M_d^{-1} Y_D^\dagger M_Q^{-1} Y_D\right]_{ij}.
\end{split}	
\end{align}
\end{subequations}

The loop contributions in Fig.~\ref{fig:flavormixloop} give the following contributions to the SMEFT operators, 
\begin{subequations}
\label{eq:matchloop}
\begin{align}
\begin{split}
\label{eq:wQQ13}
[w_{QQ}^{(1,3)}]_{ijkl}&=-\frac{f_{Q(i)}f_{Q(j)}f_{Q(k)}f_{Q(l)}}{16(4\pi)^2}\sum_{rr^\prime}\Big([Y_U]_{ir}[Y_U^\dagger]_{rj}[Y_U]_{kr^\prime}[Y_U^\dagger]_{r^\prime l}f(M_{u(r)},M_{u(r^\prime)})
\\
&\hspace{4.cm}+[Y_D]_{ir}[Y_D^\dagger]_{rj}[Y_D]_{kr^\prime}[Y_D^\dagger]_{r^\prime l}f(M_{d(r)},M_{d(r^\prime)})
\\
&\hspace{4.cm}\mp[Y_U]_{ir}[Y_U^\dagger]_{rj}[Y_D]_{kr^\prime}[Y_D^\dagger]_{r^\prime l}f(M_{u(r)},M_{d(r^\prime)})\Big)\\
&\hspace{4.cm}\mp[Y_U]_{kr}[Y_U^\dagger]_{rl}[Y_D]_{ir^\prime}[Y_D^\dagger]_{r^\prime j}f(M_{u(r)},M_{d(r^\prime)})\Big),
\end{split}
\\
[w_{uu}]_{ijkl}&=-\frac{f_{u(i)}f_{u(j)}f_{u(k)}f_{u(l)}}{4(4\pi)^2}\sum_{rr^\prime}[Y_U^\dagger]_{ir}[Y_U]_{rj}[Y_U^\dagger]_{kr^\prime}[Y_U]_{r^\prime l}f(M_{Q(r)},M_{Q(r^\prime)}),
\\
[w_{dd}]_{ijkl}&=-\frac{f_{d(i)}f_{d(j)}f_{d(k)}f_{d(l)}}{4 (4\pi)^2}\sum_{rr^\prime}[Y_D^\dagger]_{ir}[Y_D]_{rj}[Y_D^\dagger]_{kr^\prime}[Y_D]_{r^\prime l}f(M_{Q(r)},M_{Q(r^\prime)}),
\\
\begin{split}
[w_{Qu}]_{ijkl}&=\frac{f_{Q(i)}f_{Q(j)}f_{u(k)}f_{u(l)}}{4(4\pi)^2}\sum_{rr^\prime}\Big([Y_U]_{ir}[Y_U^\dagger]_{rj}[Y_U^\dagger]_{kr^\prime}[Y_U]_{r^\prime l}f(M_{u(r)},M_{Q(r^\prime)})
\\
&\hspace{4.cm}-[Y_D]_{ir}[Y_D^\dagger]_{rj}[Y_U^\dagger]_{kr^\prime}[Y_U]_{r^\prime l}f(M_{d(r)},M_{Q(r^\prime)})\Big),
\end{split}
\\
\begin{split}
[w_{Qd}]_{ijkl}&=\frac{f_{Q(i)}f_{Q(j)}f_{d(k)}f_{d(l)}}{4(4\pi)^2}\sum_{rr^\prime}\Big([Y_D]_{ir}[Y_D^\dagger]_{rj}[Y_D^\dagger]_{kr^\prime}[Y_D]_{r^\prime l}f(M_{d(r)},M_{Q(r^\prime)})
\\
&\hspace{4.cm}-[Y_U]_{ir}[Y_U^\dagger]_{rj}[Y_D^\dagger]_{kr^\prime}[Y_D]_{r^\prime l}f(M_{u(r)},M_{Q(r^\prime)})\Big),
\end{split}
\\
\begin{split}
[w_{ud}^{(1)}]_{ijkl}=&-\frac{f_{u(i)}f_{u(j)}f_{d(k)}f_{d(l)}}{4(4\pi)^2}\Big(\frac{4}{N_c}[Y_U^\dagger]_{ir} [Y_D]_{rl} [Y_D^\dagger]_{kr'}[ Y_U]_{r'j}
\\ 
&\hspace{4.cm}-2[Y_U^\dagger]_{ir}[ Y_U]_{rj}[Y_D^\dagger]_{kr'}[ Y_D]_{r'l}\Big)f(M_{Q(r)},M_{Q(r^\prime)})
\end{split}
\\
\begin{split}
[w_{ud}^{(8)}]_{ijkl}=&-\frac{2f_{u(i)}f_{u(j)}f_{d(k)}f_{d(l)}}{(4\pi)^2}[Y_U^\dagger]_{ir}[ Y_D]_{rl} [Y_D^\dagger]_{kr'}[ Y_U]_{r'j}f(M_{Q(r)},M_{Q(r^\prime)})
\end{split}
\end{align}
\end{subequations}
 where
\begin{align}\label{eq:loopfunction}
f(m_a,m_b)=\frac{\log\left(\frac{m_a^2}{m_b^2}\right)}{m_a^2-m_b^2},
\end{align}
is a loop function with $f(m,m)=1/m^2$.

The finite parts in the matching between SMEFT and LEEFT operators that contribute to meson  oscillations with $d_i\bar d_j\to d_j\bar d_i$ transitions are, at $\mu=m_W$~\cite{Bobeth:2017xry},
\begin{subequations}
\label{eq:DeltaC:finite}
\begin{align}
\begin{split}
[\Delta C_{\rm VLL}]_{ij}&=\frac{1}{16\pi^2}y_t^2\lambda_{ij}^{(t)}\bigg\{\left[L_d w_{HQ}^{(1)}L_d^\dagger\right]_{ij}H_1(x_t,m_W)-\left[L_d w_{HQ}^{(3)}L_d^\dagger\right]_{ij}H_2(x_t,m_W)
\\
&\hspace{1.6cm}+\frac{2S_0(x_t)}{x_t}\sum_m\left(\lambda^{(t)}_{im}\left[L_d w_{HQ}^{(3)}L_d^\dagger\right]_{mj}+\left[L_d w_{HQ}^{(3)\dagger}L_d^\dagger\right]_{im}\lambda^{(t)}_{mj}\right)\bigg\},
\label{eq:CVLLfinite}
\end{split}
\\
[\Delta C_{{\rm LR},1}]_{ij}&=\frac{1}{16\pi^2}y_t^2\lambda_{ij}^{(t)}\left[R_d w_{Hd}R_d^\dagger\right]_{ij}H_1(x_t,m_W),\label{eq:CVLRfinite}
\end{align}
\end{subequations}
where $x_t=\overline{m}_t^2/m_W^2$, $S_0(x)$ is the conventional Inami-Lim loop function~\cite{Inami:1980fz} and $H_{1,2}(x,\mu)$ are the remaining loop functions~\cite{Bobeth:2017xry}
\begin{subequations}
\begin{align}
H_1(x,\mu)&=\log\frac{\mu}{m_W}-\frac{x-7}{4(x-1)}-\frac{x^2-2x+4}{2(x-1)^2}\log x,\\ 
H_2(x,\mu)&=\log\frac{\mu}{m_W}+\frac{7x-25}{4(x-1)}-\frac{x^2-14x+4}{2(x-1)^2}\log x.
\end{align}
\end{subequations}
The finite contribution to $[\Delta C_{\rm VLL}]_{ij}$ in the second line of Eq.~(\ref{eq:CVLLfinite}) is from the top box diagram of the SM calculation including the contributions of the gears to the CKM matrix elements, Eq.~(\ref{eq:dgWgears}). 

Our results agree with the tree level calculation in Ref.~\cite{delAguila:2000rc} apart from a minus sign in the contributions from diagram \textit{(c)} in Fig.~\ref{fig:flavormix}. The one loop results agree with the specific cases of vector-like quarks calculated in Ref.~\cite{Bobeth:2016llm}, i.e.,  for the box-diagrams in Fig.~\ref{fig:flavormixloop} for  gears  of the $Q_L$- and $d_R$-type, and with the contributions calculated in Ref.~\cite{Ishiwata:2015cga} for gears of the $u_R$-type. 

\bibliographystyle{apsrev}
\bibliography{Clockwork_ref}

\end{document}